\documentclass{article}
\pdfoutput=1
\usepackage{jcappub}
\title{Perturbation theory approach for the power spectrum: from dark matter in real space to massive haloes in redshift space}

\author[a,b]{H\'ector Gil-Mar\'in,}
\author[b]{Christian Wagner,}
\author[b,c,d]{Licia Verde,}
\author[e]{Cristiano Porciani}
\author[b,c,d]{and Raul Jimenez}

\affiliation[a]{Institut de Ci\`encies de l'Espai (ICE), Facultat de Ci\`encies , Campus UAB (IEEC-CSIC), Bellaterra E-08193, Spain}
\affiliation[b]{Institut de Ci\`encies del Cosmos (ICC), Universitat de Barcelona (IEEC-UB), Mart\'i i Franqu\`es 1, Barcelona E-08028, Spain}
\affiliation[c]{ICREA Instituci\'o Catalana de Recerca i Estudis Avan\c{c}ats. Passeig Llu\'is Companys 23, E-08010 Barcelona, Spain}
\affiliation[d]{Theory Group, Physics Department, CERN, CH-1211, Geneva 23, Switzerland}
\affiliation[e]{Argelander Institut f\"ur Astronomie der Universit\"at Bonn, Auf dem H\"ugel 71, D-53121 Bonn, Germany}

\emailAdd{hectorgil@icc.ub.edu,\, cwagner@icc.ub.edu, \, liciaverde@icc.ub.edu, \,porciani@astro.uni-bonn.de, \,raul.jimenez@icc.ub.edu, }

\abstract{We investigate the accuracy of Eulerian perturbation theory for describing the matter and galaxy power spectra in real and redshift space in light of future observational probes for precision cosmology. Comparing the analytical results with a large suite of N-body simulations (160 independent boxes of 13.8 (Gpc/$h)^3$ volume each, which are publicly available), we find that re-summing terms in the standard perturbative approach predicts the real-space matter power spectrum with an accuracy of  $\lesssim  2\%$ for $k\leq 0.20\,h/\mbox{Mpc}$ at redshifts $z\lesssim 1.5$.  This is obtained following the widespread technique of writing the resummed propagator in terms of 1-loop contributions. We show that the accuracy of this scheme increases by considering higher-order terms in the resummed propagator.
By combining  resummed perturbation theories with several  models for the mappings from real to redshift space discussed in the literature,   the multipoles of the dark-matter power spectrum can be described with sub-percent deviations from N-body results for  $k\leq0.15\,h/\mbox{Mpc}$ at $z\lesssim 1$. As a consequence, the logarithmic growth rate, $f$, can  be recovered with sub-percent accuracy on these scales.
Extending the models to massive dark-matter haloes in redshift space, our results describe the monopole term from N-body data within $2\%$ accuracy for scales $k\leq0.15\,h/{\rm Mpc}$ at $z\lesssim0.5$;  here $f$ can  be recovered within $< 5\%$ when the halo bias is known.  We conclude that these techniques are suitable to extract cosmological information from future galaxy surveys.}
\begin{document}

\maketitle


\section{Introduction}

Galaxy clustering is a key observational probe to study the large-scale structure of the Universe. The shape of the galaxy power spectrum, bispectrum and higher-order moments,  contain information about the matter content of the Universe, about gravity and also about possible non-Gaussian initial conditions. In galaxy surveys, the galaxy distribution is distorted along the line of sight due to peculiar velocities that cause Doppler shifts, namely redshift space distortions (RSD). As these distortions depend on the growth of structure, they 
offer a complementary technique (to studies of the cosmic expansion history) to measure the matter content or to test gravity e.g., \citep{DETF,Euclidreview,rsd_fR}. On very large scales and at higher redshifts, the RSD can be described by linear theory. However, on smaller scales and at later epochs, non-linearities start to play an important role and must be taken into account to accurately estimate the cosmological parameters from observational data.

Current surveys like BOSS\footnote{Baryon Oscillation Spectroscopic Survey \url{http://www.sdss3.org/surveys/boss.php}} \cite{sdss3}, and future missions like EUCLID\footnote{\url{http://sci.esa.int/euclid/}}\cite{Euclid} will soon provide unprecedented datasets about the distribution of galaxies on large scales. In order to extract useful information from data of this quality we need more accurate theoretical models of structure formation. Standard perturbation theory (SPT) is the straightforward way of proceeding and has been the workhorse in the field for decades. However, its practical applications to statistics of the density field provide limited accuracy in both real and redshift space. A number of studies have investigated the fidelity of different theoretical models to describe the redshift space distortions and the possibility of extracting cosmological information from them  \cite{vlah,nishimichi_taruya, okumura1, okumura2, beth, de_la_torre, kwan,matsubara11,okumura_jing,jennings11,nico,alan98,licia98,taruya09,nishimichi07,bianchi}.  This work explores the potential of describing the redshift-space distortions combining different mappings between real and redshift space with resummed perturbation theories. This is the basis to model  the redshift-space distorted power spectrum of dark matter and dark-matter haloes.  We then focus on the systematic errors that these models produce when they recover the logarithmic growth rate. For future surveys with  forecasted statistical error on this quantity at the \% level, the accuracy of the analytic description of clustering must be such that residual systematic errors due to modeling is well below this level.  A fast, analytic description of clustering, calibrated on N-body simulations is an approach  fully complementary to one based entirely on N-body simulations e.g.,\cite{coyote1, coyote3}.
 
In particular, we start by describing the real-space power spectrum using 1- and 2-loop standard perturbation theory and the so-called renormalized (or resummed) perturbation theory \citep{CS06A,CS06B}. We combine these models with different methods to obtain the redshift-space power spectrum: i) the Kaiser model  \cite{kaiser}, ii) the Scoccimarro model \cite{SC04} and iii) the Taruya et al. model \cite{taruya_model} (TNS model, from now on). For dark matter, we additionally account for 
the Finger-of-God (FoG) effect produced by virial motions on small scales by introducing a phenomenological damping term. This term is not needed in the case of haloes, as we only consider isolated haloes which are not part of a larger host halo.
We compare these different methods with the results of a large suite of N-body simulations, focusing on the multipole expansion of the redshift-space power spectrum. Our suite of N-body simulations sums up to a volume of 2212 (Gpc/$h)^3$, which is much larger than the volume surveyed by any current or planned experiment, ensuring that statistical errors in the simulations are negligible. 

This paper is organized as follows. In \S \ref{theory_section} we present the basic theory of redshift-space distortions. In particular, in \S \ref{section_real_space} we review standard perturbation theory and renormalized perturbation theory, extending the latter to higher-order propagators than previously considered. In \S\ref{section_redshift_space} we present different RSD models while Finger-of-God effects are discussed in \S \ref{FoG}. In \S \ref{sims_section} we provide details of the N-body simulations and describe the halo catalogues used in this paper. In \S\ref{results_section} we compare the results of our models for the real-space power spectrum and compare them with N-body simulations for the dark-matter case. We also consider the  RSD models mentioned above, focusing on their capacity to recover the logarithmic growth rate $f$, both for dark matter and massive haloes. Finally in \S\ref{discussion_section}, we discuss and summarize the obtained results.
Appendices \ref{A} and \ref{B} contain details about standard and resummed perturbation theory providing a justification
of the formulae presented in \S\ref{section_real_space}. Note that our discussion of resummed theories does not assume a field-theory background and is supposed to be accessible to all readers.

\section{Theory}\label{theory_section}

The matter-matter real-space power spectrum $P_{\delta\delta}(k)$, the Fourier transform of the two-point correlation function, is the simplest statistic of interest one can extract from the dark matter overdensity field in Fourier space, $\delta({\bf k})$,
\begin{equation}
 \langle \delta({\bf k})\delta({\bf k'})\rangle \equiv (2\pi)^3 \delta^D({\bf k+k'}) P_{\delta\delta}(k)\,,
\label{Pk}
\end{equation}
where $\delta^D$ denotes the Dirac delta function and $\langle\dots \rangle$ the ensemble average. 
Under the assumption of an isotropic Universe, the power spectrum in real space does not depend on the direction of the $\bf k$-vector. Since we only have one observable Universe, under the hypothesis of ergodicity the average $\langle\dots \rangle$ can be  taken over all different directions  for each $\bf k$-vector. 

The mapping between the radial coordinate in real space and the radial coordinate in redshift space is given by the Hubble flow and the Doppler effect due to the peculiar velocities, $\bf v$. Since only the radial distance is computed from the measured redshift, the two angular coordinates remain the same in both real and redshift space. In this paper, we adopt the distant observer approximation, i.e., we assume that all line-of-sights are virtually parallel to each other. If we identify this direction with the third axis of our coordinate system, the mapping from real-space coordinates $\bf x$ to redshift-space coordinates $\bf s$ reads
\begin{equation}
{\bf s}={\bf x}+\frac{v_3({\bf x})}{H(a)a}\bf{\hat{x}_3} ,
\label{zs_mapping}
\end{equation}
where $H(a)$ is the Hubble parameter at the scale factor $a$ and $\bf\hat{x}_3$ denotes the unit vector of the third axis. 
Using the scaled velocity field ${\bf u}\equiv -{\bf v}/[H(a) a f(a)]$ where $f$ is the logarithmic derivative of the linear growth factor with respect to the scale factor, $f\equiv d\ln D(a) / d\ln a$, we can write this mapping as
\begin{equation}
{\bf s}={\bf x}-f u_3({\bf x}){\bf\hat{x}_3} .
\end{equation}

The 2-point correlation function in redshift space is then defined by
\begin{equation}
\xi_2^s({\bf r})= \langle\delta({\bf s}+{\bf r})\delta({\bf s})\rangle
\end{equation}
and its Fourier transform,
\begin{equation}
\label{Ps}P_{\delta\delta}^s(k,\mu)=\int d^3 {\bf r}\, \xi^s_2({\bf r})\exp(-i{\bf k}\cdot{\bf r}), 
\end{equation}
is the power spectrum in redshift space. Here, $\mu\equiv \hat{\bf k}\cdot\hat{\bf k}_3$ is the cosine of the angle between the vector $\bf k$ and the line-of-sight. The redshift space power spectrum, can also be computed from Eq. \ref{zs_mapping}. In this mapping the mass must be conserved, which implies  $[1+\delta^{(s)}({\bf s})]d^3{\bf s}=[1+\delta({\bf r})]d^3{\bf r}$. Thus the transformation from $\delta({\bf r})$ to $\delta^{(s)}({\bf s})$ at linear order in $\delta$ and $\bf v$ reads,
\begin{equation}
\delta^{(s)}({\bf s})=\delta({\bf x})-\frac{\nabla_3 v_3({\bf x})}{H(a)a} ,
\end{equation}
where $\nabla_3$ is short for $\frac{\partial}{\partial x_3}$.
The two-point correlation function in Fourier space, $\langle \delta^{(s)}({\bf k})\delta^{(s)}({\bf k}')\rangle$ is then \cite{taruya_model},
\begin{eqnarray}
\label{rsd_exact}P_{\delta\delta}^s(k,\mu)&=&\int \frac{d^3{\bf r}}{(2\pi)^3} e^{i {\bf k}\cdot{\bf r}}\langle e^{-ik\mu f \Delta u_3}\left[\delta({\bf x})+f\nabla_3 u_3({\bf x})][\delta({\bf x}')+f\nabla_3 u_3({\bf x}')\right]\rangle,
\end{eqnarray}
where ${\bf r}={\bf x}-{\bf x}'$ and $\Delta u_z=u_z({\bf x})-u_z({\bf x}')$. Note that we have written $u$ instead of $v$ to make the dependence on $f$ explicit. In Eq. \ref{rsd_exact} the enhancement and damping effect of the redshift space distortions on the power spectrum are manifest. The enhancement due to RSD, also known as Kaiser effect, is produced by the  the $ +f\nabla_3 u_3$ terms in Eq. \ref{zs_mapping}, that increase the overdensity $\delta({\bf x})$. These terms, represent the coherent distortions by the peculiar velocities along the line-of-sight direction, and are controlled by the growth factor parameter $f$. On the other hand, the damping effect comes from the exponential factor in Eq. \ref{rsd_exact}. This term is mainly due to the small scale velocity dispersion around the most clustered regions, and produces the suppression of power at small scales in the power spectrum.

\subsection{Perturbation theory in real space}\label{section_real_space}
In order to describe the non-linear matter power spectrum in redshift space we first need a theory that is able to provide an accurate description of the power spectrum in real space.  There are several models that attempt to do this task: the halo model (see \cite{halo_model} for a review), HALOFIT \citep{halofit}, cosmological standard perturbation theory \cite{bernardeau,fry94,nina,jeong_komatsu1,jeong_komatsu2,sc97,sc96,catelan95,jain_bert,matsubara2000} , and other perturbation theories approaches based on Lagrangian perturbation theory \citep{LPT,hivon95,okamura,valegas,carlson12}, time renormalization \citep{pietroni,anselmi}, Eulerian renormalized (or resummed) perturbation theories \citep{CS06A,CS06B,wang_szalay,anna,regpt,multipoint,bernardeau12} and closure theory \citep{closure} (see \cite{Carlson} for comparison of some of these theories).

In this paper we focus on two approaches: standard perturbation and resummed perturbation theory, both in Eulerian space.

{\bf Standard perturbation theory} (SPT hereafter) consists of expanding the statistics of interest as a sum of infinite terms, where every term correspond to a $n$-loop correction. For the power spectrum the SPT prediction is written as (see appendix \ref{A} for the explicit formulae of SPT terms),
\begin{equation}
\label{spt} P^{\rm{SPT}}(k)=P^{(0)}(k)+P^{(1)}(k)+P^{(2)}(k)+\dots.
\end{equation}
The 0-loop term correction is just the linear power spectrum,  $P^{(0)}(k)=P^{\rm lin}(k)$.  The 1-loop term is expressed as a sum of 2 different subterms,
\begin{equation}
 P^{(1)}(k)=2P_{13}(k)+P_{22}(k),
\end{equation}
where the subscripts $i$ and $j$ refer to the perturbative order of the terms $\delta({\bf k})$ used
in eq. \ref{Pk} to compute the power spectrum $P_{ij}(k)$. 
In this case, both $P_{13}$ and $P_{22}$ requires a 2-dimensional integration (after exploiting rotational invariance). The 2-loop term is the sum of three different subterms,
\begin{equation}
P^{(2)}(k)=2P_{15}(k)+2P_{24}(k)+P_{33}(k).
\end{equation}
In this case, all these three terms, require a 5-dimensional integration (after exploiting rotational invariance). One can keep going to higher-order terms. However, the 3-loop correction term requires the computation of 8-dimensional integrals and the 4-loop correction term 11-dimensional integration. For practical and computational reasons, one does not usually go beyond the 2-loop correction terms. Thus, one has to truncate Eq. \ref{spt} series at some loop order. Truncating it at $P^{(1)}$ and at $P^{(2)}$ term is what is respectively called in this paper, 1L-SPT and 2L-SPT.

{\bf Renormalized perturbation theory} (RPT hereafter) attempts to reorganize the perturbative series expansion of SPT and resum some of the terms into a function that can be factorized out of the series. This function is usually called {\it the resummed propagator} and we refer to it as ${\cal N}$. In order to make the resummation possible, all the kernels of the $P^{(\ell)}$ terms have to be expressed as a product of kernels that correspond to full-mode-coupling terms and full-propagator terms (see appendix \ref{B} for details). The full-mode-coupling kernels are those kernels contained in $P_{nn}(k)$ terms that contain a coupling of the form, ${\bf k}-{\bf q}_1-\ldots-{\bf q}_{n-1}$ (see Eq. \ref{n_term}). The full-propagator kernels are those contained in $P^{(\ell)}$ terms of the form $P_{1n}$ with no-mode coupling term (see Eq. \ref{P13} and \ref{P15} as examples). The resulting expression of resumming terms in this way is (see appendix \ref{B} and Refs. \cite{CS06A,CS06B} for a full derivation),
\begin{equation}
\label{RPT}P^{\rm RPT-{\cal N}_i}(k,z)=\left[P^{\rm lin}(k,z)+P_{22}(k,z)+P_{33}^{2L}(k,z)+\ldots+ P_{nn}^{(n-1)L}(k,z) +\ldots \right]{\cal N}_i(k)^2,
\end{equation}
where the term $P_{nn}^{(n-1)L}$ is the part of the $P_{nn}$ term that describes a full-mode coupling (see Eq. \ref{n_term}). In spite of the resummation, Eq. \ref{RPT} contains an infinite series as Eq. \ref{spt} and has to be truncated after a certain number of loops. However, some of the infinite terms of Eq. \ref{spt} have now been reorganized into the ${\cal N}_i$ function. We will refer as 1L-RPT-${\cal N}_i$ and 2L-RPT-${\cal N}_i$ the truncation of Eq. \ref{RPT} at 1- and 2-loop, respectively. Finally, the form of the function ${\cal N}_i$ depends on the way we approximate the kernels in the resummation process. In the case the kernels are approximated according to the Zel'dovich approximation (see Eq. \ref{zeldovich}), they are expressed as a product of 0-loop propagators and the resulting function ${\cal N}_0$ is,
\begin{equation}
\label{theory_N0}{\cal N}_0(k)\equiv\exp\left[-\frac{1}{2}k^2\sigma_v^2\right],
\end{equation}
with,
\begin{equation}
\sigma_v^2\equiv\frac{4\pi}{3}\int \frac{dq}{(2\pi)^3}\, P^{\rm lin}(q).
\end{equation}
When the kernels are approximated as a product of 1-loop propagator kernels (see Eq. \ref{N1_kernela}-\ref{N1_kernelc}) the resulting ${\cal N}_1$ function is,
\begin{equation}
\label{theory_N1}{\cal N}_1(k)\equiv\exp\left[P_{13}(k)/P^{\rm lin}(k)\right].
\end{equation}
This is the expression presented in \cite{MPTbreeze} (see Eq. \ref{f_sc} for the correspondence).
Expressing the kernels as a product of 2-loop propagators, (see Eq. \ref{N2_kernela}-\ref{N2_kernelf}) yields the function ${\cal N}_2$, 
\begin{equation}
\label{theory_N2}{\cal N}_2(k)\equiv \cosh\left[  \sqrt{ \frac{2P_{15}(k)}{P^{\rm lin}(k)} }     \right] + \frac{P_{13}(k)}{P^{\rm lin}(k)}\sqrt{\frac{P^{\rm lin}(k)}{2P_{15}(k)}}\sinh\left[  \sqrt{ \frac{2P_{15}(k)}{P^{\rm lin}(k)} }     \right].
\end{equation}
Note that the angular part of the propagator terms, $P_{13}$, $P_{15}$, $P_{17}$,\footnote{$P_{17}$ is required for the 3-loop expansion of the resummed propagator, ${\cal N}_3$ (see Eq. \ref{N3}).}\ldots is analytically integrable for any shape of the power spectrum. Thus, the 2-dimensional integration of $P_{13}$ can be reduced to a 1-dimensional integration (see Eq. \ref{f_sc}). In the same way, the 5- and 8-dimensional integrations of the terms $P_{15}$ and $P_{17}$, are reducible to 2- and 3-dimensional integration. However this is a hard task due to the symmetrized kernels, which are constructed as a sum of $5!=120$ and $7!=5040$ terms respectively (see Eq. \ref{sym_kernels}). Because of that, we stop at 2-loop. In this paper, the ${\cal N}$-function is computed considering $P_{15}$ as a 5-dimensional integral. If more accuracy is needed, the 3-loop resummed propagator written in Eq. \ref{N3} can be used. These extensions in the resummed propagator could be easily incorporated in the current public codes for the RPT \citep{MPTbreeze,regpt,Carlson}.

\subsection{Perturbation theory in redshift space} \label{rsd_section}\label{section_redshift_space}
In order to describe the non-linear matter power spectrum in redshift space (Eq. \ref{Ps}) we need a model that, given the  non-linear power spectrum in real space, is able to ``map it'' to the power spectrum in redshift space. There are several models that attempt to do this task.
While in principle the same perturbation theory approach used for the dark matter  could be employed to model also the velocity and the density velocity coupling yielding therefore a real-to-redshift space mapping, it has become clear in the literature that  the redshift space  clustering and in particular the redshift space  power spectrum  is not well described perturbatively.  In fact, highly non-linear scales are superimposed to linear scales by the real-to-redshift space mapping, as realized in the seminal papers by \cite{kaiser,davis_peebles}. Some of this effect is  even visually apparent in the galaxy distribution as the so-called ``Fingers of God" effect. In this paper we consider physically motivated, but phenomenological models, we study the Kaiser model \cite{kaiser}, the Scoccimarro model\cite{SC04} and the TNS model \cite{taruya_model}. All these models propose a functional form of $P^s(k)$ that depends on real-space statistics. 

The simplest model, is the {\bf Kaiser model} proposed by Nick Kaiser 25 years ago \citep{kaiser},
\begin{equation}
\label{kaiser_eq}P^s(k,\mu)=(b(k)+f^2\mu)^2P_{\delta\delta}(k),
\end{equation}
where $b(k)$ is a possibly scale-dependent biasing function, which relates the observable tracers to the underlying dark matter distribution.
This expression is obtained when Eq. \ref{rsd_exact} is treated linearly. Because of that, in principle $P_{\delta\delta}$ should refer to the {\it matter-matter} {\it linear} power spectrum. However, the prescription $P_{\delta\delta}^{\rm lin}\rightarrow P^{\rm nl}_{\delta\delta}$ has been demonstrated to work better. With this recipe, the Kaiser model is usually known as `non-linear Kaiser'. In this paper we refer to Eq. \ref{kaiser_eq} with a non-linear $P_{\delta\delta}$ simply as Kaiser model.

The {\bf Scoccimarro model} proposes that the redshift-space power spectrum is given by \citep{SC04},
\begin{equation}
\label{SC}P^s(k,\mu)=\left[b(k)^2 P_{\delta\delta}(k)+2b(k)\mu^2 f P_{\delta \theta}(k)+ f^2\mu^4 P_{\theta \theta}(k)\right],
\end{equation}
where $P_{\delta\theta}$ and $P_{\theta\theta}$ are the {\it velocity-matter} and {\it velocity-velocity} power spectra in real space, respectively, and are defined by
\begin{eqnarray}
 \langle \delta({\bf k})\theta({\bf k'})\rangle &\equiv& (2\pi)^3 \delta^D({\bf k+k'}) P_{\delta\theta}(k)\,,\\
 \langle \theta({\bf k})\theta({\bf k'})\rangle &\equiv& (2\pi)^3 \delta^D({\bf k+k'}) P_{\theta\theta}(k)\,,
\end{eqnarray}
where $\theta({\bf k})\equiv[-i {\bf k}\cdot{\bf v({\bf k})}]/[af(a)H(a)]$.
Note that Eq. \ref{SC} tends to \ref{kaiser_eq} when $P_{\delta\theta}$ and $P_{\theta\theta}$ tend to $P_{\delta\delta}$. This is the case for the linear regime in SPT.

The {\bf TNS model} \citep{taruya_model} takes into account the cross interaction due to linear and non-linear processes.  This produces two extra terms to Eq. \ref{SC},
\begin{equation}
\label{taruya_eq}P^s(k,\mu)=\left[b(k)^2 P_{\delta\delta}(k)+2b(k)\mu^2 f P_{\delta \theta}(k)+ f^2\mu^4 P_{\theta \theta}(k)+A(k,\mu,b)+B(k,\mu,b)\right].
\end{equation}
where the $A$ and $B$ terms, arise from the interaction between the enhancement terms and the damping terms in Eq. \ref{rsd_exact}. Also \cite{zhang} have provided a complementary explanation and a possible generalization of the $A$ and $B$ terms. However, in this paper we use the basic formalism presented by  \cite{taruya_model},
\begin{eqnarray}
\label{taruya_a}A(k,\mu,b)&=& (k\mu f)\int \frac{d^3{\bf q}}{(2\pi)^3}\frac{q_z}{q^2}\left\{B_\sigma({\bf q}, {\bf k}-{\bf q}, -{\bf k})-B_\sigma({\bf q},{\bf k}, -{\bf k}-{\bf q})\right\}, \\
\label{taruya_b}B(k,\mu,b)&=&(k\mu f)^2\int\frac{d^3{\bf q}}{(2\pi)^3}F({\bf q})F({\bf k}-{\bf q}),
\end{eqnarray}
where,
\begin{equation}
F({\bf q})\equiv\frac{q_z}{q^2}\left\{b(q)P_{\delta\theta}(q)+f\frac{q_z^2}{q^2}P_{\theta\theta}(q)\right\},
\end{equation}
and
\begin{eqnarray}
\nonumber(2\pi)^3\delta_D({\bf k}_{123})B_\sigma({\bf k}_1,{\bf k}_2,{\bf k}_3)\equiv\left\langle \theta({\bf k}_1)\left\{b(k_2)\delta({\bf k}_2)+f\frac{k_{2z}^2}{k_2^2}\theta({\bf k}_2)\right\}\left\{b(k_3)\delta({\bf k}_3)+f\frac{k_{3z}^2}{k_3^2}\theta({\bf k}_3)\right\} \right\rangle,\\
\end{eqnarray}
with ${\bf k}_{123}\equiv{\bf k}_1+{\bf k}_2+{\bf k}_3$. Since the functions $A$ and $B$ require an integration over the whole range of momenta $q$, we cannot use RPT predictions to compute them. Furthermore, $A$ requires the crossed bispectra between $\delta$ and $\theta$, whose computation at 1- and 2-loop requires more effort than in the power spectrum case. Since we expect $A$ and $B$ to be small compared to $P_{\delta\delta}$, $P_{\delta\theta}$ and $P_{\theta\theta}$ \citep{taruya_model}, in this paper we compute $A$ and $B$ assuming the leading terms for the power spectrum and bispectrum inside the integrals of Eq. \ref{taruya_a} and \ref{taruya_b}. For the rest of the power spectrum terms of Eq. \ref{kaiser_eq}, \ref{SC} and \ref{taruya_eq} we use the perturbative approaches described in section \ref{section_real_space}.

None of these models is able to account for non-linear effects such as Fingers of God. These effects have to be included {\it ad hoc} through a function that damps the power spectrum in redshift space at small scales.

\subsection{Fingers of God}\label{FoG}
The effects of small-scale velocities are not completely included in the models presented in section \ref{rsd_section}. Both the Kaiser and Scoccimarro model ignore them as we have already commented. Only the TNS model takes them into account, but only as a cross term with large-scale squashing. Thus, the effect of these small-scales velocities has to be inserted as a multiplicative damping function into the models described in section \ref{rsd_section} \citep{fry84,davis_peebles,jackson}, 
\begin{equation}
\label{fog_changing}P^s(k,\mu;z)\rightarrow P^s(k,\mu;z)D^2_{\rm FoG}(k,\mu;z).
\end{equation}
The most used prescriptions for that, are the Gaussian and the Lorentzian functions that both depend on a redshift-dependent parameter, $\sigma(z)\equiv\sigma_0D(z)$,
\begin{eqnarray}
\label{lorentzian}D^{\rm{Lor}}_{\rm FoG}(k,\mu,z;\sigma_0)&=&\frac{1}{1+0.5[k\mu\sigma_0 D(z)f(z)]^2}\quad\mbox{Lorentzian},\\
D^{\rm{Gau}}_{\rm FoG}(k,\mu,z;\sigma_0)&=&\exp\left\{ -0.5[k\mu\sigma_0 D(z)f(z)]^2 \right\}\quad\mbox{Gaussian}.
\end{eqnarray}
where $\sigma_0\equiv\sigma(z=0)$. Theoretically, it has been suggested \cite{SC04} that $\sigma(z)$ can be computed analytically as,
\begin{equation}
\label{sigmav}\sigma^2(z)=\frac{4\pi}{3}\int \frac{dq}{(2\pi)^3}\, P_{\theta\theta}(q,z).
\end{equation}
Such a parameter is physically motivated as it tries to enclose the effect of the velocity dispersion of dark matter particles.  Similar modeling was discussed e.g. in \cite{PD94} and in \cite{SC04}, although the two have different interpretations. In any case, the fit and theoretical  numerical  value for $\sigma$ need not to coincide with the actual  value of the  velocity dispersion. The theoretical value is computed in the linear approximation and the fit value is obtained using a Gaussian approximation  while the actual velocity dispersion is highly non-gaussian   even on large scales.  More discussion on this in \cite{SC04}. In most of the cases, this modeling has shown a very poor agreement with N-body simulation results. Therefore, in this paper we always treat $\sigma_0$ as a free parameter to be fit from N-body simulations.

\section{Simulations}\label{sims_section}

The simulations used in this paper model the structure formation on very large scales within a flat $\Lambda$CDM cosmology consistent with current observational data. The adopted cosmological parameters are: $\Omega_\Lambda=0.73$, $\Omega_m=0.27$ $h=0.7$, $\Omega_bh^2=0.023$, $n_s=0.95$ and $\sigma_8(z=0)\approx0.8$. Our suite of simulations consists of 160 independent runs with $768^3$ particles in a box of length $2.4 \, h^{-1}\mbox{Gpc}$. Hence, each box has a volume of $V_{\rm 1 box}=13.8\, ({\rm Gpc}/h)^3$ and in total, we simulate about $2,200\, ({\rm Gpc}/h)^3$.

The initial conditions were generated at the starting redshift $z=19$ by displacing the particles according to the second-order Lagrangian PT from their initial grid points. The initial power spectrum of the density fluctuations was computed with CAMB \cite{camb}. The simulations were performed with the GADGET-2 code \citep{gadget}
taking only the gravitational interaction into account. 

In this paper, we consider snapshots at $z=0,\, 0.5,\, 1,\, 1.5$. In order to obtain the dark-matter field from particles we discretize each box using $512^3$ grid cells. Thus the size of the Cartesian mesh is 4.68 $h^{-1}$Mpc. We assign mass to the cells using the cloud-in-cell prescription. 
Using a much higher resolution simulation, we checked that the power spectrum derived from the simulation data is accurate at the 1\% level up to $k < 0.2\,h\mbox{Mpc}^{-1}$.

To identify the dark-matter haloes, we used the Amiga Halo Finder \citep{amiga1,amiga2}, which defines a halo by the bound dark-matter particles inside a spherical overdensity equal to the so-called virial overdensity. 
We only consider haloes which are at least resolved by 40 particle. This leads to a minimum halo mass of $10^{14}M_\odot/h$.

The errors associated to the statistical quantities measured from the simulations are obtained from the dispersion among the 160 runs: we report the error on the mean of the independent runs. We make the dark matter and halo power spectra publicly available and also provide the multipoles measured from the simulations used in this paper for possible comparisons\footnote{\url{http://icc.ub.edu/~hector/Hector\_Gil\_Marin/Public.html}}.

\section{Results}\label{results_section}
In this section we compare and test the theoretical formalism presented in \S \ref{theory_section} (for both real and redshift space) with the N-body simulation data. We start with perturbation theory in real space, and we follow the theoretical predictions of different models for redshift-space power spectrum: Kaiser (Eq. \ref{kaiser_eq}), Scoccimarro (Eq. \ref{SC}) and TNS model (Eq. \ref{taruya_eq}). In particular, we consider here 1- and 2-loop SPT, and 2-loop RPT with ${\cal N}_1$ and ${\cal N}_2$. We focus on the multipole prediction of these models but also on the accuracy on recovering the logarithmic growth rate $f$. We explore this for both dark matter and for massive dark matter haloes.

\subsection{Performance of the perturbation theory approach: comparison to N-body simulations}\label{section_real_space2}
In this subsection we test the formalism presented in section \S \ref{section_real_space}, both SPT and RPT, and we compare them with the outcome of the N-body simulations.

\begin{figure}

\centering
\includegraphics[clip=false, trim= 20mm 0mm 30mm 0mm,scale=0.31]{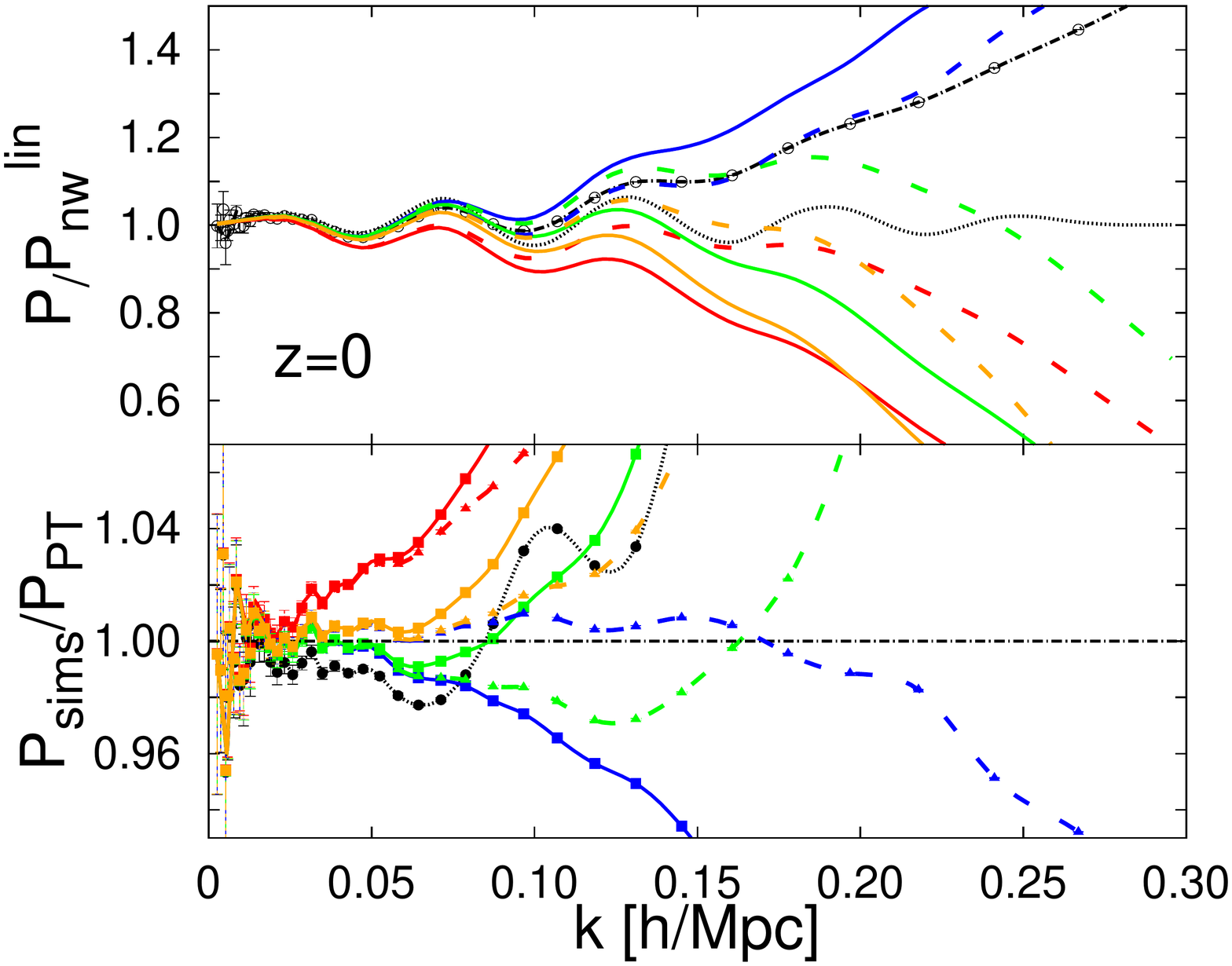}
\includegraphics[clip=false, trim= 20mm 0mm 30mm 0mm,scale=0.31]{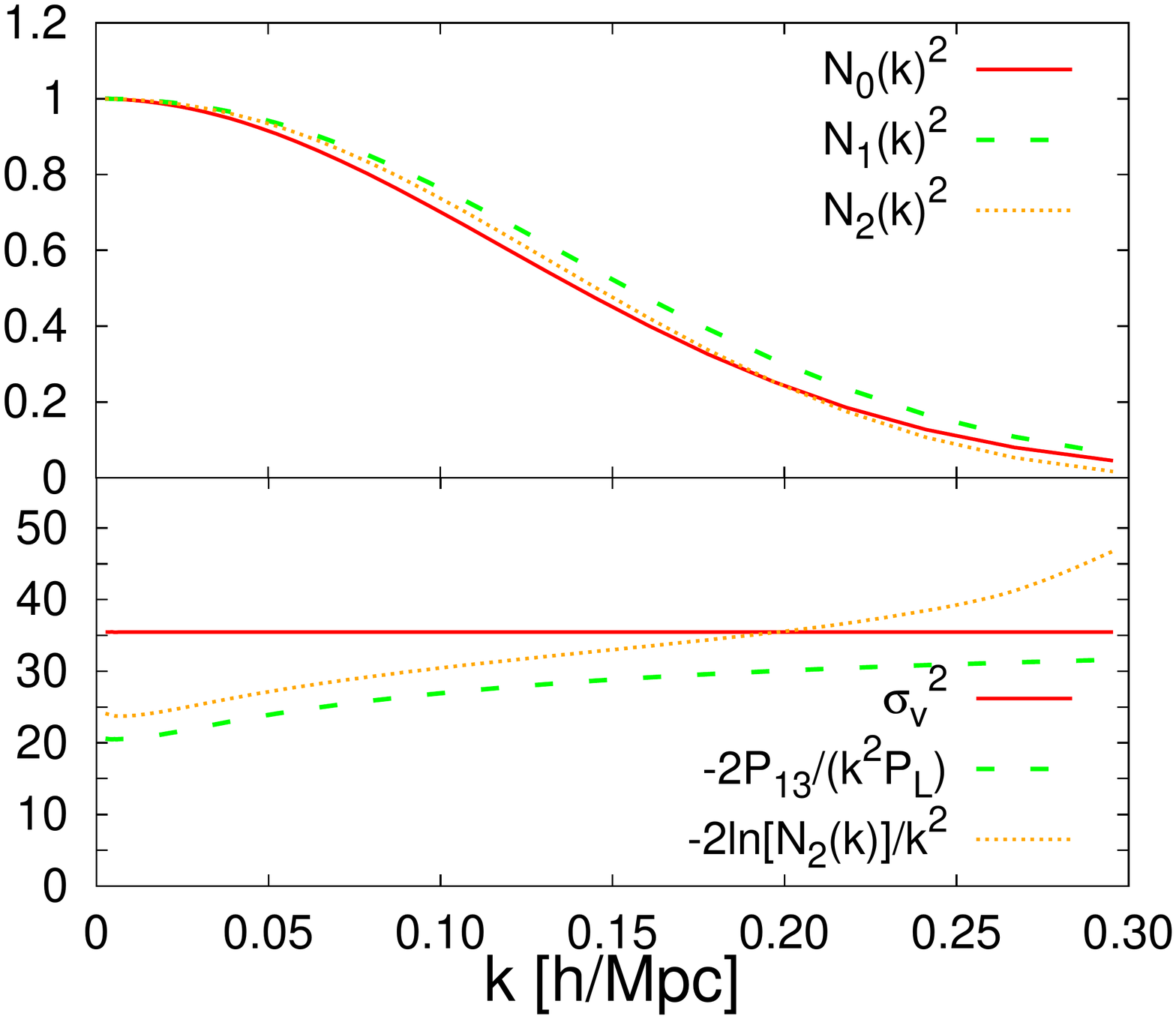}

\caption{{\it Left top subpanel}: power spectrum at $z=0$ normalized to the linear non-wiggle power spectrum to reduce the dynamic range. N-body data in black circles and dot-dashed line. Different theoretical models are also shown: linear prediction (black dotted lines), SPT model (blue lines), RPT-${\cal N}_0$ model (red lines),  RPT-${\cal N}_1$ model (green lines) and RPT-${\cal N}_2$ model (orange lines); for 1-loop truncation (solid lines) and 2-loop truncation (dashed lines). {\it Left bottom subpanel}: ratio between the power spectrum of N-body simulation and different PT models with the same color notation. {\it Right top subpanel}: ${\cal N}_0(k)^2$ (red solid line),  ${\cal N}_1(k)^2$ (green dashed line) and ${\cal N}_2(k)^2$ (orange dotted line) at $z=0$. {\it Right bottom subpanel}: $\sigma_v^2$ (red solid line), $-2P_{13}(k)/[k^2P^{\rm lin}(k)]$ (green dashed line) and $-2\ln[{\cal N}_2(k)]/k^2$ (orange dotted line) in units of $(\mbox{Mpc}/h)^2$.}
\label{FigAz0}
\end{figure}

\begin{figure}

\centering
\includegraphics[clip=false, trim= 20mm 0mm 30mm 0mm,scale=0.31]{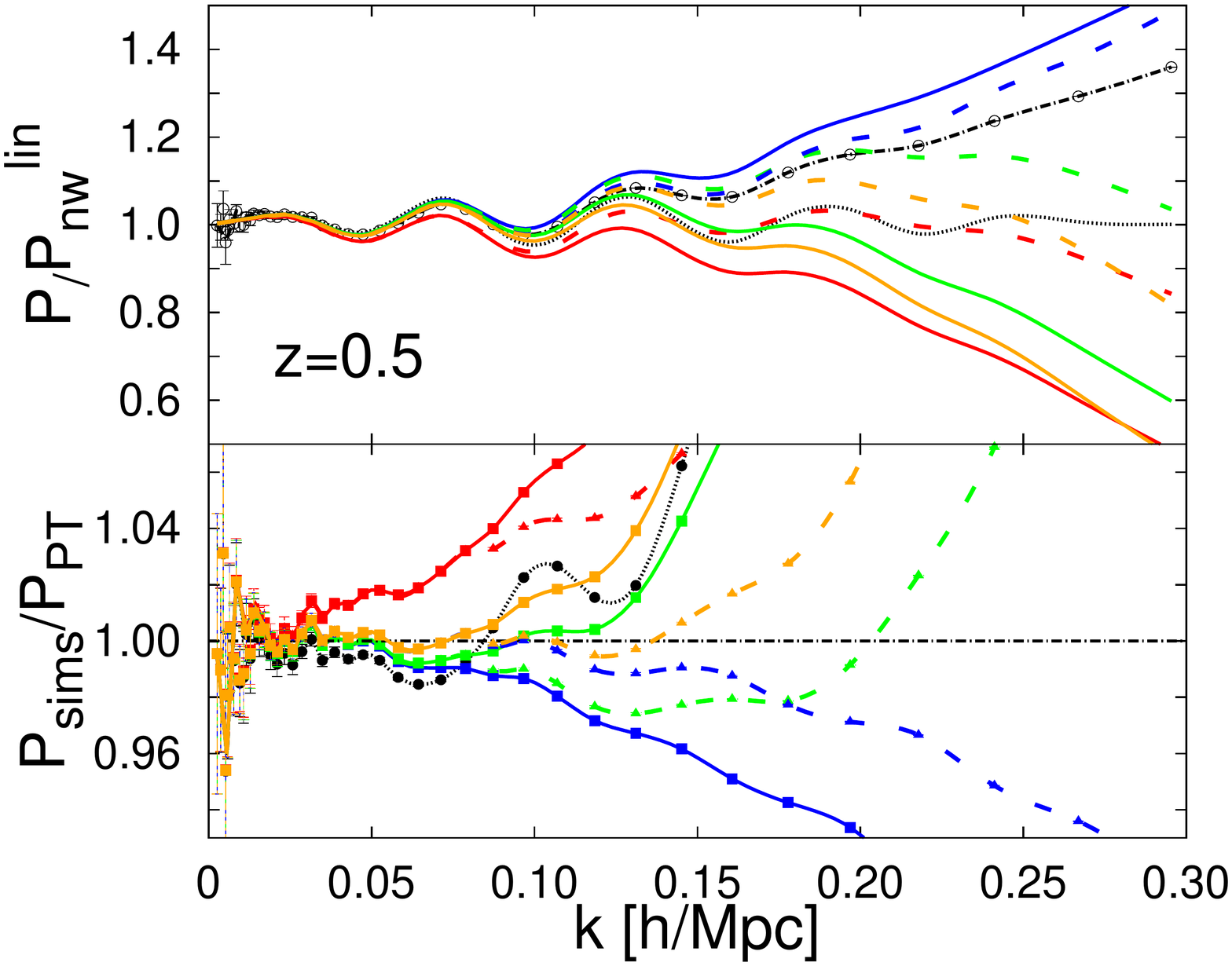}
\includegraphics[clip=false, trim= 20mm 0mm 30mm 0mm,scale=0.31]{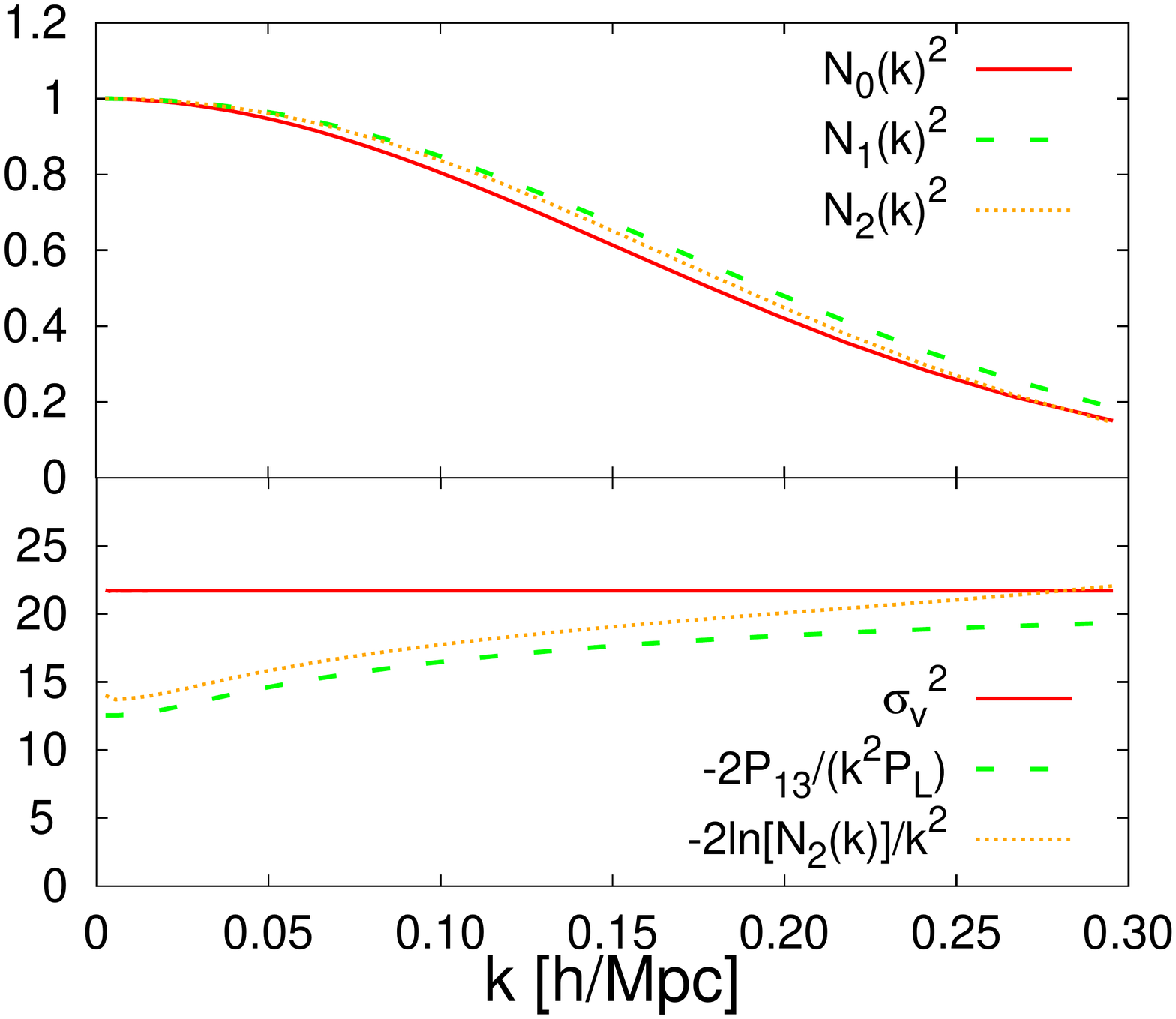}

\includegraphics[clip=false, trim= 20mm 0mm 30mm 0mm,scale=0.31]{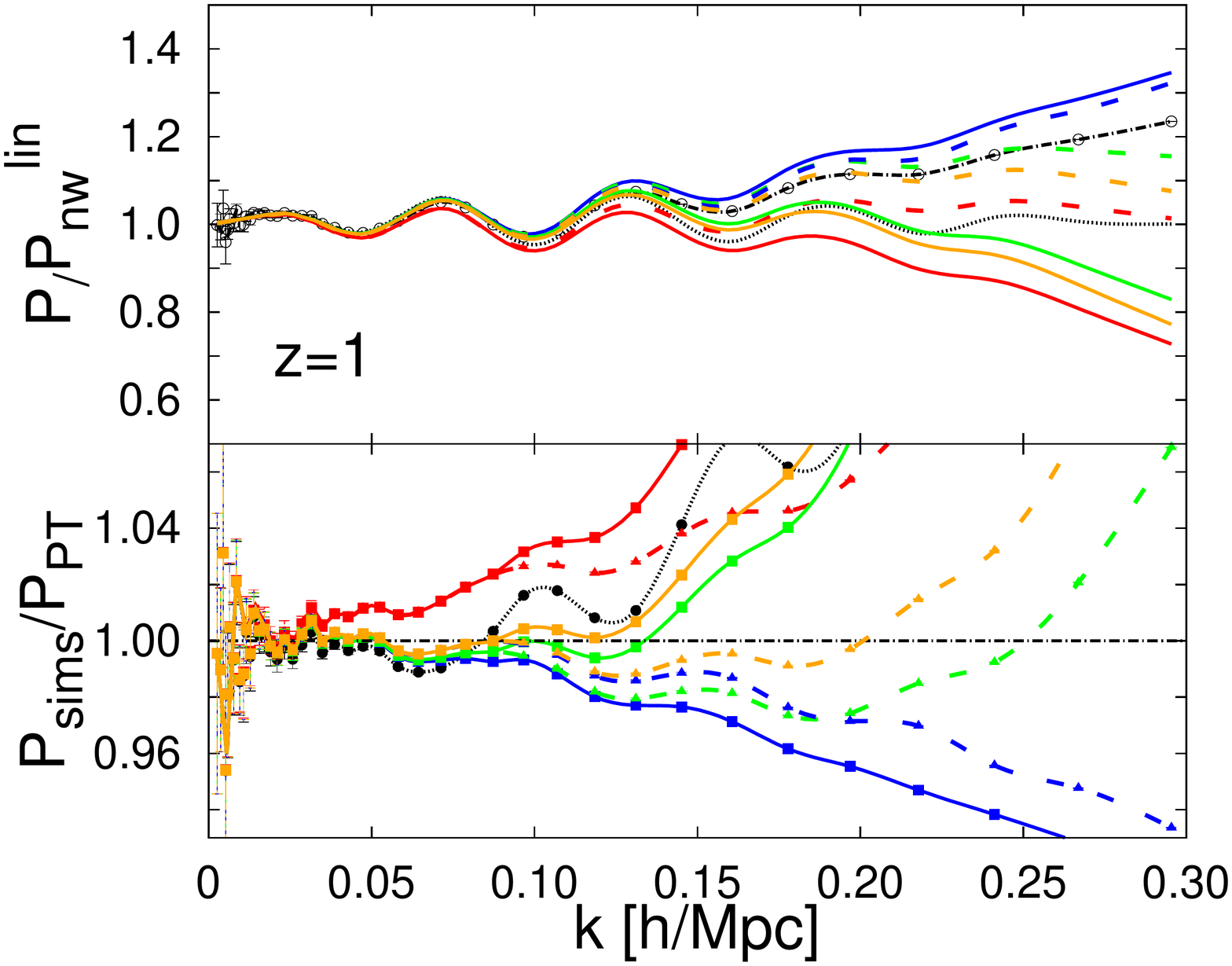}
\includegraphics[clip=false, trim= 20mm 0mm 30mm 0mm,scale=0.31]{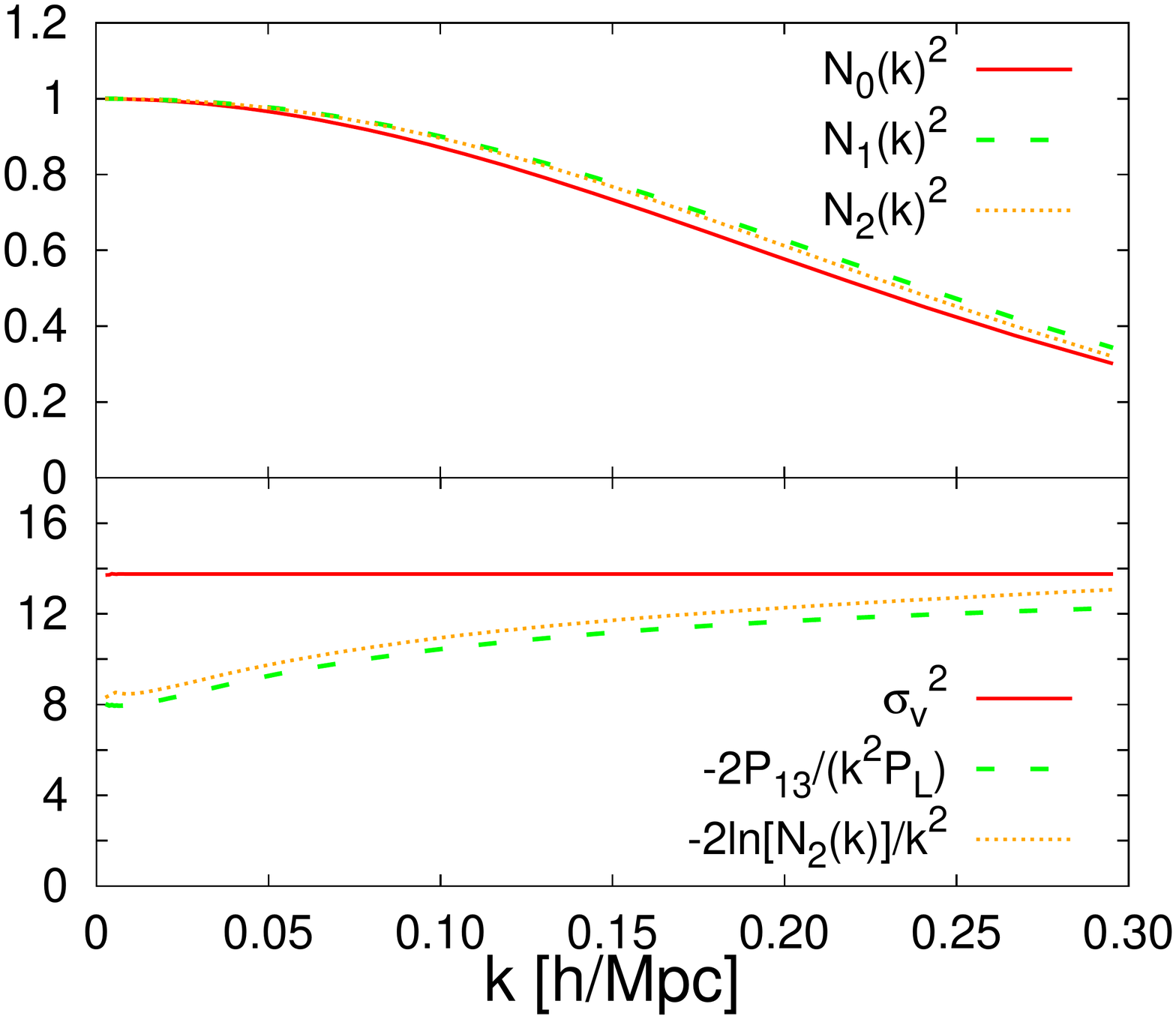}

\includegraphics[clip=false, trim= 20mm 0mm 30mm 0mm,scale=0.31]{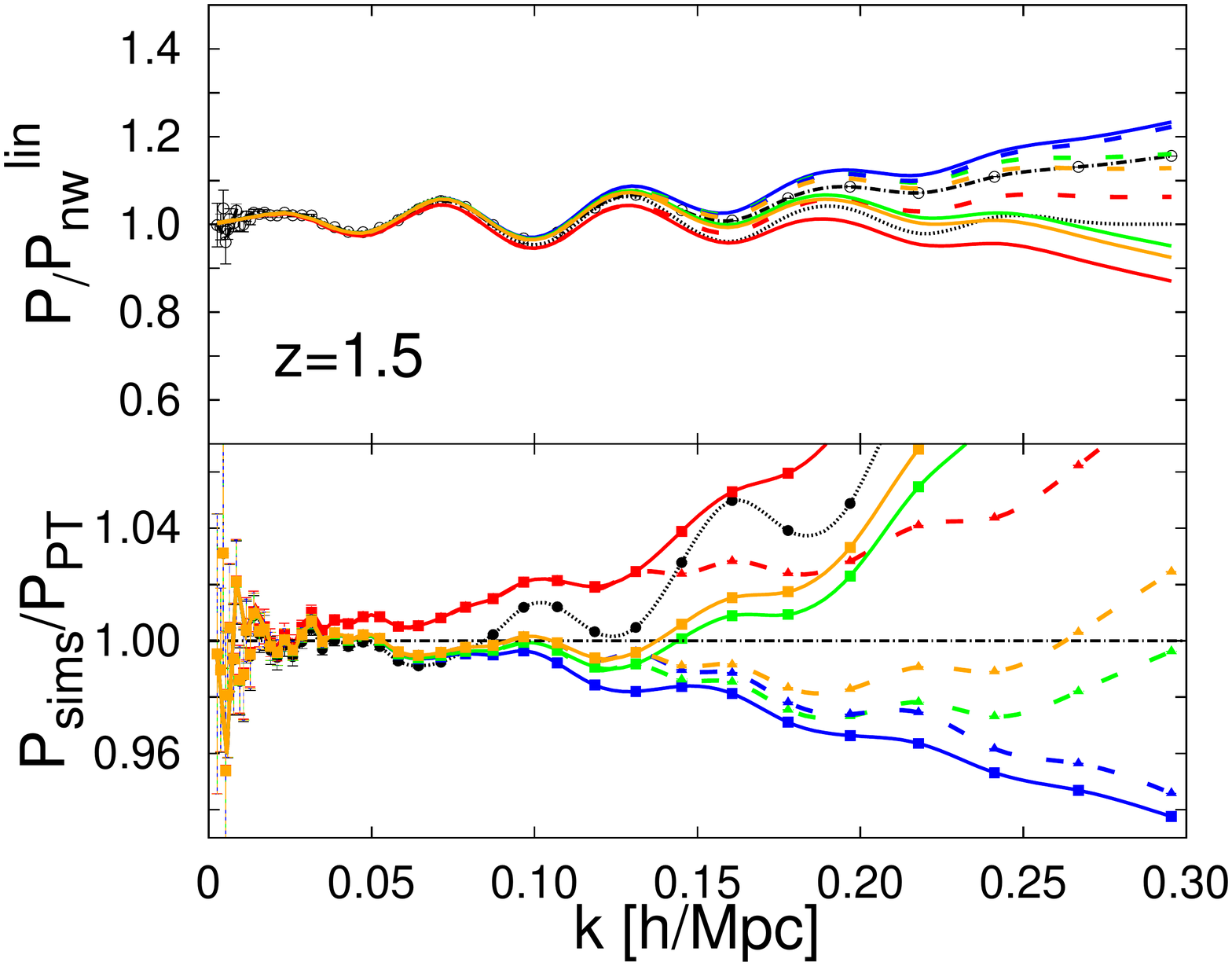}
\includegraphics[clip=false, trim= 20mm 0mm 30mm 0mm,scale=0.31]{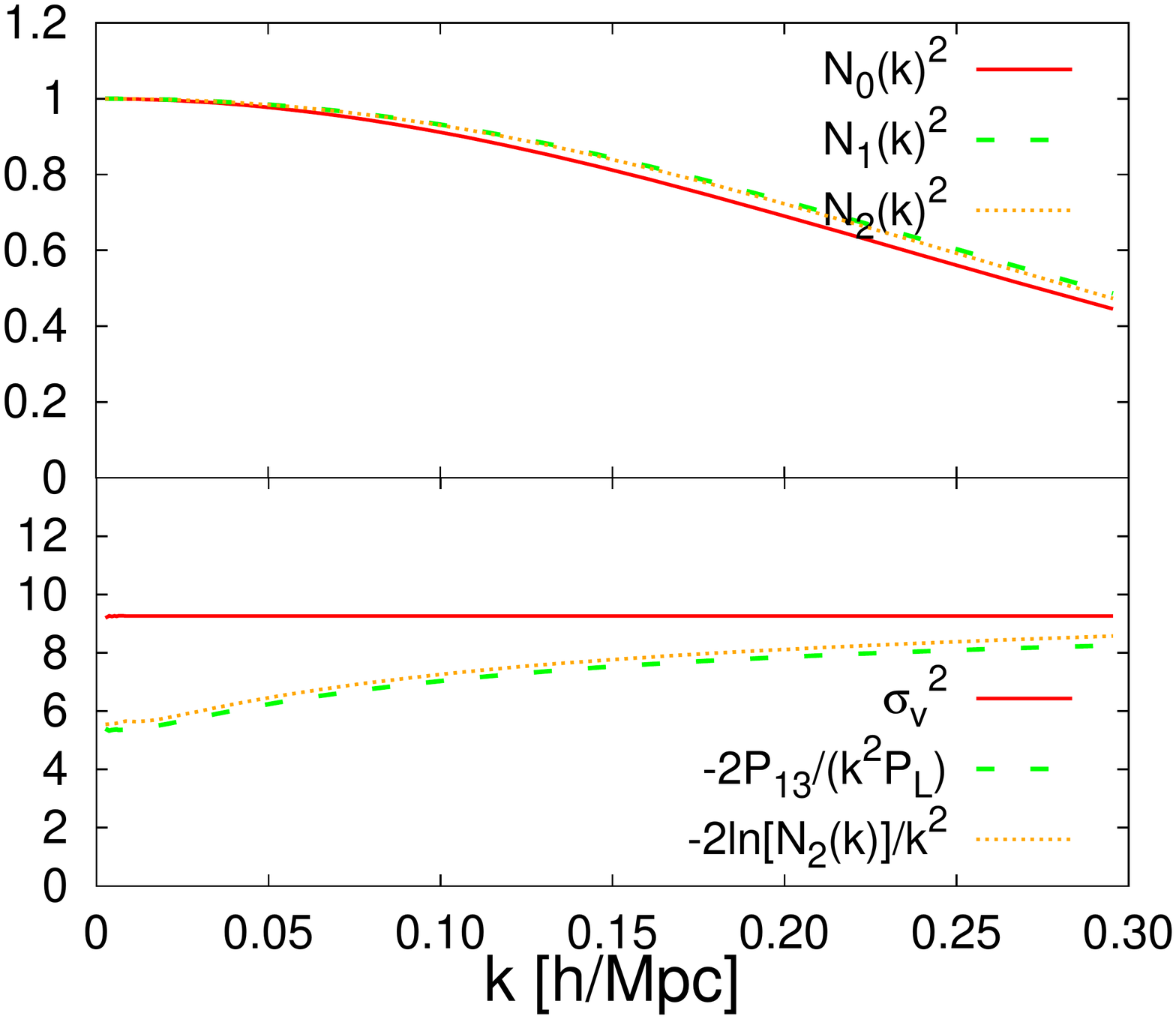}

\caption{Same notation that in Fig. \ref{FigAz0} but for $z=0.5$ (top panels), $z=1.0$ (middle panels) and $z=1.5$ (bottom panels).}
\label{FigAz15}
\end{figure}

In Fig. \ref{FigAz0} - \ref{FigAz15} (left panels) we compare the power spectra obtained with different flavors of perturbation theory against the N-body simulation results at $z=0,0.5,1.0$, and $1.5$. Each power spectrum has been normalized to the linear non-wiggle power spectrum to reduce the dynamical range. In the left top subpanels the power spectrum of the N-body simulations is shown as black circles linked with a black dot-dashed line, the linear theory prediction is shown as a dotted black line, SPT in blue lines, RPT-${\cal N}_0$ model in red lines, RPT-${\cal N}_1$ model in green lines and RPT-${\cal N}_2$ model in orange lines; where the infinite series of Eq. \ref{spt} and \ref{RPT} are truncated at 1-loop (solid lines) and 2-loop (dashed lines). In the bottom left subpanel the ratio of these models with N-body simulation data is shown with the same color and line notation. In that case, circle symbols correspond to linear prediction, square symbols to 1-loop truncation and triangle symbols to 2-loop truncation.

We see that SPT with 1-loop truncation over-predicts the N-body power spectrum at all redshifts, and can only make an accurate prediction of the power spectrum ($\leq1\%$ deviation) at very large scales: $k\leq0.05\,h/\rm{Mpc}$ for $z=0$ and $k\leq0.10\,h/\rm{Mpc}$ for $z=1$. Going to 2-loop correction improves considerably the behavior of SPT, but one has to deal with the terms $P_{33}$, $P_{24}$ and $P_{15}$ that require a 5-dimensional integration. In this case, 2-loop SPT makes a very good prediction of the N-body power spectrum at $z=0$ up to relatively small scales ($k\leq0.20\,h/{\rm Mpc}$), but at  $z=0.5$ starts over-predicting it, and at $z=1$ and $z=1.5$ the over-prediction is a few percent at intermediate scales ($k\gtrsim0.1\,h/{\rm Mpc}$). 

The RPT-${\cal N}_0$ model for both 1- and 2-loop truncation behaves accurately only at very large scales where it shows $\lesssim1\%$ deviation respect to N-body simulations both at all redshifts, but breaks down at relatively large scales: $k\simeq0.03\,h/\mbox{Mpc}$ for 1-loop at $z=0$ and $k\simeq0.07\, h/\mbox{Mpc}$ for 2-loop at $z=0$. 

The RPT-${\cal N}_1$ model presents at large scales a very good agreement with the N-body simulation results with the advantage that the breakdown happens at smaller scales: $k\simeq0.1\,h/\mbox{Mpc}$ for 1-loop and $k\simeq0.15$ for 2-loop at $z=0$; at $k\simeq0.15\,h/\mbox{Mpc}$ for 1-loop and at $k\simeq0.25$ for 2-loop at $z=1$. However, the for 2-loop truncation over-predicts the N-body power spectrum with a systematic $\sim2\%$ deviation at intermediate scales at all redshifts. This is due to the limitation in the resummed propagator ${\cal N}_1$, which is expressed as 1-loop expansion terms (in terms of $P_{13}$).

Finally, the RPT-${\cal N}_2$ model presents a modest improvement over previous models on large scales at $z=0$, but breaks down already at $k\simeq0.07\,h/\rm{Mpc}$ and $k\simeq0.10\,h/\rm{Mpc}$ for 1-loop and 2-loop, respectively. However at $z>0$, this model presents a better behavior, with $<2\%$ accuracy, breaking down at $k\simeq0.20\,h/\rm{Mpc}$ at $z=1$ and fixing the systematic $\sim2\%$ over-prediction observed for the RPT-${\cal N}_1$ model. The accuracy of this model at $z\ge0.5$ is then better than the 2-loop RPT-${\cal N}_1$ and 2-loop SPT.

In general, we see that the model that best describes the N-body data at $z\le1.5$ is the RPT-${\cal N}_2$ model at 2-loop truncation. The 2L-RPT-${\cal N}_1$ model shows also good results but presents a  $\sim2\%$-systematic over-prediction at intermediate scales. However at $z=0$, RPT-${\cal N}_1$ is able to reach smaller scales than RPT-${\cal N}_2$, which breaks down at relatively large scales. Indeed, 2L-RPT-${\cal N}_2$ works better than 2L-RPT-${\cal N}_1$ and 2L-SPT  at $z\ge0.5$ but not at $z=0$.

In Fig. \ref{FigAz0} - \ref{FigAz15} (right panels) the behavior of the damping functions of RPT-${\cal N}_i$ models is shown for the same redshift range. In particular, in the right top subpanel we show the scale dependence of the damping functions of Eq. \ref{RPT}:  ${\cal N}_0(k)^2=\exp(-k^2\sigma_v^2)$ (red solid line), ${\cal N}_1(k)^2=\exp\left[2P_{13}(k)/P^{\rm lin}(k)\right]$ (green dashed line) and ${\cal N}_2(k)^2$ (see Eq. \ref{theory_N2}) (orange dotted line). We see that at large scales, all ${\cal N}_i$ functions converge to 1, and it is at intermediate and smaller scales ($k>0.03\, h/\mbox{Mpc}$) where the differences between these three models are significant. For the scales of interest, we see that ${\cal N}_0(k)<{\cal N}_2(k)<{\cal N}_1(k)$. Thus, as we include higher-order propagators in the computation of ${\cal N}_i$, these functions oscillate about the `true' damping function. This approach explains why RPT-${\cal N}_0$ under-predicts N-body data and why RPT-${\cal N}_1$ slightly over-predicts it. Every loop correction in the resummed propagator tends to the true value, but in a oscillatory way. For $z\geq0.5$, the 2-loop correction in the resummed propagator seems to be sufficient for a $\leq1\%$ prediction. However at $z=0$, the convergence in the resummed propagator is still not reached for the 2-loop correction in $\cal N$. In that case, higher-order loop corrections would be necessary to reach the $\leq1\%$ deviation. In the right bottom panel we show the effective $\sigma_v^2$, i.e. $-2\ln[{\cal N}_i(k)]/k^2$, for the different orders in the resummed propagator: 0-loop (red solid line), 1-loop (blue dashed line) and 2-loop (orange dotted line). In that case, we see that at large scales the three models diverge, whereas at small scales all models seem to converge (at least for $z\ge0.5$). 

In other words, it may happen that a lower-order approximation  appears to work better than a higher-order approximation which, in principle, should be more accurate. This is due to a fortuitous cancellation of the truncation errors in the propagator and in the damping function. This cancellation does not hold for all redshifts  and/or all cosmologies. The performance of an analytical  approximation scheme must be quantified looking at different redshifts (or different cosmologies).

The formalism presented here deals with the $F_n$ kernels, that correspond to the $\delta$-field. However, this formalism is also perfectly valid for the computation of $P_{\delta\theta}$ and $P_{\theta\theta}$, only changing appropriately the $F_n$ kernels by the $G_n$ kernels as in SPT.

\subsection{Dark matter multipoles}
In order to obtain information about the growth rate $f$ from the redshift-space distortions, it is convenient to work with the expansion in Legendre moments, $P_{\ell}$,  defined as,
\begin{equation}
\label{real_Ps}P_{\ell}(k)=(2\ell+1)\int_0^1 d\mu\, P^s(k,\mu)L_{\ell}(\mu),
\end{equation}
where $L_{\ell}$ are the Legendre polynomials of order $\ell$. For the first three non-vanishing $P_\ell$,
\begin{eqnarray}
L_0(x)&=&1,\\
L_2(x)&=&\frac{1}{2}(3x^2-1), \\
L_4(x)&=& \frac{1}{8}(35x^4-30x^2+3).
\end{eqnarray}
According to linear theory (Kaiser model with $P_{\delta\delta}^{\rm{lin}}$) only the monopole ($\ell=0$), the quadrupole ($\ell=2$) and the hexadecapole ($\ell=4$) are different from 0. In that case, for an unbiased tracer these three moments read,
\begin{eqnarray}
\label{monopole_eq}P_0(k)&=&P^{\rm lin}(k)\left(1+\frac{2}{3}f+\frac{1}{5}f^2\right), \\
\label{quadrupole_eq}P_2(k)&=&P^{\rm lin}(k)\left(\frac{4}{3}f+\frac{4}{7}f^2 \right),\\
\label{hexadecapole_eq}P_4(k)&=&P^{\rm lin}(k)\left( \frac{8}{35}f^2  \right).
\end{eqnarray}
Hence, knowing the dark matter power spectrum in both real and redshift space, one can directly measure the growth rate $f$ from any of these multipoles. It is interesting to note that the ratio between any of these multipoles does not depend (at large scales) on the real-space power spectrum and the ratio tends to a constant that only depends on $f$ when $k\rightarrow0$. However, non-linearities produce deviations from these formulae. Depending on the ability of modeling the non-linearity in the redshift space distortions, we will be able to use information from non-linear scales to estimate $f$ with accuracy.



In this section we focus on checking the quality of the different theoretical models in predicting the multipole power spectrum of dark matter in redshift space. We focus on the models described in section \ref{section_redshift_space}, using as real-space inputs, the PT-theory approaches described in section \ref{section_real_space}. Here, we assume that $f$ is known and we only fit the FoG parameter $\sigma_0$ assuming a Lorentzian damping function (Eq. \ref{lorentzian}), although no significant difference is observed when a Gaussian damping function is assumed. We allow $\sigma_0$ to depend on $z$ and on $k_{\rm max}$\footnote{$k_{\rm max}$ is the maximum $k$ used for the fit} and find the best-fit value by minimizing 
\begin{equation}
\chi^2=\sum_{k=k_0}^{k_{\rm max}}\left[\frac{P^\ell_{\rm sims}(k)-P_{\rm theo}^\ell(k)}{\sigma_{\rm P, sims}(k)}\right]^2 .
\end{equation}
Here, the subscript ``sims" refers to the simulations and ``theo" to the theoretical models  described above. $\sigma^2_{\rm P,sims}(k)$ is the variance of the multipoles computed from the 160 simulations and $k_o$ is the minimum $k$ considered, which is set by the size of the simulation box but has little effect on the final results. Note that we neglect for simplicity any covariance between different $k$-bins, which is a good approximation for small $k$ and broad bins.

\begin{figure}
\centering

\includegraphics[clip=false, trim= 10mm 0mm 30mm 0mm,scale=0.27]{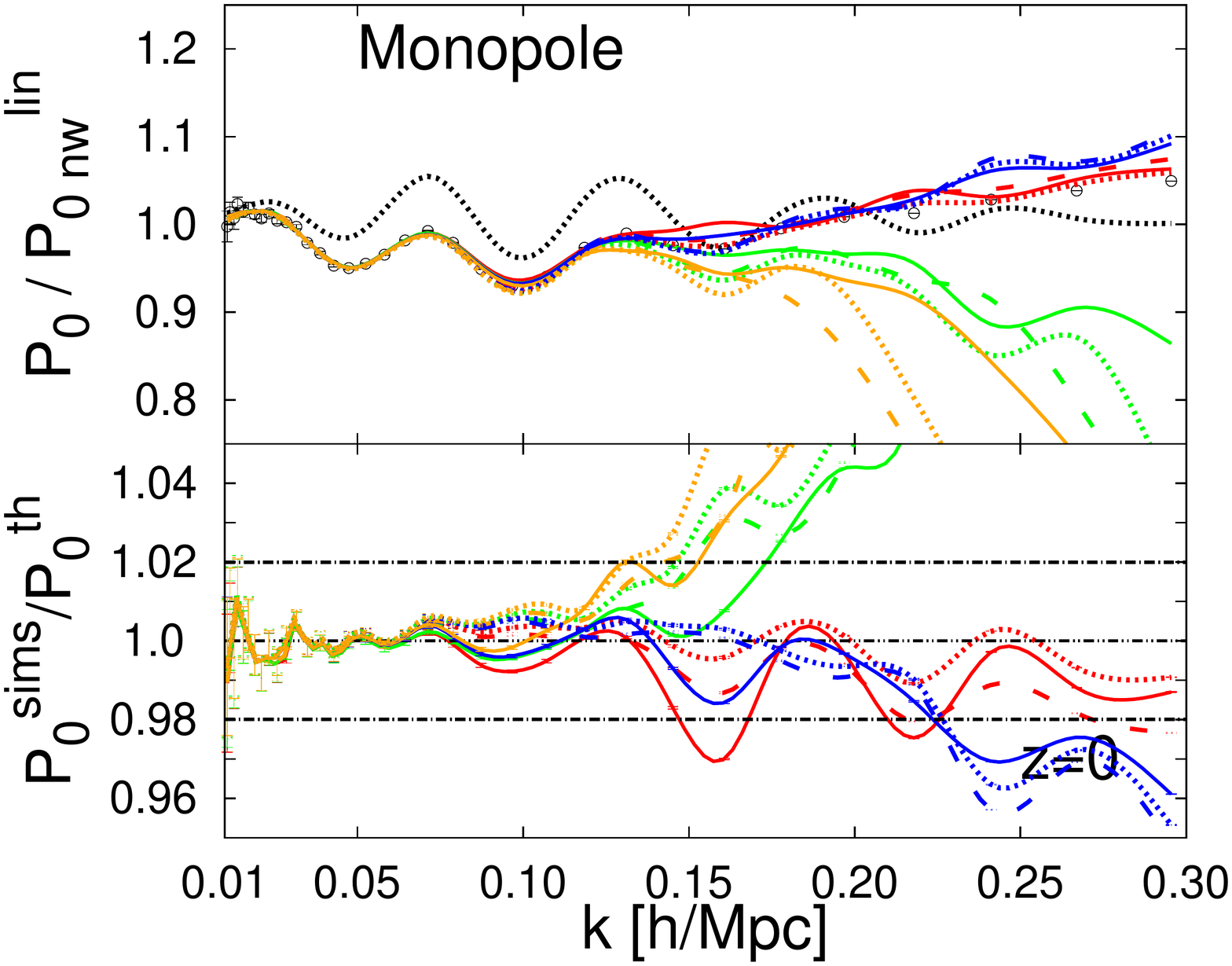}
\includegraphics[clip=false, trim= 10mm 0mm 30mm 0mm,scale=0.27]{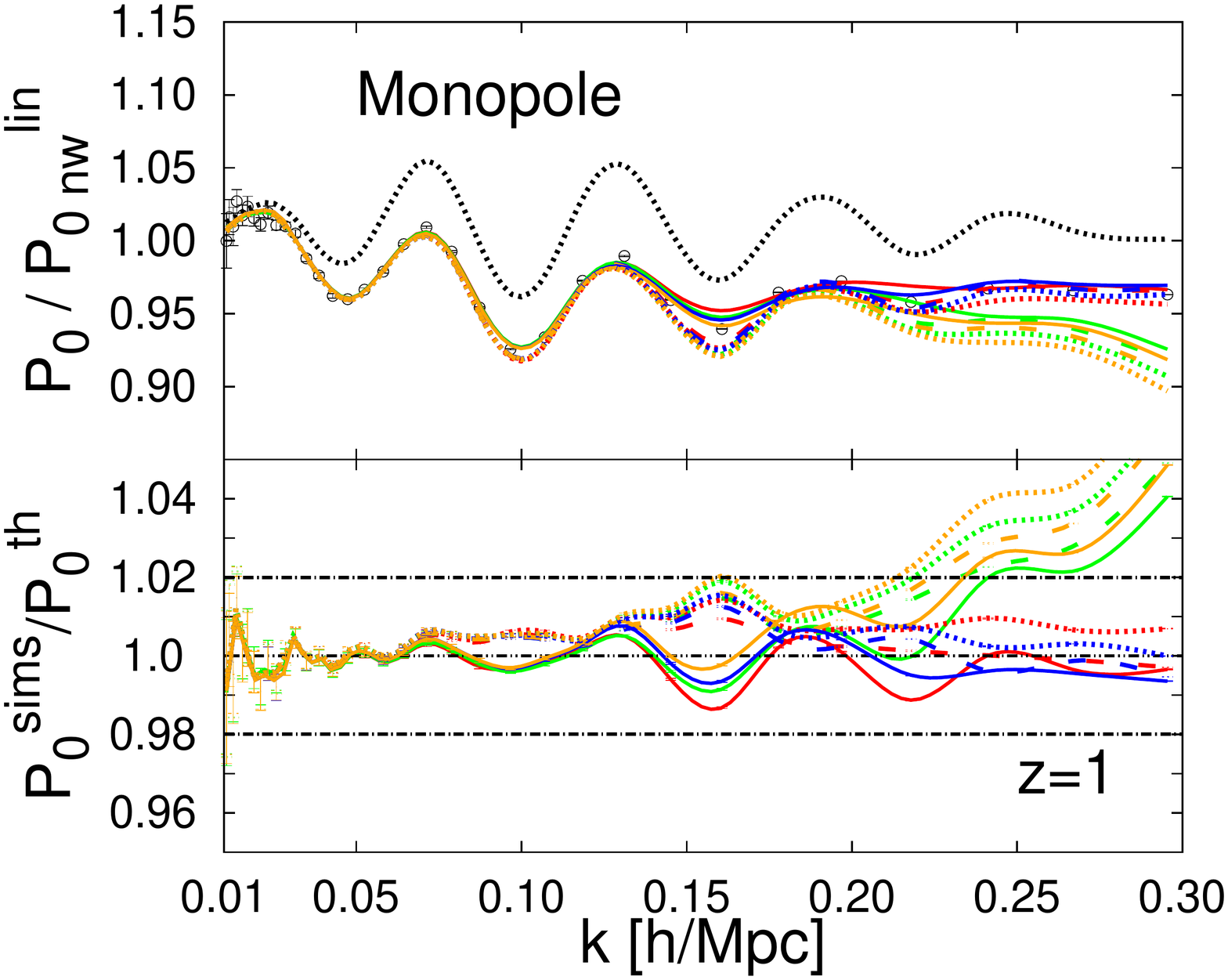}

\includegraphics[clip=false, trim= 10mm 0mm 30mm 0mm,scale=0.27]{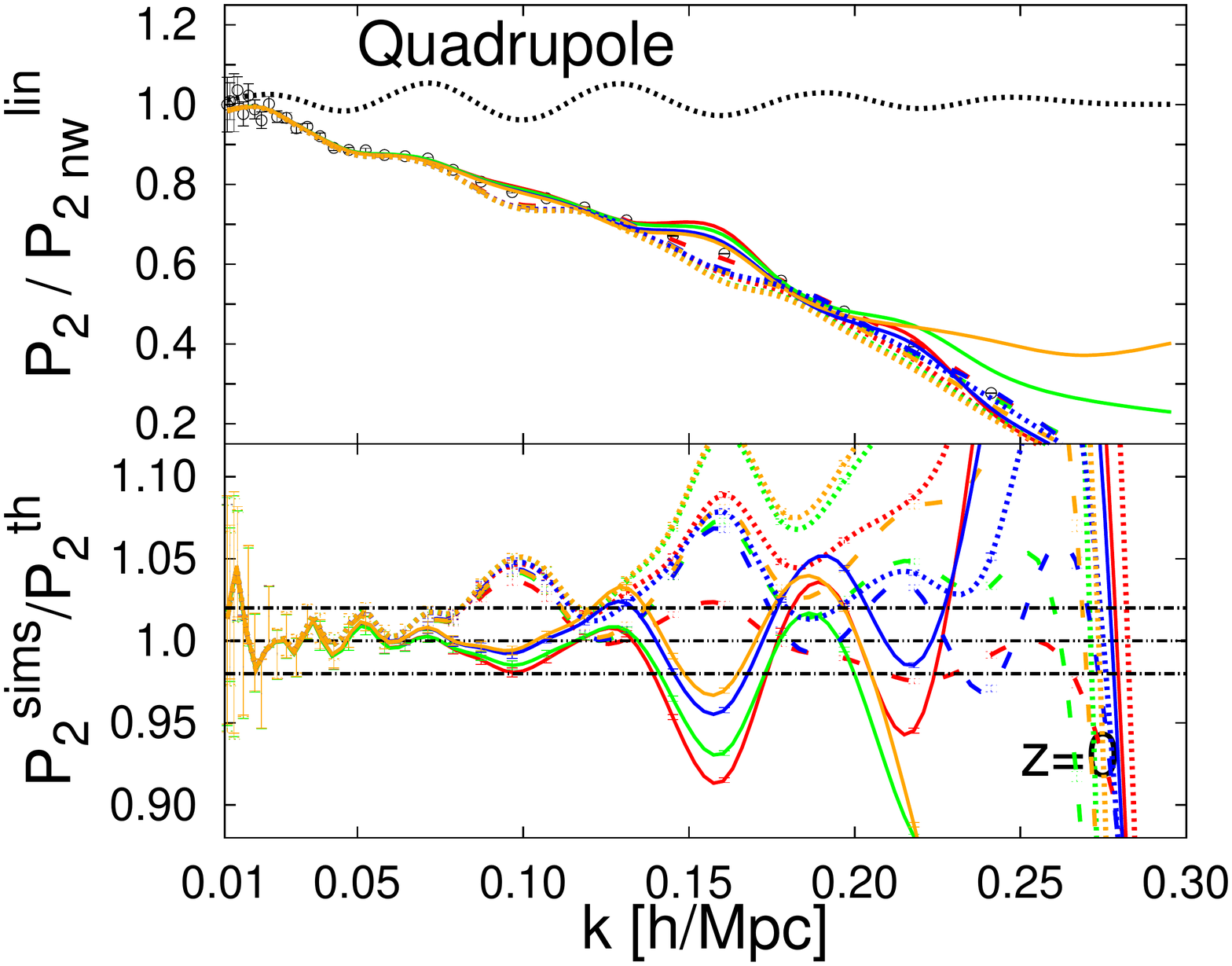}
\includegraphics[clip=false, trim= 10mm 0mm 30mm 0mm,scale=0.27]{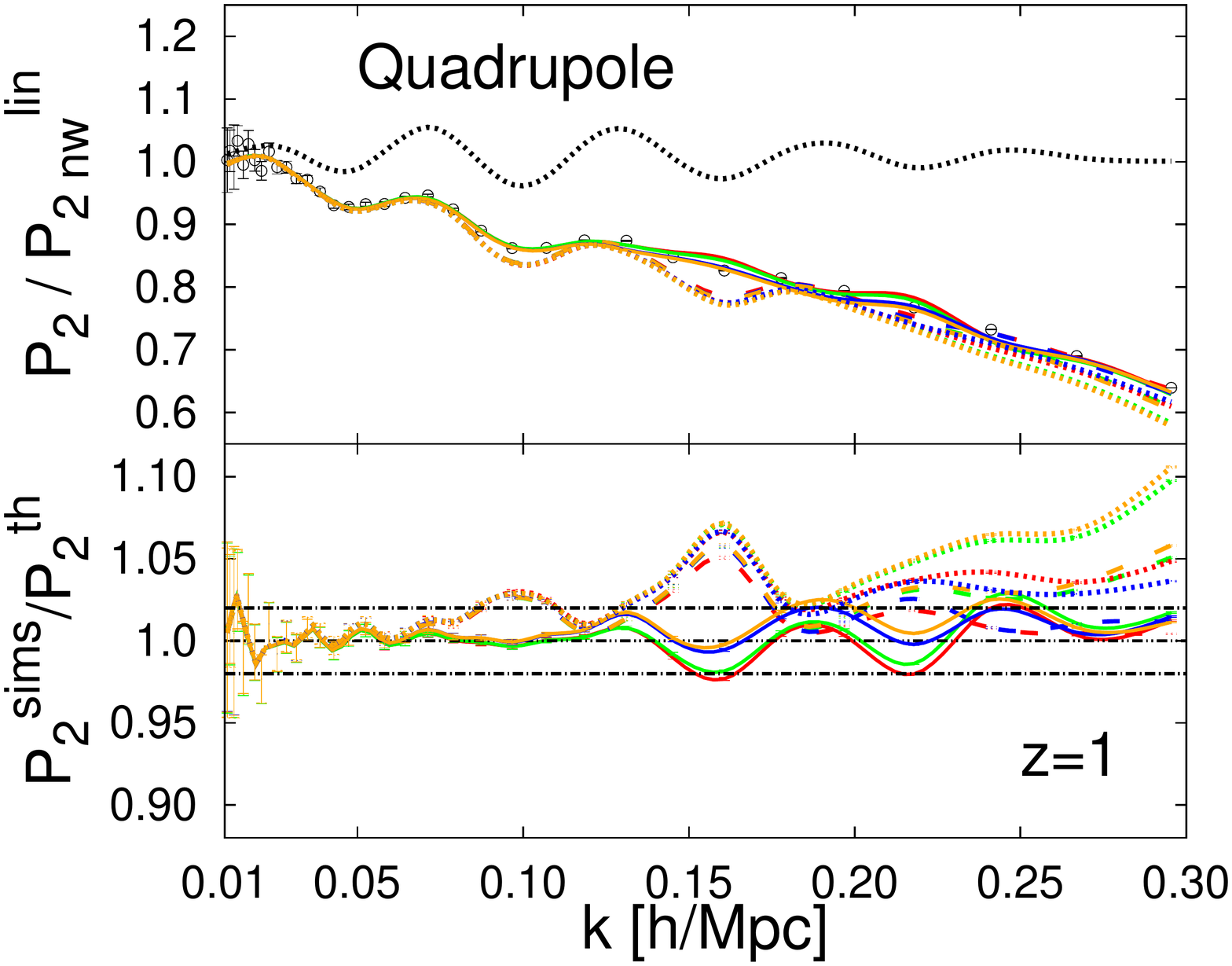}

\includegraphics[clip=false, trim= 10mm 0mm 30mm 0mm,scale=0.27]{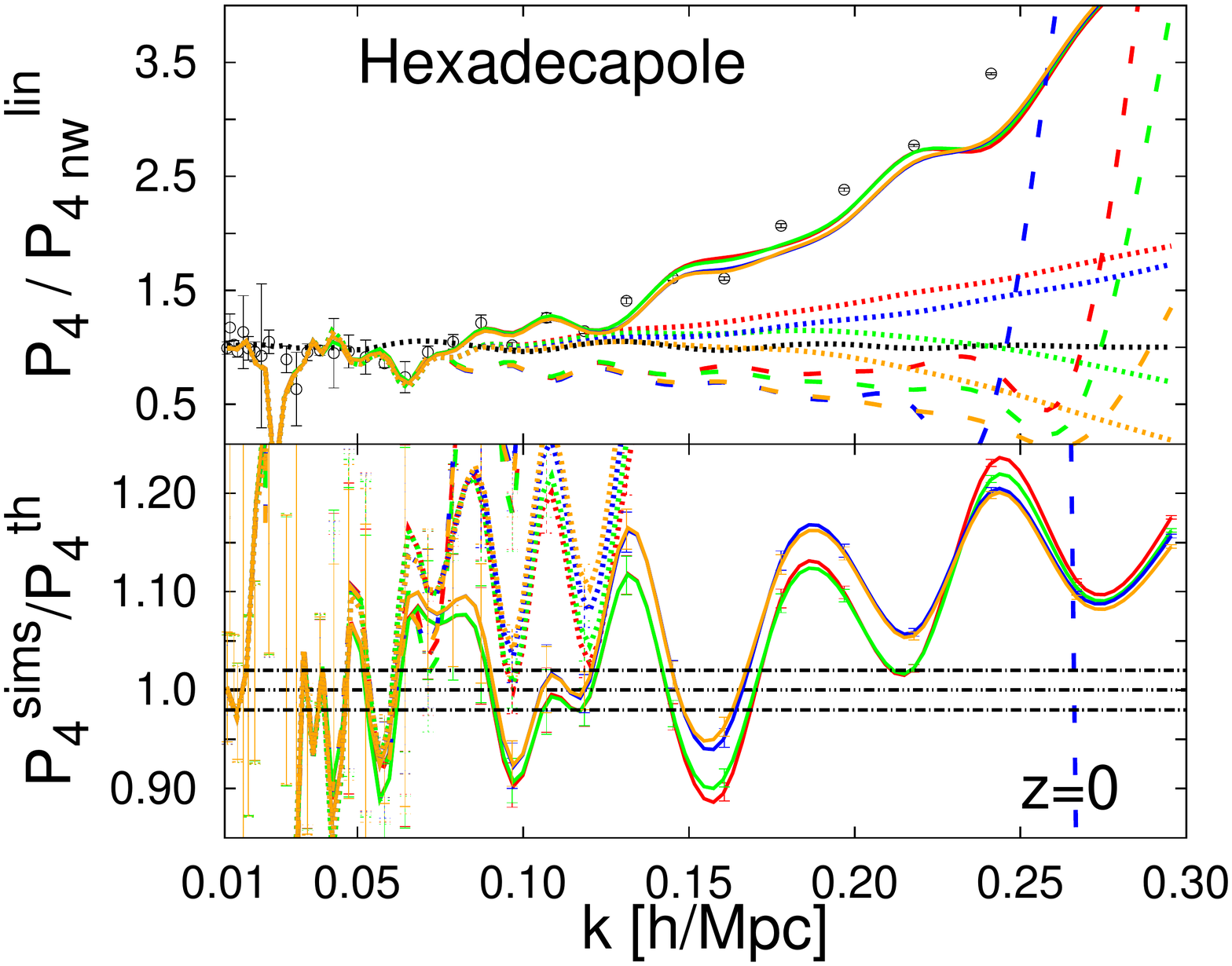}
\includegraphics[clip=false, trim= 10mm 0mm 30mm 0mm,scale=0.27]{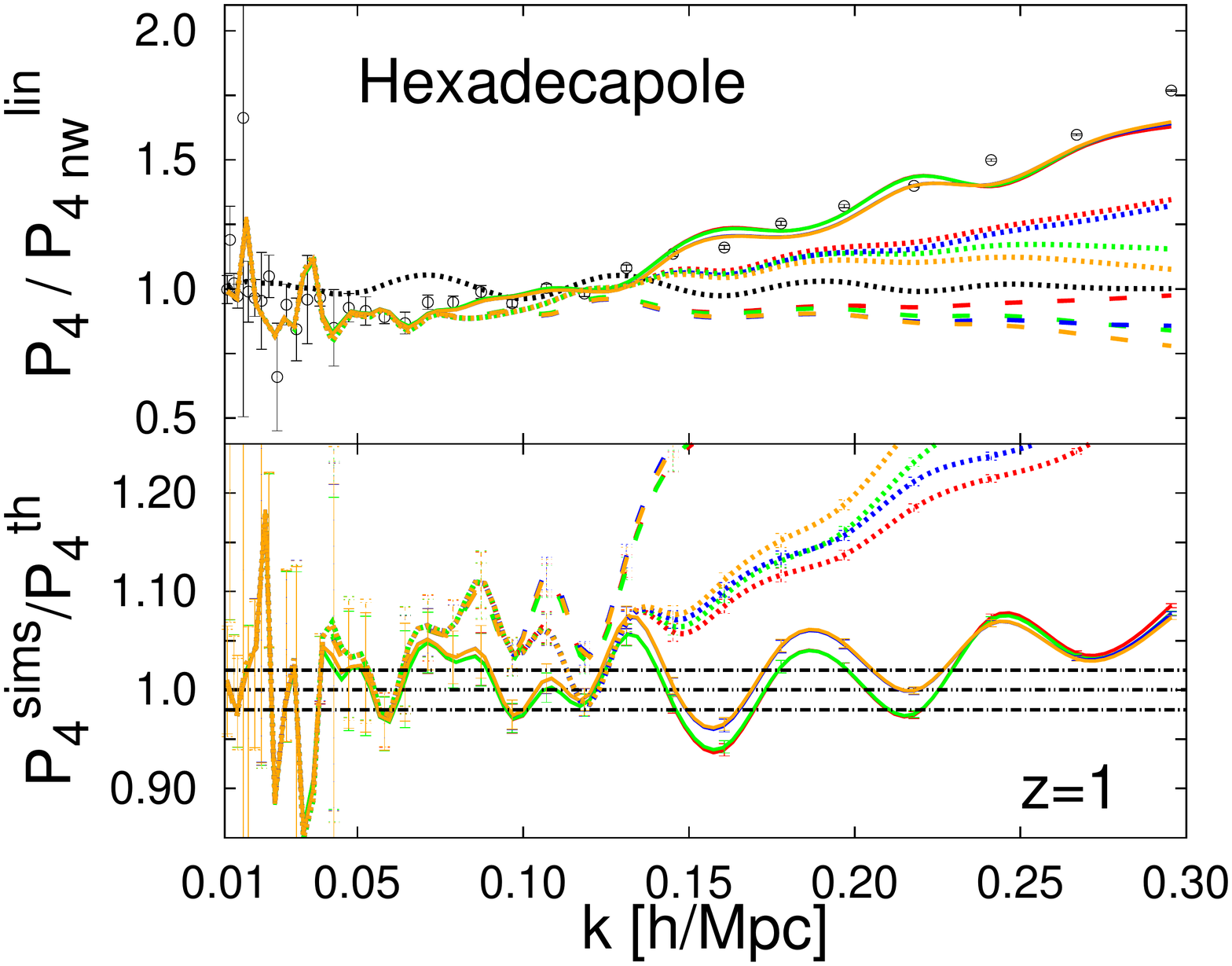}

\caption{Multipoles corresponding to dark matter power spectrum: monopole (top panels), quadrupole (middle panels) and hexadecapole (bottom panels), for $z=0$ (left panels) and $z=1$ (right panels), where $f$ is fixed to the true value, and $\sigma_0$ is the only free parameter, fit to N-body data. In the upper subpanels the value of the corresponding multipole is shown, normalized to the linear, no-wiggles value  to reduce the dynamical range. In the lower subpanels the ratio between the N-body simulation data and different perturbation theory predictions is shown. Dotted lines correspond to the Kaiser model, dashed lines to Scoccimarro model and solid lines to the TNS model. Different perturbation theory models are shown: linear prediction (black dotted lines), 1L-SPT (red lines), 2L-SPT (blue lines), 2L-RPT-${\cal N}_1$ (green lines) and 2L-RPT-${\cal N}_2$ (orange lines). In bottom subpanels the $2\%$ deviation is marked with black dot-dashed horizontal lines.}
\label{multipoles_dark_matter}

\end{figure}

In Fig. \ref{multipoles_dark_matter}, we show the measurements of the monopole (top panels), quadrupole (middle panels) and hexadecapole (bottom panels) from N-body simulations (black empty circles) in top sub-panels. In all cases, the multipoles have been normalized to the non-wiggle linear prediction to reduce the dynamical range. In the top subpanels, different theoretical predictions are shown: the Kaiser model (dotted lines), the Scoccimarro model (dashed lines) and the TNS model (solid lines). The chosen real-space power spectrum for each of these models is: linear prediction (black dotted lines), 1L-SPT (red lines), 2L-SPT (blue lines), 2L-RPT-${\cal N}_1$ (green lines) and 2L-RPT-${\cal N}_2$ (orange lines).

In all cases, the TNS model combined with both SPT and RPT predictions is the model that describes best the N-body results. Using the TNS model it is possible to achieve $<1\%$ accuracy for the monopole up to $k\leq0.12\,h/\rm{Mpc}$ at $z=0$ and $k\leq0.17\,h/\rm{Mpc}$ at $z=1$. The models are also very accurate for the quadrupole: at $z=0$ we can describe N-body data up to scales of $k=0.12\,h/{\rm Mpc}$ and at $z=1$ up to $k=0.30\,h/{\rm Mpc}$ with a deviation $\lesssim2\%$. For the hexadecapole, the agreement is more modest: at $z=0$ we can only achieve $\sim10\%$ accuracy up to scales of $k=0.20\,h/{\rm Mpc}$ at $z=0$ and $\sim5\%$ at $z=1$.

Both the Scoccimarro and Kaiser model provide a reasonably good approximation on large scales but both fail to give an accurate description on mildly non-linear scales where baryon acoustic oscillations (BAO) are located.
 The difference between the TNS model and the other models is more evident for the quadrupole and hexadecapole possibly suggesting that non-linearities become more important for higher-order multipoles. 

The imprint of the  BAO in the multipoles is clearly visible: note that the Scoccimarro and Kaiser models slightly over-predict the BAO amplitude, especially for the quadrupole, while the TNS model does better, although a trend towards the under-prediction is observed. All models correctly predict the BAO location. These considerations  might be relevant  for recovering in an unbiased way  the angular and radial BAO information (separately) from  forthcoming surveys.

We conclude that the TNS model with the RPT and SPT models studied here, has the ability of describing the redshift space power spectrum monopole and quadrupole at $z=0$ and $z=1$ within $1-2\%$ for $k \lesssim 0.2$ and  the hexadecapole  within about $\simeq 5\%$.

We do not observe a crucial difference between 1- and 2-loop SPT. Also, no significant difference between using ${\cal N}_1$ and ${\cal N}_2$ for 2-loop RPT is detected.  This indicates that on these mildly non-linear scales at redshifts $\lesssim 1.5$ the accuracy of  the  modeling of redshift space distortions is more important than that  of the non-linear evolution of the real-space dark matter power spectrum.

Because of that, for simplicity we focus on 1-loop SPT and 2-loop RPT with ${\cal N}_1$ when measuring $f$ from dark-matter multipoles in the next subsection.

\begin{figure}
\centering

\includegraphics[clip=false, trim= 10mm 0mm 30mm 0mm,scale=0.27]{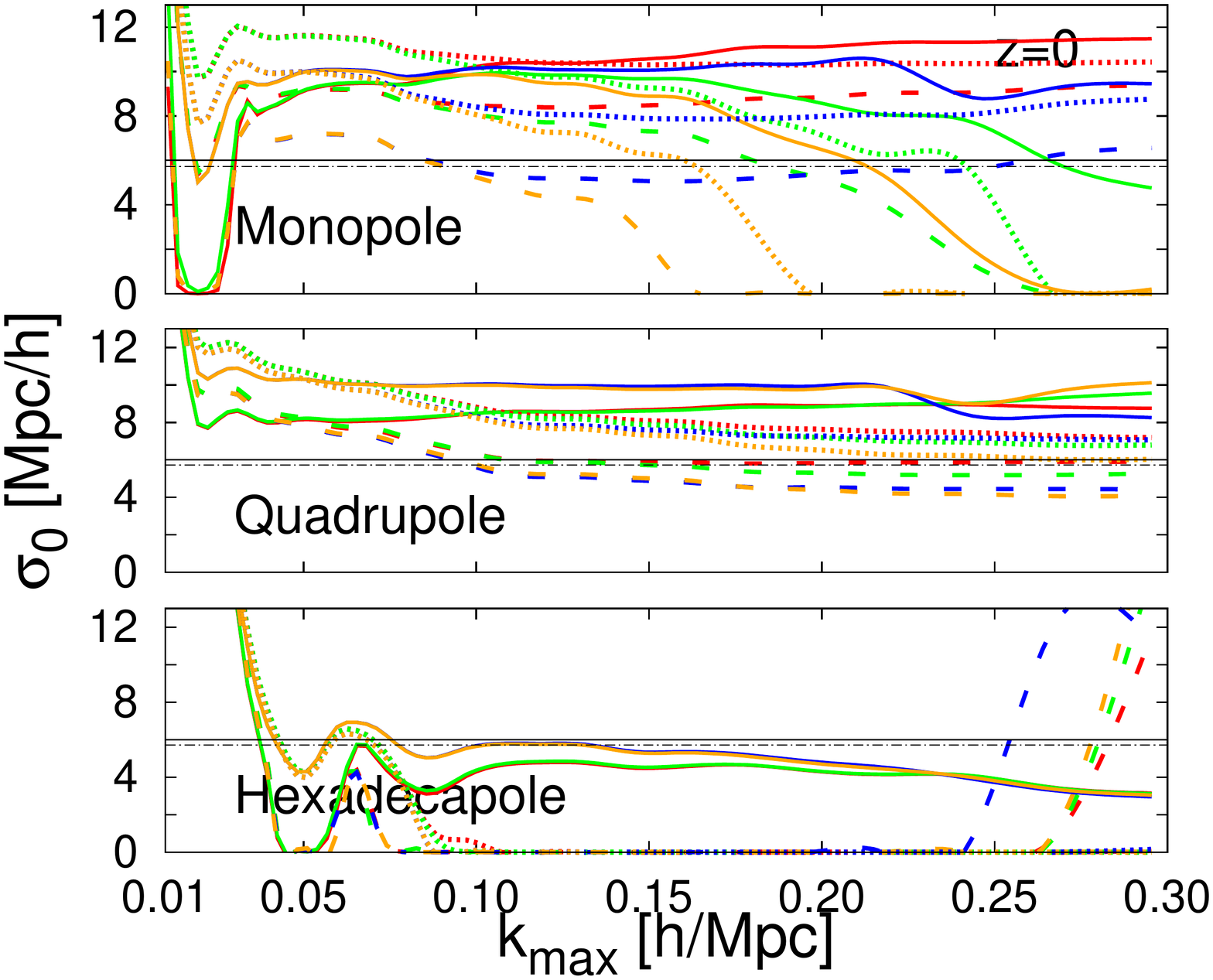}
\includegraphics[clip=false, trim= 10mm 0mm 30mm 0mm,scale=0.27]{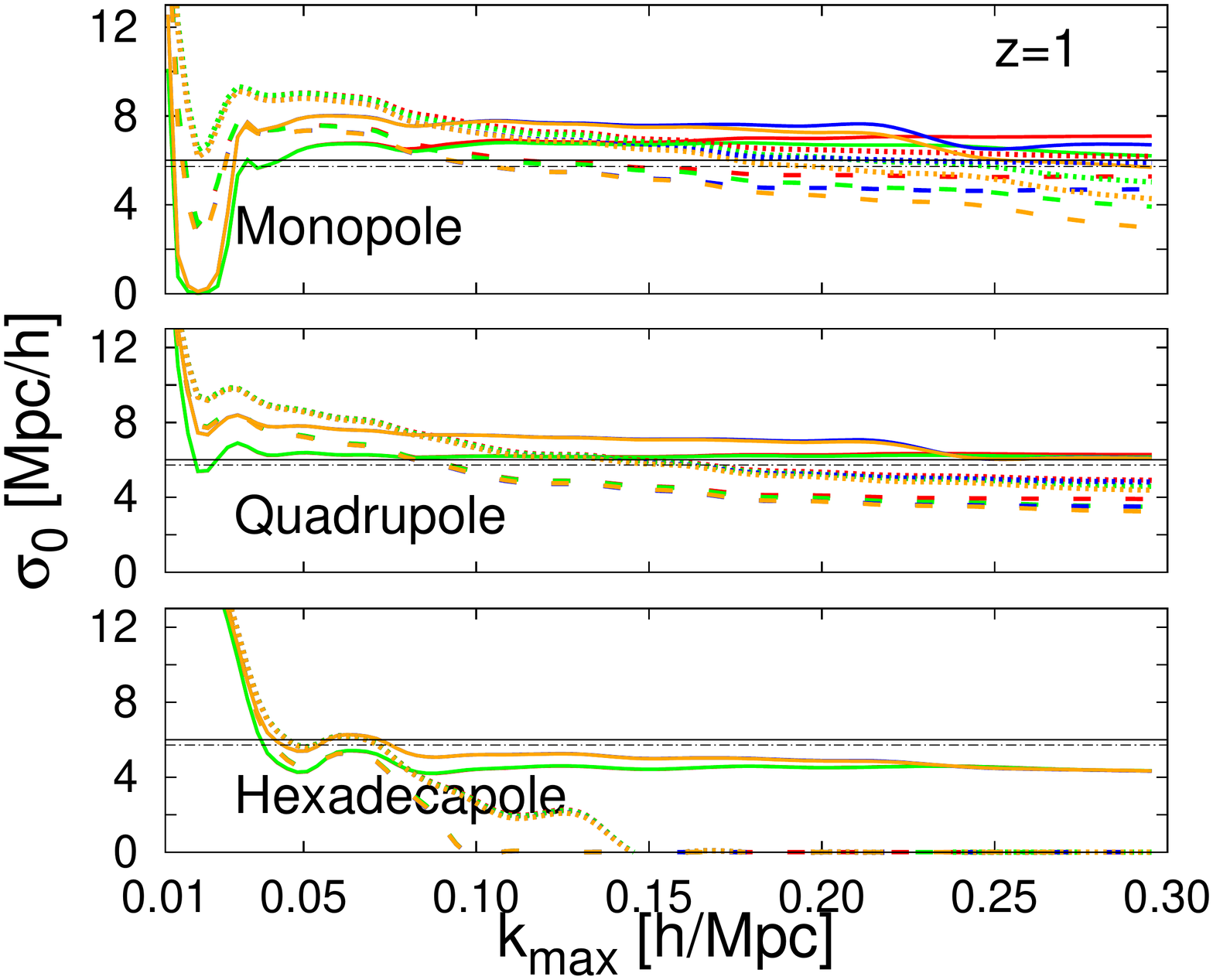}

\caption{Best-fit values of $\sigma_0$ as a function of $k_{\rm max}$ corresponding to multipoles shown in Fig. \ref{multipoles_dark_matter} with the same color notation. As indicated, top, middle and bottom panels stands for monopole, quadrupole and hexadecapole fits alone. Left and Right panels show the result at $z=0$ and $z=1$ respectively. Also theoretical predictions for $\sigma_0$ are shown according Eq. \ref{sigmav}: with $P_{\theta\theta}$ as input using 1L-SPT prediction in solid black line and using $P_{\rm lin}$ as input in dot-dashed black line. Error-bars are not shown for the sake of clarity, but are negligible for $k>0.05\,h/{\rm Mpc}$.}
\label{sigma_0_multipoles}

\end{figure}
In Fig. \ref{sigma_0_multipoles} we show the best-fit value for $\sigma_0$ corresponding to the fits shown in Fig. \ref{multipoles_dark_matter} using the same color notation for the different models. Additionally we show the theoretical predictions of Eq. \ref{sigmav} using as an input the 1L-SPT prediction for $P_{\theta\theta}$ (solid black line) and $P_{\rm lin}$ (dot-dashed line).  As indicated, top, middle and bottom panels correspond to the monopole, quadrupole and hexadecapole, whereas left panels show the result for $z=0$ and right panels for $z=1$. 

We note that at $z=1$ all the models produce a best-fit $\sigma_0$ which is close the the theoretical predictions, although an overestimate is observed for the monopole and underestimation for the hexadecapole, being the quadrupole the case which is closer to the theoretical prediction. For $z=0$, the discrepancy between theory and best-fit value is larger. We will analyze again the agreement between theory and best-fit $\sigma_0$ in the next section, when will allow $f$ also to vary.

\subsection{Estimating $f$ from dark matter multipoles}
In the last section we have shown that the TNS model was able to describe well the multipoles when one free parameter was allowed to vary in order to account for the FoG effect. In this section we want to check the ability of these models to recover $f$ from the dark-matter field. In this case, we will allow both $f$ and $\sigma_0$ to freely vary. As before, in order to find the $f$ and $\sigma_0$ best fit parameters, we minimize the $\chi^2$ value.

\begin{figure}
\centering

\includegraphics[clip=false, trim= 10mm 0mm 20mm 0mm,scale=0.28]{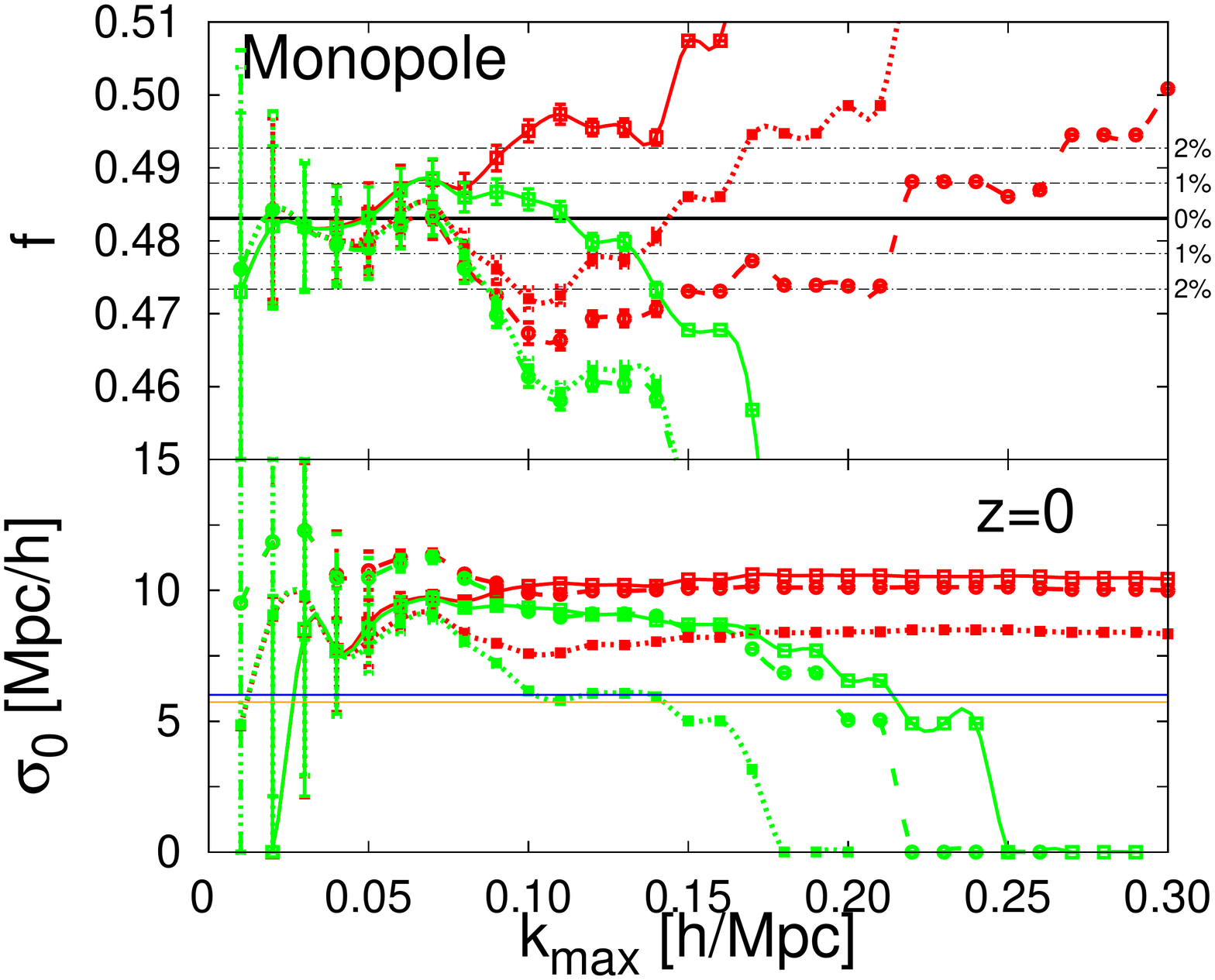}
\includegraphics[clip=false, trim= 10mm 0mm 20mm 0mm,scale=0.28]{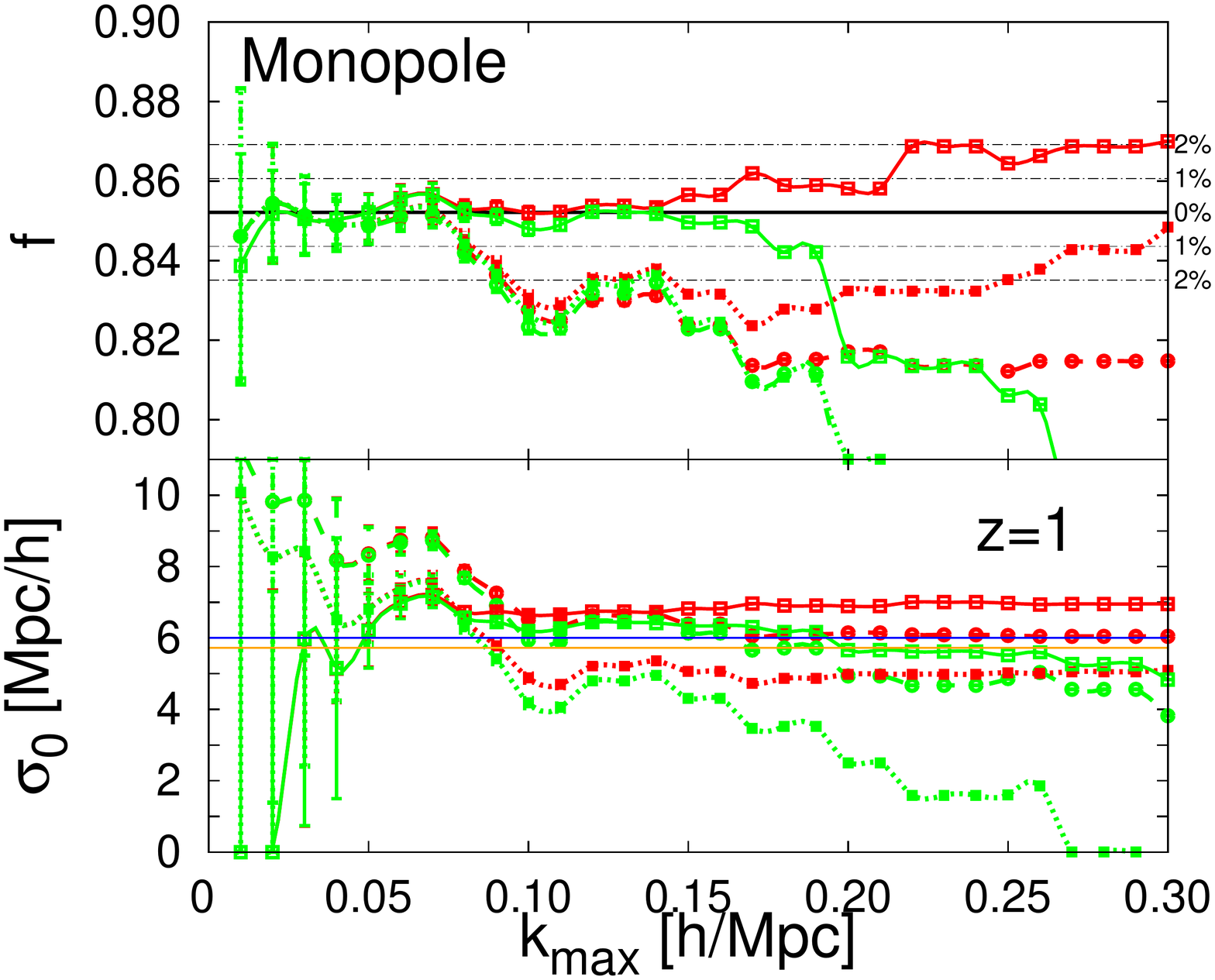}

\includegraphics[clip=false, trim= 10mm 0mm 20mm 0mm,scale=0.28]{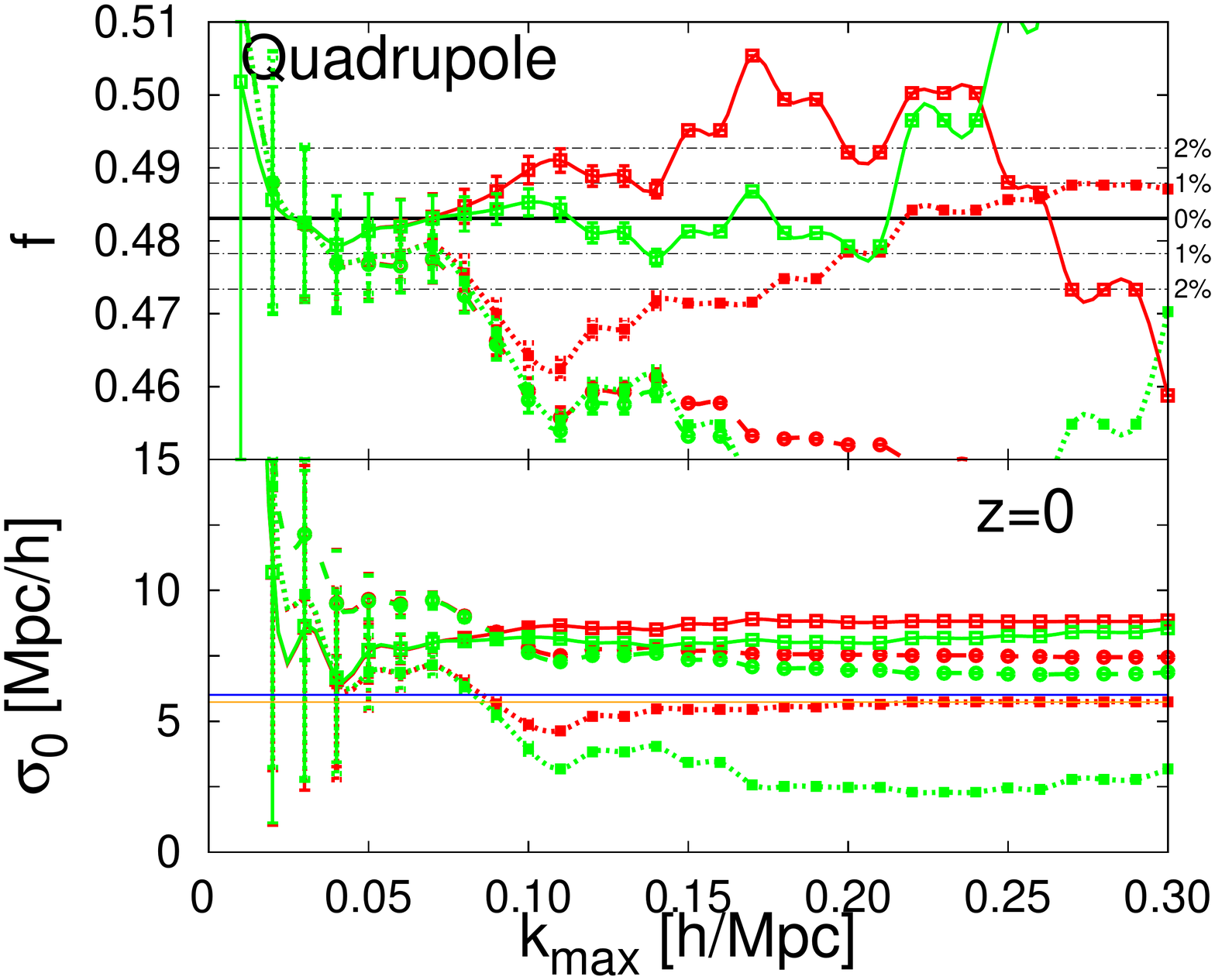}
\includegraphics[clip=false, trim= 10mm 0mm 20mm 0mm,scale=0.28]{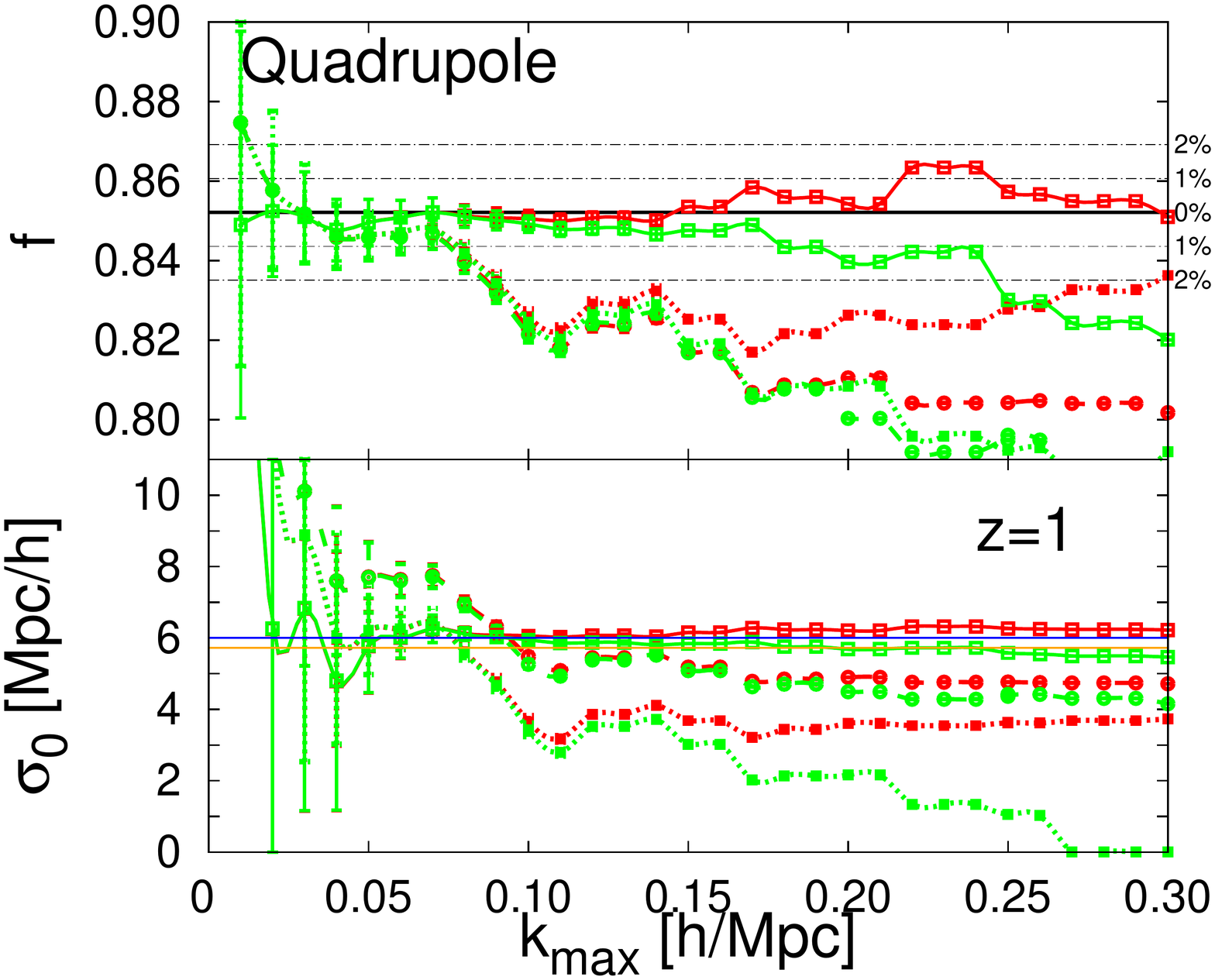}

\includegraphics[clip=false, trim= 10mm 0mm 20mm 0mm,scale=0.28]{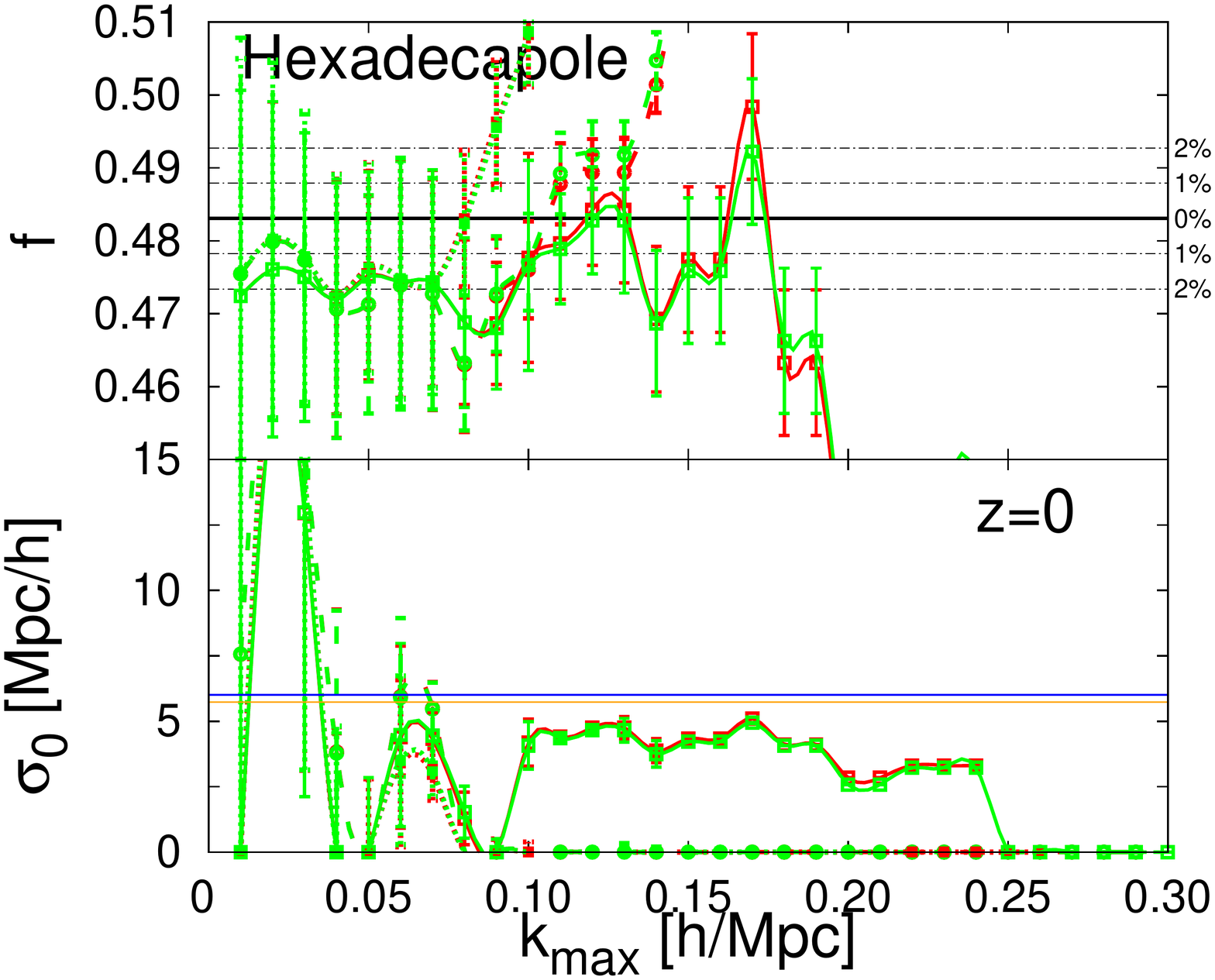}
\includegraphics[clip=false, trim= 10mm 0mm 20mm 0mm,scale=0.28]{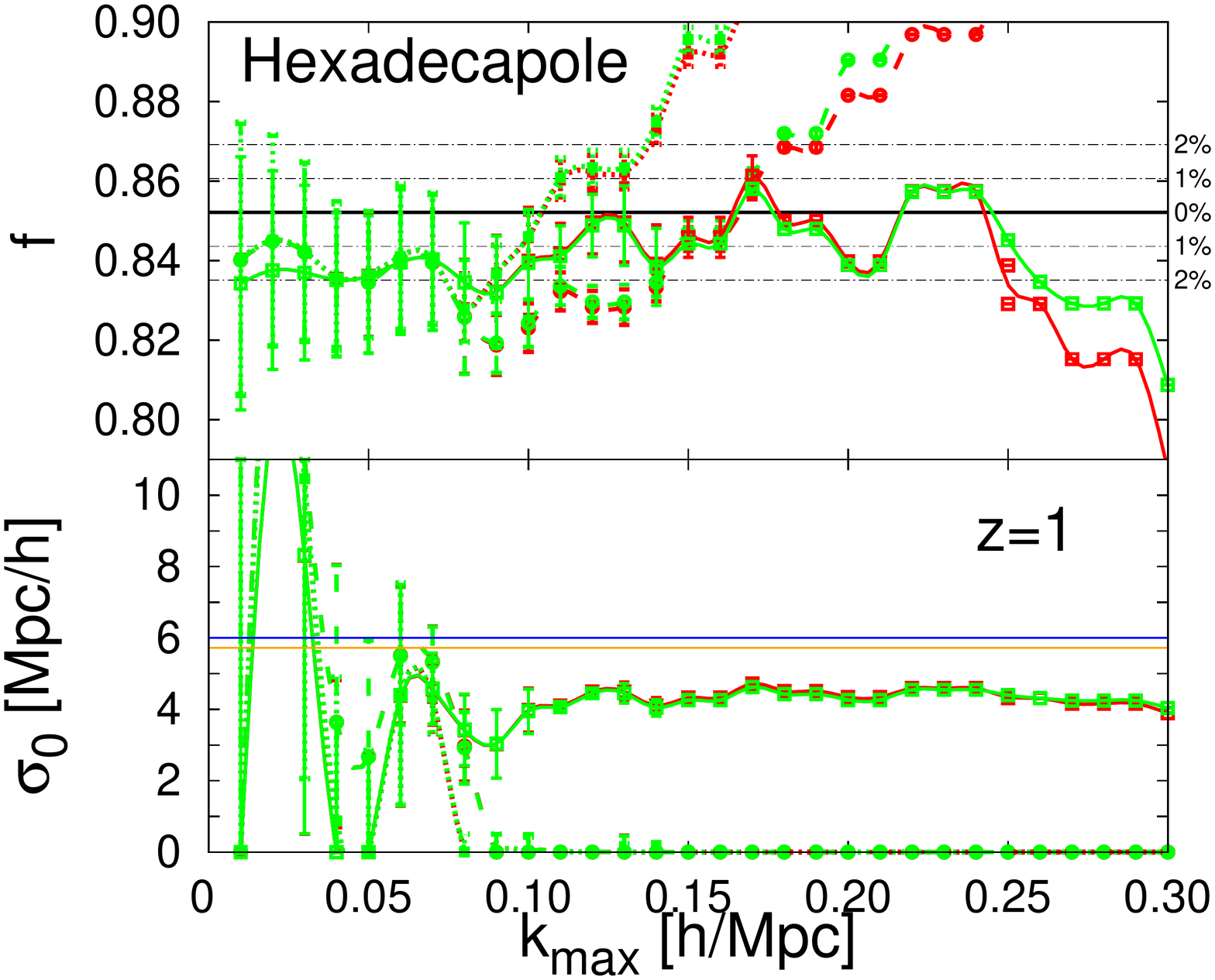}

\caption{Estimates for $f$ (top subpanels) and $\sigma_0$ (bottom subpanels) from the dark matter multipoles of the N-body data: monopole (top panels), quadrupole (middle panels) and hexadecapole (bottom panels); for $z=0$ (left panels) and $z=1$ (right panels). Results from different theoretical models are shown: dotted lines are Kaiser model, dashed lines Scoccimarro model and solid lines TNS model. Green lines are 2-loop-RPT-${\cal N}_1$ and red lines 1-loop SPT.  In top subpanels, the true value of $f$ is represented in a horizontal black-solid line, whereas $1\%$ and $2\%$ deviations are shown in the horizontal black dot-dashed lines, as indicated. In the bottom subpanels the theoretical predictions for $\sigma_0$ (Eq. \ref{sigmav}) are shown in blue (with $P_{\theta\theta}$ as input using 1L-SPT prediction) and in orange (with $P_{\rm lin}$ as input). }
\label{f_dm}
\end{figure}

 In Fig. \ref{f_dm} we show the obtained values for $f$ (top subpanels) and $\sigma_0$ (bottom subpanel) as a function of the maximum scale used in the fitting, namely $k_{\rm{max}}$. We show the results at different redshift: $z=0$ (left panels) and $z=1$ (right panels). Top, middle and bottom panels show the derived values for $f$ and $\sigma_0$ for each of the multipoles: monopole, quadrupole and hexadecapole as indicated. As in Fig. \ref{multipoles_dark_matter}, dotted lines stands for Kaiser model, dashed lines for Scoccimarro model and solid lines for TNS model. For simplicity, we only show the results corresponding to 1L-SPT (red lines) and 2L-RPT-${\cal N}_1$ (green lines). 2L-SPT and 2L-RPT-${\cal N}_2$ yield similar results. In top subpanels, horizontal solid black line shows the true value for $f$, whereas black dot-dashed horizontal lines show the $1\%$ and $2\%$ deviation, as labeled. In the bottom subpanels, the horizontal lines stands for the theoretical predictions of $\sigma_0$ according to Eq. \ref{sigmav} when $P_{\theta\theta}$ (blue line) and $P^{\rm lin}$ (orange line) are used as inputs. In the case of $P_{\theta\theta}$ we use 1L-SPT prediction. All the error bars correspond to $1-\sigma$ errors for $f$ and $\sigma_0$, and have been computed from the contour in the $f$-$\sigma_0$ space that corresponds to $\Delta\chi^2=2.3$. Since our N-body sample consists of 160 realizations ($\sim13^3 [{\rm Gpc}/h]^3$ volume), the dispersion in the measured monopole, quadrupole and hexadecapole is small. Therefore the corresponding error bars for the recovered parameters are also small, especially in the case of the monopole and quadrupole. However, one should be aware that these statistical errors in $f$ and $\sigma_0$ are not comparable to the expected errors from future surveys. These errors just provide information about the uncertainties and shortcomings of the models. Note that  for most of the studied models, these errors are much smaller than the expected ones from any galaxy survey. 
 
From Fig. \ref{f_dm}, we see that the TNS model using both 1L-SPT and 2L-RPT-${\cal N}_1$ is the only model able to recover the value of $f$ to $1\%$ accuracy: for the monopole up to $k\simeq0.15\,h/{\rm Mpc}$ for $z=0$ and up to $k\simeq0.20\,h/{\rm Mpc}$ for $z=1$; for the quadrupole up to $k\simeq0.20\,h/{\rm Mpc}$ for $z=0$ and up to $k\simeq0.25\,h/{\rm Mpc}$ for $z=1$; and for the hexadecapole up to $k\simeq0.15\,h/{\rm Mpc}$ for $z=0$ and up to $k\simeq0.20\,h/{\rm Mpc}$ for $z=1$. However, in the case of the hexadecapole, the statistical errors are too large to be able to claim $1\%$ accuracy of the model predictions.
Also note the difference between the theoretical value of $\sigma_0$ and the best-fit value obtained from the N-body data: at $z=1$ for both monopole and quadrupole, the best-fit $\sigma_0$ value for TNS model + 2L-RPT-${\cal N}_1$ yields a very similar result as Eq. \ref{sigmav}. However at $z=0$ there is a large discrepancy between these two values.

\subsection{Halo biasing and stochasticity}
\label{secbias}
So far, we have been able to recover the $f$ parameter with high accuracy using the dark matter multipoles. However, galaxy redshift surveys consist of galaxies residing in dark matter haloes, which are biased and stochastic tracers of the underlying dark matter field. Furthermore, since we only consider massive isolated haloes, we do not expect any FoG effects in the halo redshift-space statistics. 

In this paper, we model the biasing, i.e. the relation between the halo overdensity and the the dark matter density contrast as,
\begin{equation}
\delta_{\rm h}({\bf k})= b(k)\delta({\bf k})+\epsilon({\bf k}),
\label{delta_field}
\end{equation}
where $b(k)$ is the scale-dependent bias function and $\epsilon$ describes a stochastic field. The stochastic field $\epsilon$ stands for any physical or statistical process that produces a non-deterministic relation between the dark matter and the halo field. This includes the shot noise due to the discrete nature of haloes\footnote{In N-body simulations, dark matter particles represent also a discrete field. However, the number density of particles is large enough to render this effect negligible.}. For a Poisson process the shot noise is inversely proportional to the mean number density, namely  $n_{\rm h}$,
\begin{equation}
P_{\rm{Poisson}}=\frac{1}{n_{\rm h}}.
\end{equation}
However, the formalism used here allows for other stochastic processes. Assuming the $\epsilon$ field to be uncorrelated with $\delta$, namely, $\langle\delta\epsilon\rangle=0$, the bias function can be written as,
\begin{equation}
\label{bias_eq}b(k)=\frac{\langle\delta_{\rm h}({\bf k}) \delta({\bf k}')\rangle\delta^D({\bf k}+{\bf k}')(2\pi)^3}{\langle\delta({\bf k})\delta({\bf k}')\rangle\delta^D({\bf k}+{\bf k}')(2\pi)^3}\equiv\frac{P_{\rm mh}(k)}{P_{\rm mm}(k)},
\end{equation}
where the second equality stands for the numerator and denominator independently. Here the subscripts ``h" and ``m" stand for haloes and  dark matter respectively.
The power spectrum of the $\epsilon$ field can be computed combining Eq. \ref{delta_field} and \ref{bias_eq},
\begin{equation}
P_{\epsilon\epsilon}(k)\equiv\langle\epsilon({\bf k})\epsilon({\bf k}')\rangle \delta^D({\bf k}+{\bf k}')(2\pi)^3=\langle\delta_{\rm h}({\bf k})\delta_{\rm h}({\bf k}')\rangle\delta^D({\bf k}+{\bf k}')(2\pi)^3-\frac{P_{\rm mh}(k)^2}{P_{\rm mm}(k)}.
\end{equation}
Finally, we define the (shot noise free) halo-halo power spectrum $P_{\rm hh}$ as,
\begin{equation}
P_{\rm hh}(k)\equiv\langle\delta_{\rm h}({\bf k})\delta_{\rm h}({\bf k}')\rangle\delta^D({\bf k}+{\bf k}')(2\pi)^3-P_{\epsilon\epsilon}(k).
\end{equation}
Note that by this definition the following equality holds,
\begin{equation}
P_{\rm hh}(k)=b(k)P_{\rm mh}(k)=b(k)^2P_{\rm mm}(k) .
\end{equation}
Hence, in this biasing scheme, the bias functions that relate the halo-matter and halo-halo power spectra to the matter-matter power spectrum are the same. This is a reasonable approximation at least on large scales \citep{hgm}, where the bias becomes linear.
\begin{figure}
\centering

\includegraphics[clip=false, trim= 10mm 0mm 20mm 0mm,scale=0.28]{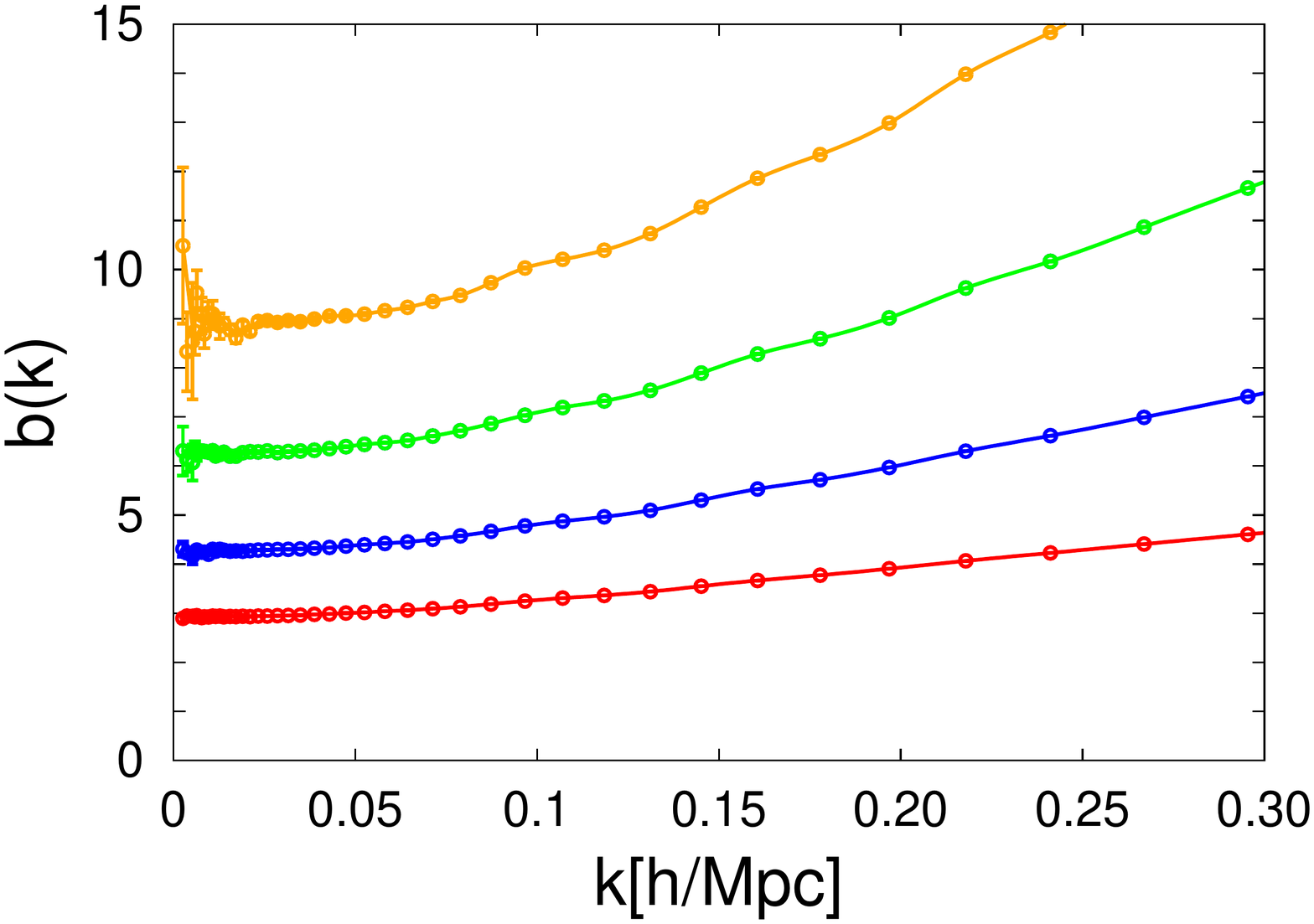}
\includegraphics[clip=false, trim= 10mm 0mm 20mm 0mm,scale=0.28]{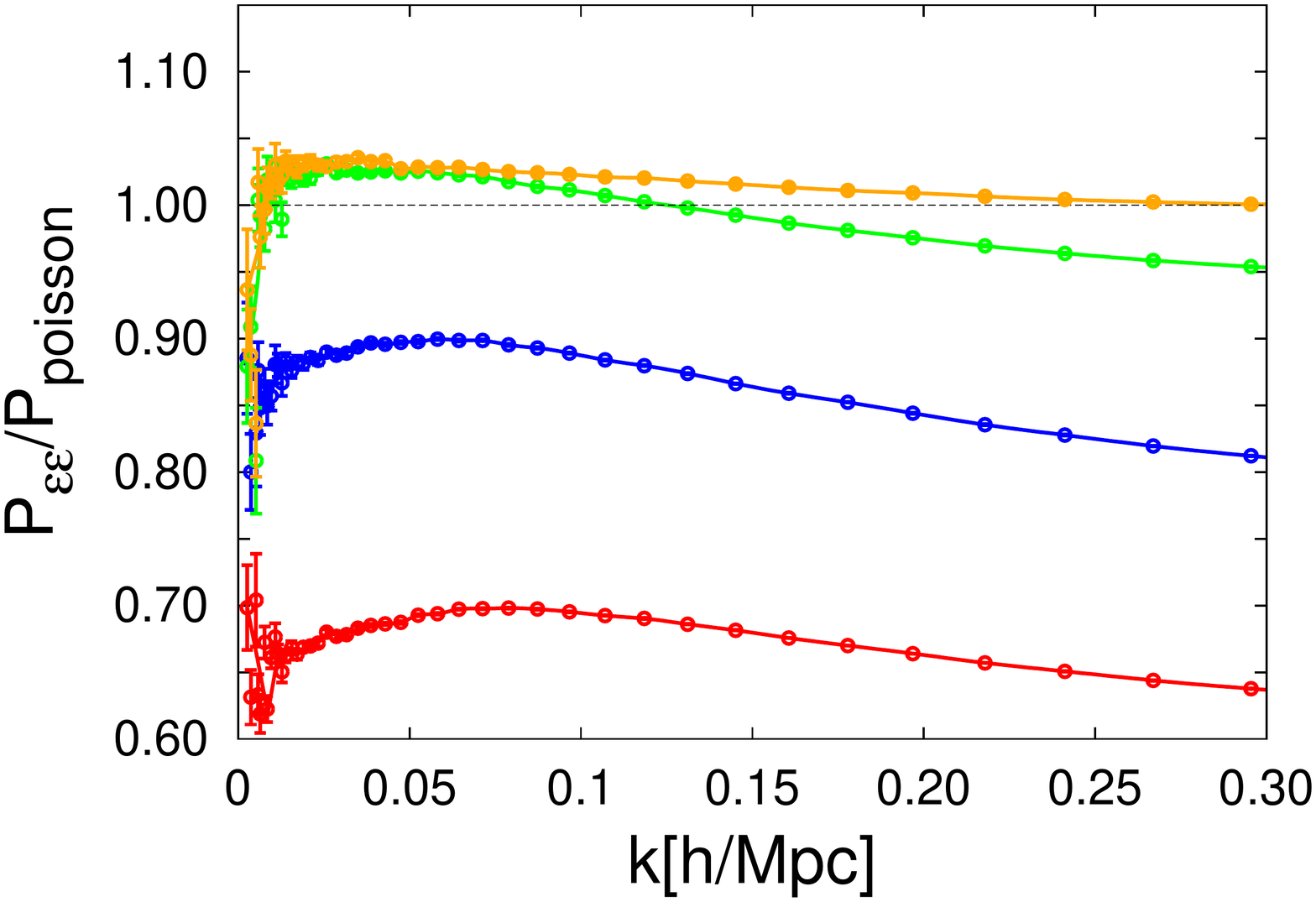}

\caption{{\it Left panel}: Halo bias for different redshifts snapshots with $M_{\rm cut}=10^{14}M_\odot/h$. {\it Right panel}: $\epsilon$-field power spectrum normalized to the Poisson prediction, associated to haloes with $M_{\rm cut}=10^{14}\,M_\odot/h$. In both panels, red lines are for $z=0$, blue lines are for $z=0.5$, green lines are for $z=1$ and orange lines are for $z=1.5$. }
\label{bias_noise}
\end{figure}

In Fig. \ref{bias_noise} we show the scale dependence of the halo bias (left panel) and the $\epsilon$-field power spectrum (right panel) measured from the halo catalogues at different redshifts. Remember that the minimum halo mass is $M_{\rm cut}=10^{14} M_\odot/h$, which ensures that all the haloes have at least $\sim40$ particles. The number density of haloes with this mass cut at different redshifts is shown in Table \ref{table_haloes}.
\begin{table}[htdp]
\begin{center}
\begin{tabular}{c|cccc}
 redshift & 0 & 0.5 & 1.0 & 1.5 \\
 \hline
 $n_{\rm h}$ in $(h/{\rm Mpc})^3$ & $1.79\times 10^{-5}$ & $7.02\times 10^{-6}$ & $1.66\times 10^{-6}$ & $2.33\times 10^{-7}$
\end{tabular}
\end{center}
\caption{Number density of haloes at different redshifts with a minimum mass of $M_{\rm cut}=10^{14}M_\odot/h$.}
\label{table_haloes}
\end{table}%
Fig. \ref{bias_noise} shows that the bias increases with $z$ and with $k$. The $P_{\epsilon\epsilon}$ increases with $z$ and slightly decreases  with $k$. For $M_{\rm cut}=10^{14}M_\odot/h$, the $P_{\epsilon\epsilon}$ is sub-Poissonian at low redshifts, $65\%$ at $z=0$ and $90\%$ at $=0.5$, but  turns out to be very close to the Poissonian prediction at high redshifts, $z=1$ and $z=1.5$. 

\begin{figure}
\centering
\includegraphics[scale=0.28]{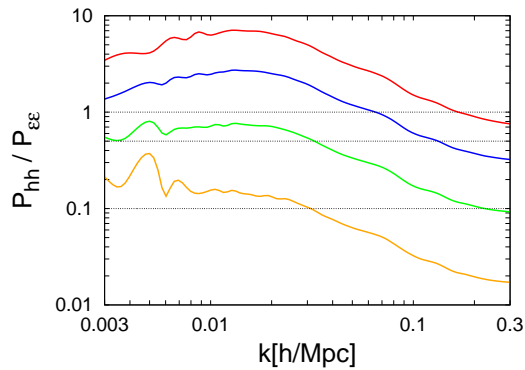}
\caption{Signal-to-noise for haloes with mass above $M_{\rm cut}=10^{14}M_\odot/h$ for $z=0$ (red line), $z=0.5$ (blue line), $z=1.0$ (green line) and $z=1.5$ (orange line). For reference, the values corresponding to 1, 0.5 and 0.1 of the signal-to-noise value have been plotted in horizontal black dotted lines.}
\label{S2N}
\end{figure}

In Fig. \ref{S2N} we show the signal-to-noise ratio (or $P_{hh}/P_{\epsilon\epsilon}$) for the haloes studied here at different redshifts: $z=0$ (red line), $z=0.5$ (blue line), $z=1.0$ (green line) and $z=1.5$ (orange line). Horizontal dotted lines mark the values $P_{\rm hh}/P_{\epsilon\epsilon}=1.0$, 0.5 and 0.1 as a reference. In Table \ref{table_noise} the scale at which the signal-to-noise ratio reaches these values is written for the same $z$-snapshots. We see that only for $z=0$ and $z=0.5$ the signal-to-noise ratio is above 0.5 at scales $k<0.1h\,{\rm Mpc}$, whereas for $z=1.0$ is above 0.5 only at very large scales, $k<0.03\,h/{\rm Mpc}$ and never for $z=1.5$. Conservatively, in this paper we consider only the scales where $P_{hh}/P_{mm}\gtrsim0.5$ to be suitable for extracting information. Hence, we do not study the halo power spectra at $z=1.5$ and $z=1.0$ because only at very large scales (where the behavior is linear) the signal-to-noise ratio satisfies this condition.


\begin{table}[htdp]
\begin{center}
\begin{tabular}{c|cccc}
 $z$ & 0 & 0.5 & 1.0 & 1.5 \\
 \hline
 $P_{\rm hh}/P_{\epsilon\epsilon}=1.0$ & $0.168\,h/{\rm Mpc}$ & $0.065\,h/{\rm Mpc}$ & --- & --- \\
 $P_{\rm hh}/P_{\epsilon\epsilon}=0.5$ & --- &$0.132\,h/{\rm Mpc}$  & $0.033\,h/{\rm Mpc}$  & --- \\
  $P_{\rm hh}/P_{\epsilon\epsilon}=0.1$ & --- & ---  & $0.225\,h/{\rm Mpc}$  & $0.031\,h/{\rm Mpc}$\\
\end{tabular}
\end{center}
\caption{  Scale where the noise starts to be comparable to the signal: $P_{\rm hh}/P_{\epsilon\epsilon}=1.0,\, 0.5,\, {\rm and }\,\, 0.1$ for haloes with $M_{\rm cut}=10^{14}M_\odot/h$ at $z=0$, 0.5, 1.0 and 1.5.}
\label{table_noise}
\end{table}%

\subsection{High-bias halo power spectrum multipoles}
The  multipoles  for the halo power spectrum are defined by Eq. \ref{real_Ps} just changing $P^s_{\rm mm}(k,\mu)$ by $P^s_{\rm hh}(k,\mu)$. In this case, we define the halo-power spectrum in redshift space by
\begin{equation}
P^s_{\rm hh}(k,\mu)=\langle \delta_{\rm h}^s({\bf k})\delta_{\rm h}^s({\bf k}')\rangle \delta^D({\bf k}+{\bf k}')(2\pi)^3-P_{\epsilon\epsilon}(k),
\end{equation}
where we assume that $P_{\epsilon\epsilon}$ does not depend on $\mu$. For the monopole term, the shot noise subtraction is important. However, since we are assuming that $P_{\epsilon\epsilon}$ does not depend on $\mu$, it is irrelevant for higher-order moments. Indeed, a $\mu$-independent offset on $P^s_{\rm hh}(k,\mu)$ has no effect in the quadrupole and hexadecapole, only in the monopole. As in Eq. \ref{monopole_eq}-\ref{hexadecapole_eq}, at very large scales, the halo multipoles are written as,
\begin{eqnarray}
\label{monopole_bias_eq}P_0(k)&=&P^{\rm lin}(k)\left(b(k)^2+\frac{2}{3}b(k)f+\frac{1}{5}f^2\right), \\
\label{quadrupole_bias_eq}P_2(k)&=&P^{\rm lin}(k)\left(\frac{4}{3}b(k)f+\frac{4}{7}f^2 \right),\\
\label{hexadecapole_bias_eq}P_4(k)&=&P^{\rm lin}(k)\left( \frac{8}{35}f^2  \right).
\end{eqnarray}
Note that for the case of biased tracer, $f$ and $b$ are degenerate in the linear regime when we treat $b^2P^{\rm lin}$ as input or when any ratio between different multipoles is used to constrain $f$. In this case, only the ratio among them, $\beta(k)\equiv f/b(k)$ can be measured. However, as for the dark matter case, non-linearities cause deviations from these formulae. Depending on the ability of modeling the redshift space distortions, we will be able to use information from non-linear scales to estimate $f$ with accuracy but also, in the case of biased tracers, we might be able to break the degeneracy between the bias and $f$. For both the Kaiser and the Scoccimarro model, $f$ and $b$ always appear in the $\beta$ combination, only for the TNS model this degeneracy is not exact.

\begin{figure}
\centering
\includegraphics[clip=false, trim= 10mm 0mm 30mm 0mm,scale=0.28]{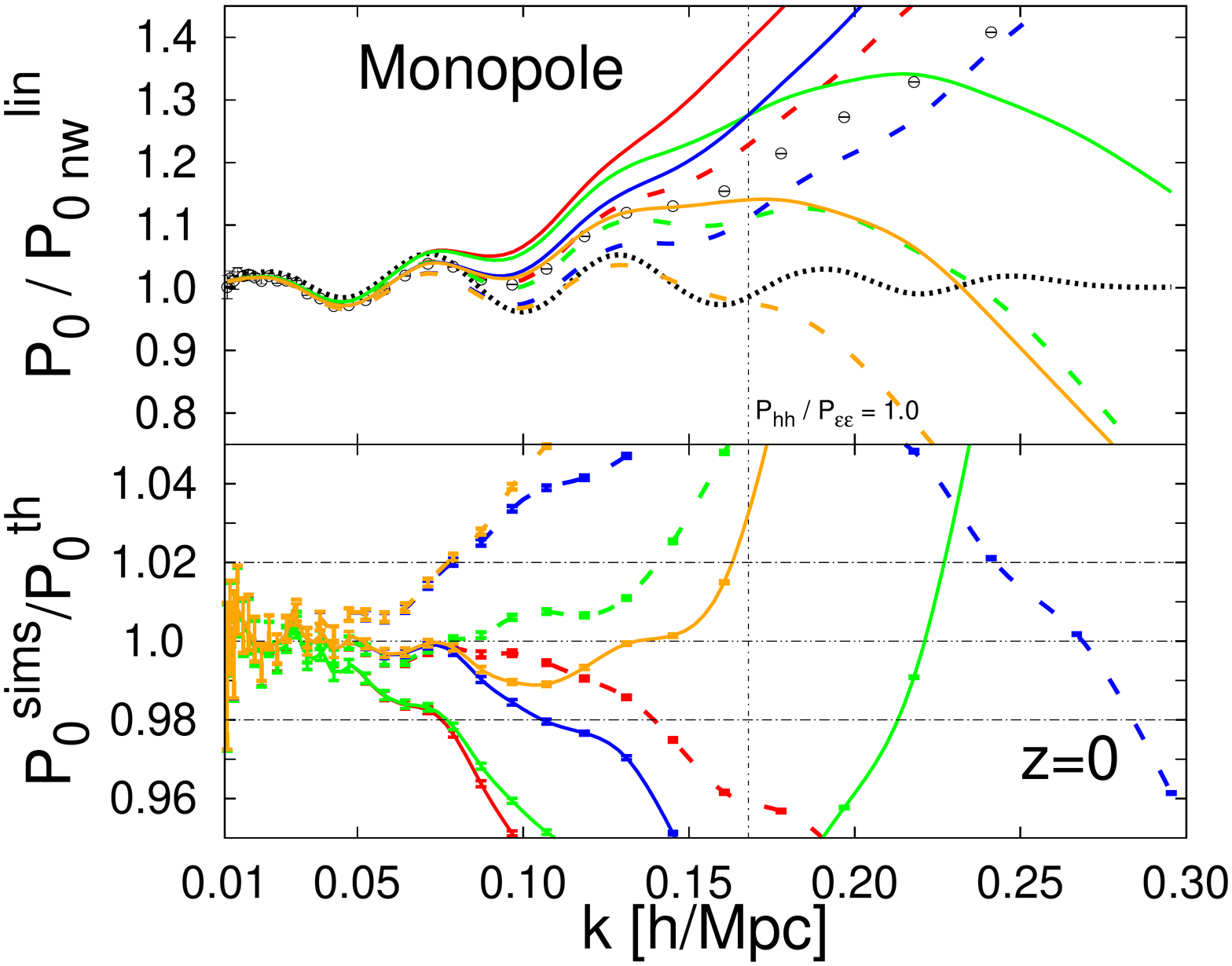}
\includegraphics[clip=false, trim= 10mm 0mm 30mm 0mm,scale=0.28]{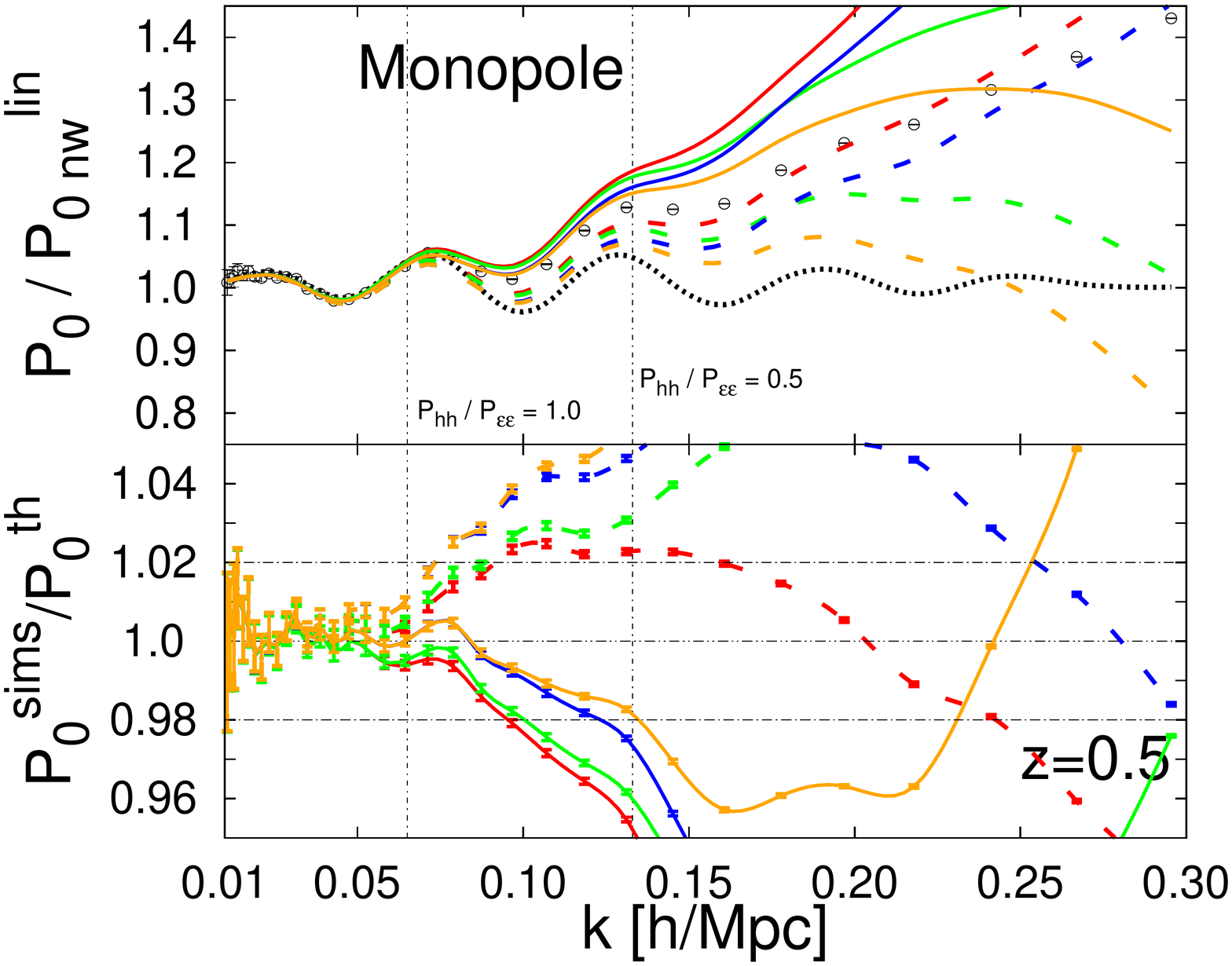}

\includegraphics[clip=false, trim= 10mm 0mm 30mm 0mm,scale=0.28]{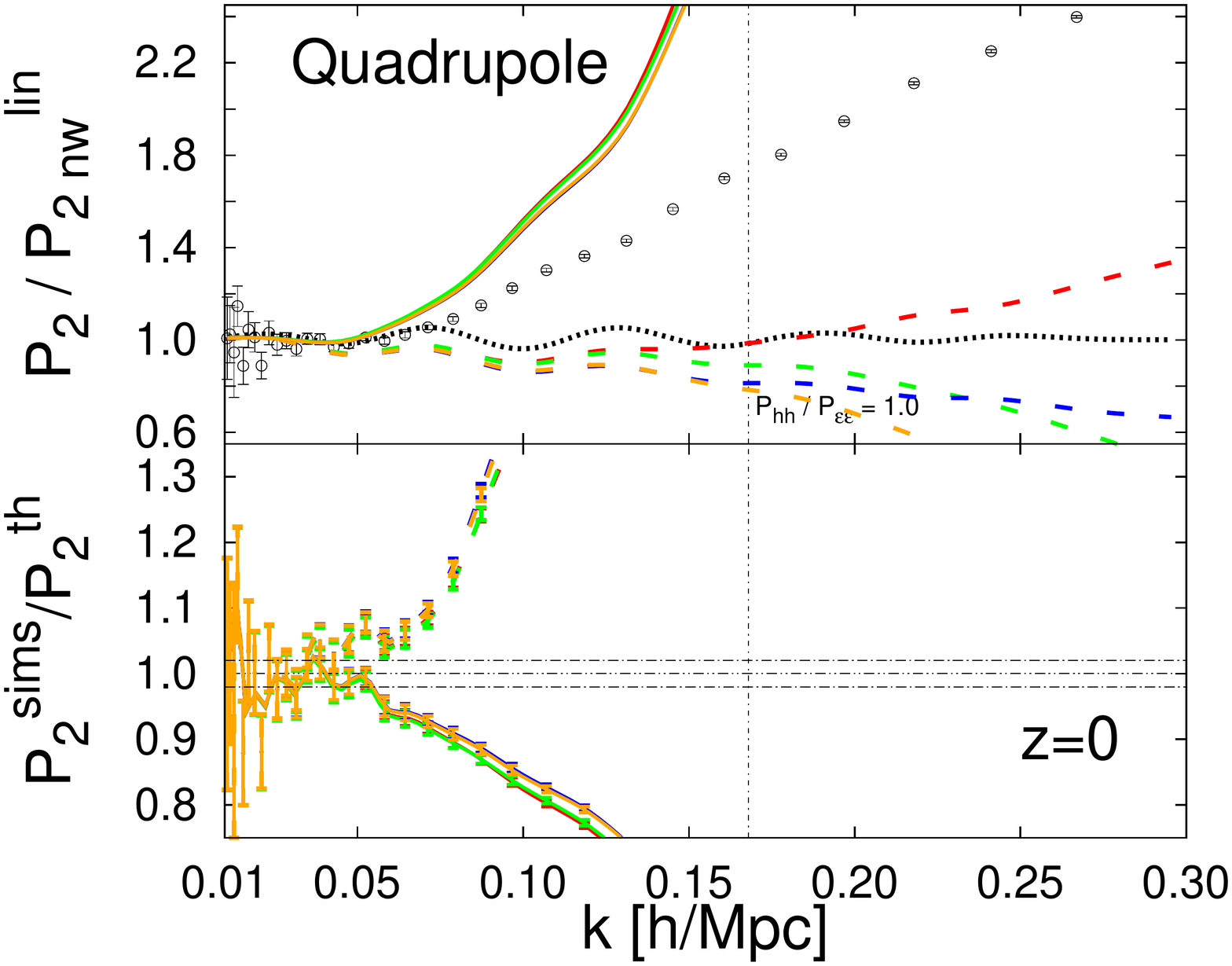}
\includegraphics[clip=false, trim= 10mm 0mm 30mm 0mm,scale=0.28]{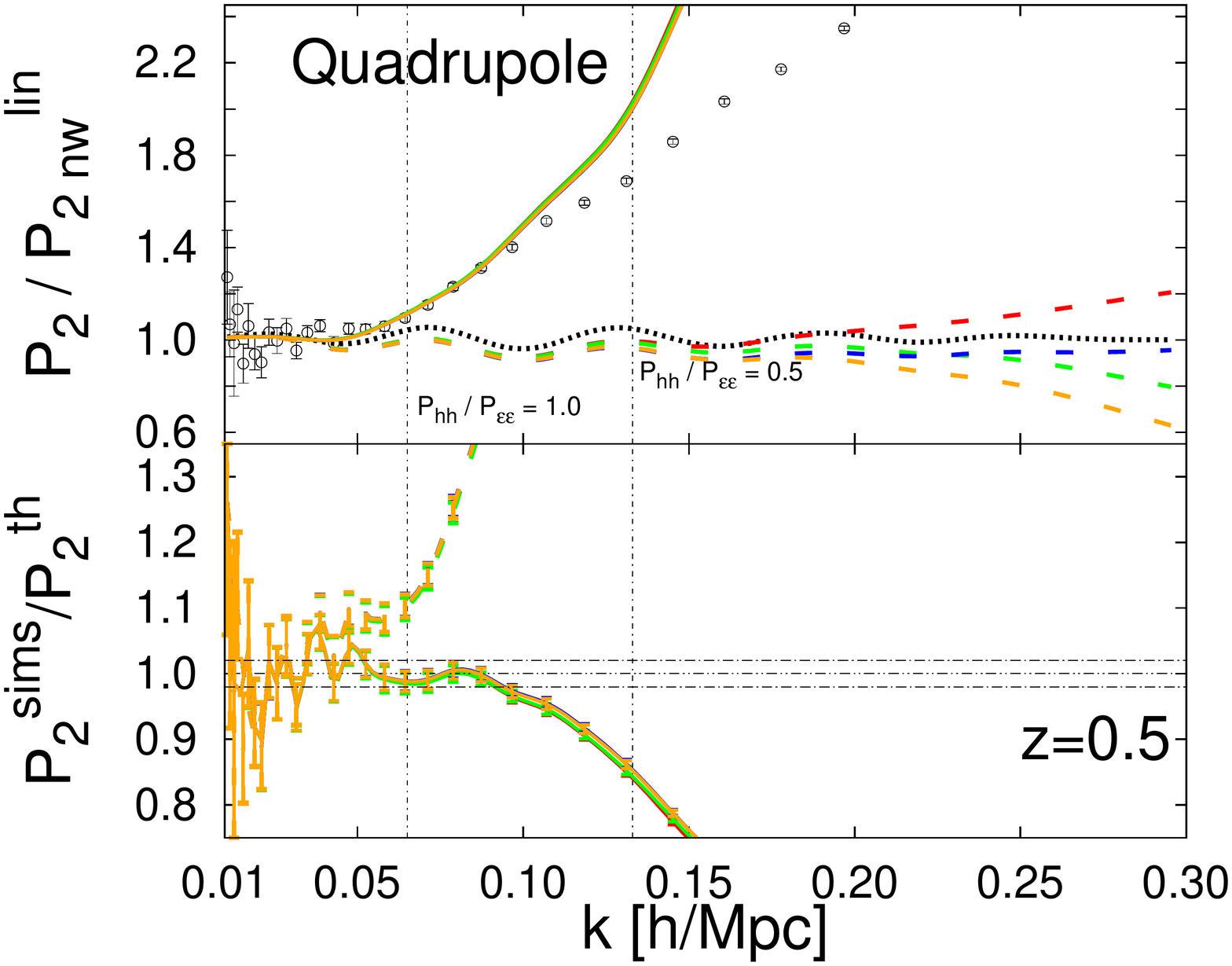}

\caption{Multipoles corresponding  the  halo-halo power spectrum: monopole (top panels), quadrupole (bottom panels), for $z=0$ (left panels) and $z=0.5$ (right panels). In the upper subpanels the value of the corresponding multipole is shown, normalized to the corresponding non-wiggle linear prediction to reduce the dynamical range. In the lower subpanels the ratio between the N-body simulation data and different PT-predictions is shown. Dashed lines corresponds to Scoccimarro model and solid lines to the TNS model. Different PT models are shown: linear prediction (black dotted lines), 1L-SPT (red lines), 2L-SPT (blue lines), 2L-RPT-${\cal N}_1$ (green lines) and 2L-RPT-${\cal N}_2$ (orange lines). In bottom subpanels the 2\% deviation is marked with black dot-dashed horizontal lines. Vertical dot-dashed lines mark the regions where $P_{\rm hh}/P_{\epsilon\epsilon}=1.0$ and $P_{\rm hh}/P_{\epsilon\epsilon}=0.5$ as labeled.}
\label{multipoles_haloes}
\end{figure}

Fig. \ref{multipoles_haloes} shows the halo monopole (top panels) and halo quadrupole (bottom panels) for $z=0$ (left panels) and $z=0.5$ (right panels). Both $f$ and $b(k)$ have been set to their true values. The color and line notation is the same as in Fig. \ref{multipoles_dark_matter}: dashed lines stands for Scoccimarro model and solid lines for TNS model. The different real-space power spectrum inputs are: linear prediction (black dotted lines), 1L-SPT (red lines), 2L-SPT (blue lines), 2L-RPT-${\cal N}_1$ (green lines) and 2L-RPT-${\cal N}_2$ (orange lines). Since we do not expect Fingers of God for isolated haloes, the theoretical models have no free parameters in this case.

From Fig. \ref{multipoles_haloes}, we see considerably different results compared to the dark matter case (see Fig. \ref{multipoles_dark_matter}). The accuracy of the modeling is  reduced, especially for the quadrupole at $z=0$.
 In the case of the halo-halo monopole, we see that the different PT models make very  different predictions,  while this was not the case for the dark matter.
  In particular we see that TNS + 2L-RPT-${\cal N}_2$ is the only model able to make sub-percent predictions at $z=0$ up to $k=0.15\,h/{\rm Mpc}$ and at $z=0.5$ up to $k=0.10\,h/{\rm Mpc}$ for the monopole. Any other PT theory + TNS yields worse results. We also see that Scoccimarro + 1L-SPT and Scoccimarro + 2L-RPT-${\cal N}_1$ provide a good description  at $z=0$ but not at $z=0.5$. This is due to an (accidental) cancellation of two terms that go in opposite directions. The Scoccimarro model does not take into account the $A$ and $B$ functions of the TNS model, that add a positive contribution to $P^s$ (see Fig. \ref{Taruya_terms} in the next section). On the other hand, in Fig. \ref{FigAz0} we have seen that 1L-SPT and 2L-RPT-${\cal N}_1$ over-predict the true value for $P_{\delta\delta}$ at $z=0$ (but also for $P_{\delta\theta}$ and $P_{\theta\theta}$). For  the dark matter power spectrum, both effects are small and negligible. However, for massive haloes, the high bias increases these effects. At $z=0.5$, the TNS approach is clearly modeling better the monopole than the Scoccimarro model, which under-predicts the N-body data by $\sim10\%$.
  
   In the case of the halo-halo quadrupole, we see that all the PT models make similar predictions, and that the main difference is due to the RSD model chosen but the modeling breaks down at relatively large, almost linear, scales. We see that the TNS model is able to describe well the quadrupole only up to $k=0.05\,h/{\rm Mpc}$ at $z=0$ and up to $k=0.10\,h/{\rm Mpc}$ at $z=0.5$. 

In general we see that for the monopole, all the models describe the N-body data better at $z=0$ than at $z=0.5$. This seems counter-intuitive, because at higher redshifts, non-linearities are less important and perturbation theory should work better. However,  it could be explained by the fact that at $z=0.5$ the signal-to-noise ratio is considerably less than at $z=0$ (see Table \ref{table_noise} and Fig. \ref{S2N}). For instance, recall that at $z=0$, $P_{\rm hh}/P_{\epsilon\epsilon}=1.0$ at $k=0.168\,h/{\rm Mpc}$ whereas at $z=0.5$ this happens already at $k=0.065\,h/{\rm Mpc}$. Also the bias of the selected haloes grows with redshift, making more apparent any systematic error in the model (of both RSD and  biasing).  It is reasonable to expect (but it remains to be tested) that the model performance improves for lower mass --thus less rare and less biased-- haloes.


\subsection{Simultaneously estimating $f$  and $b$ from halo multipoles}
In this section we show how well the $f$ parameter can be recovered from the halo-monopole N-body data. Since in the last section we have seen that none of the models studied here is able to reproduce the quadrupole data for haloes with sufficient accuracy at the mildly non-linear scales we are interested in, we do not try to recover $f$ from $P_2$. Instead we focus on the degeneracy between $f$ and the bias in the monopole. According to Eq. \ref{monopole_eq}, these two parameters are perfectly degenerate when $P(k)b^2$ is set from observations or when the $P_2/P_0$ ratio is used to compute $f$ and $b$. This is the case for the Scoccimarro and Kaiser models without Finger-of-God effects. However, non-linear terms, namely the $A$ and $B$ functions of the TNS model, are expected to break this degeneracy at non-linear scales even when $P(k)b^2$ is fixed: from Eq. \ref{taruya_a}-\ref{taruya_b}, we see that $f$ and $b$ do not appear always in the same combination in the $A$ and $B$ functions. In particular $A$ can be expressed as $b^3A(k,\mu,\beta)$ and $B$ as $b^4B(k\,\mu,\beta)$. Since in this paper the input is the dark matter power spectrum, the degeneracy between $b$ and $f$ is not perfect, and there is the possibility of recovering these parameters separately with certain accuracy even for the Kaiser and Scoccimarro models. For simplicity, we do not model or fit the scale dependence of the bias function $b(k)$, but instead assume that we do know the scale dependence of the bias, and try to recover the growth rate $f$ and the bias amplitude, $A_b$ defined to be 1,
\begin{equation}
b(k)\rightarrow A_b b(k).
\end{equation}
\begin{figure}
\centering

\includegraphics[clip=false, trim= 10mm 0mm 10mm 0mm,scale=0.28 ]{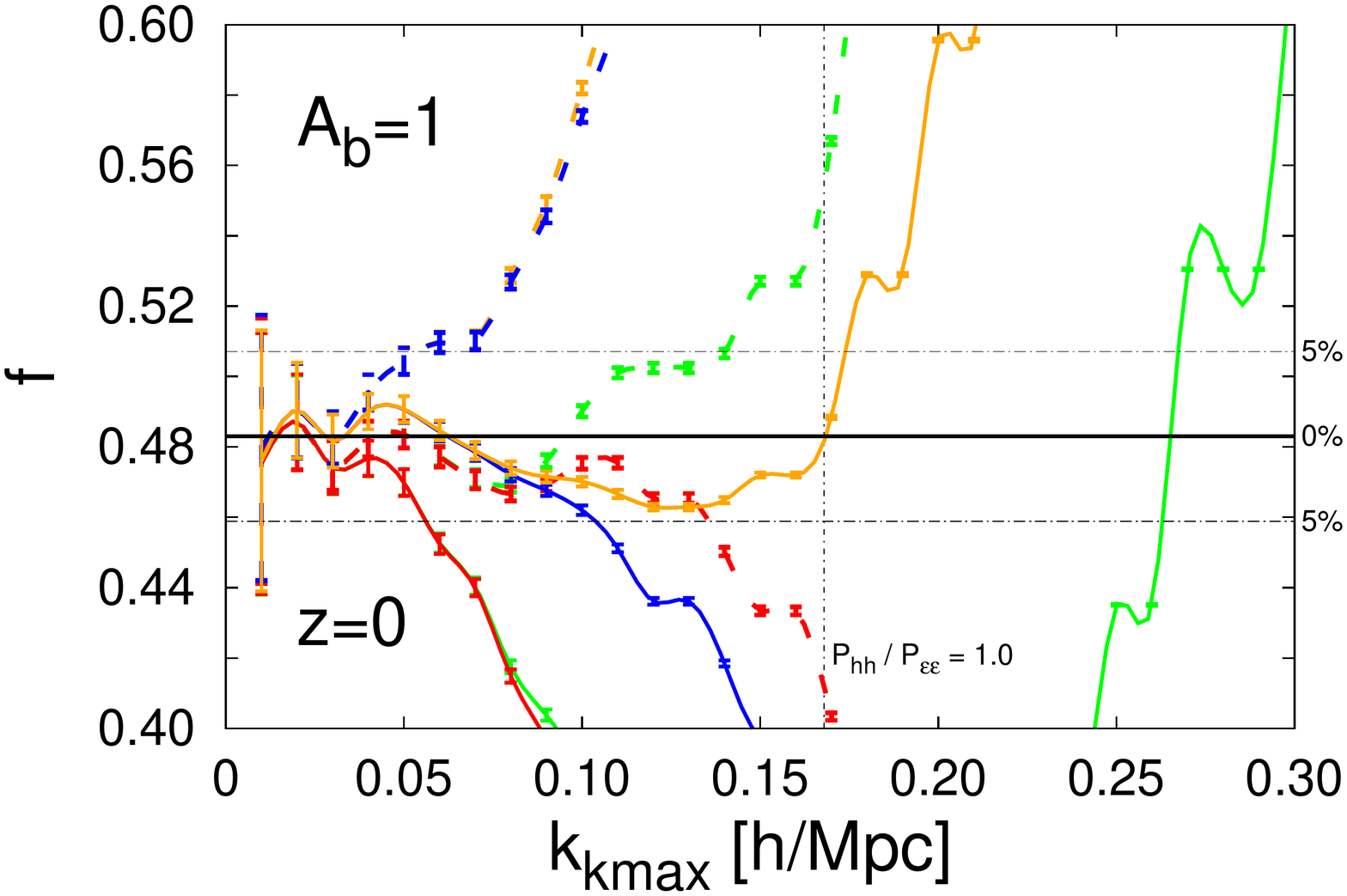}
\includegraphics[ clip=false, trim= 10mm 0mm 10mm 0mm,scale=0.28]{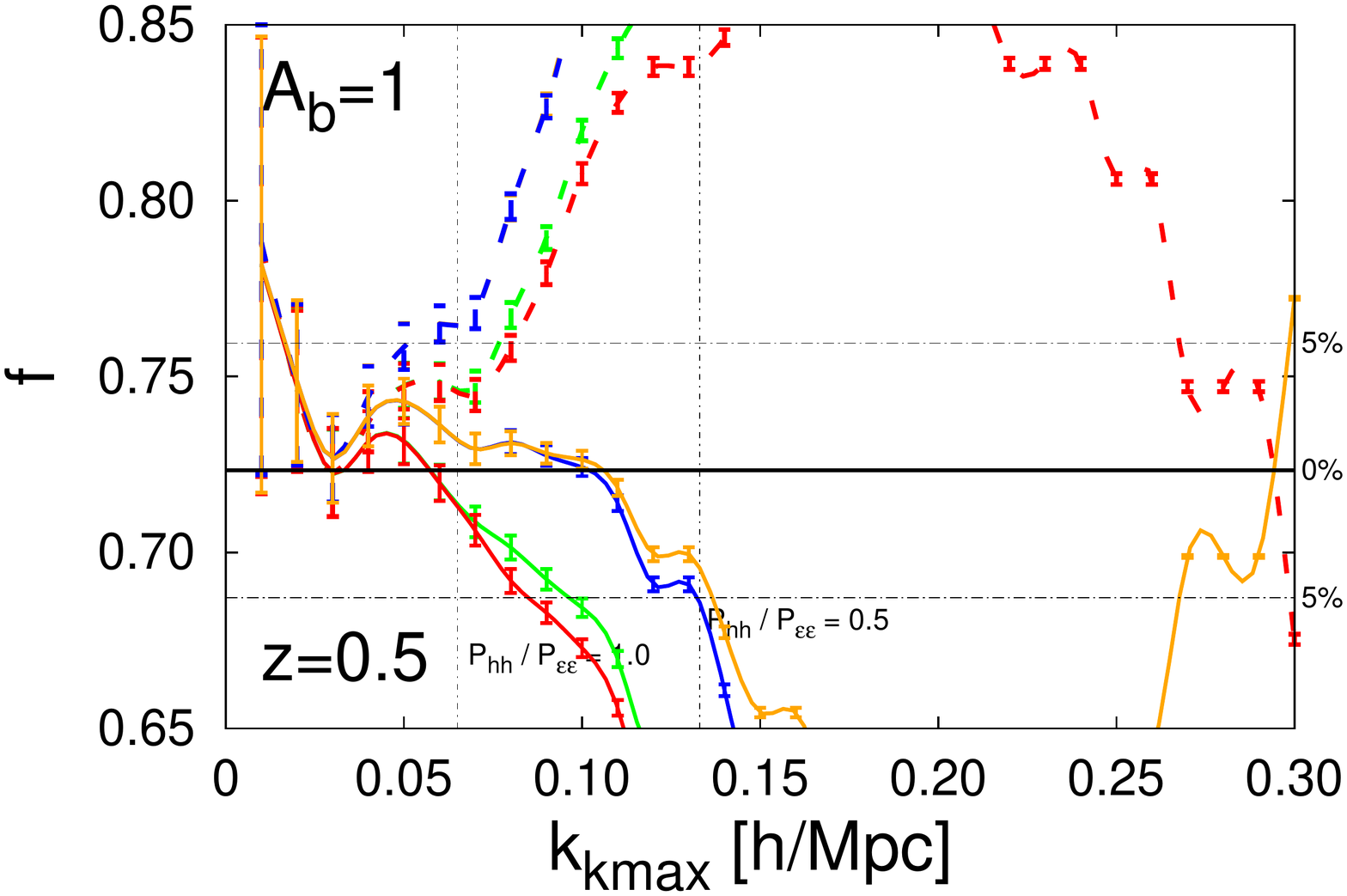}

\caption{Best-fit values for $f$ derived from the halo monopole for $z=0$ (left panel) and $z=0.5$ (right panel), assuming that the bias amplitude $A_b$ is perfectly known. Results from different theoretical models are shown: dashed lines correspond to Scoccimarro model and solid lines to the TNS model. Different PT models are shown: 1L-SPT (red lines), 2L-SPT (blue lines), 2L-RPT-${\cal N}_1$ (green lines) and 2L-RPT-${\cal N}_2$ (orange lines) whereas 5\% deviations are shown in the horizontal black dot-dashed lines, as indicated. Errors correspond to $1-\sigma$ error, or $\Delta\chi^2=1.0$.}
\label{refind_beta_haloes}
\end{figure}
Realistic approaches need an analytical modeling of the scale dependence of the bias, that in principle one could expand perturbatively e.g.,  \cite{wang_szalay,zheng}.  Here we focus on the modeling  of the dark matter and redshift space distortions and thus we measure the bias directly from the simulations.

First, we assume that we also know that $A_b=1$ when we fit $f$. The fitting results are shown in Fig. \ref{refind_beta_haloes} using the same color and line notation as in Fig. \ref{multipoles_haloes}. As before, the error bars correspond to the interval defined by $\Delta\chi^2$=1.0.

As we have commented for the dark matter case, these errors are much smaller than the ones which could be obtained form a real survey. They contain information of 160 realizations of a large volume, the total volume is close to that enclosed by an all-sky survey up to $z=25$. 

For both $z=0$ and $z=0.5$ the models that do best at recovering $f$ are TNS + 2L-SPT (solid blue line) or TNS + 2L-RPT-${\cal N}_2$ (solid orange line), especially the latter one. In particular, at $z=0$, the TNS model with 2L-RPT-${\cal N}_2$ implementation is able to recover the $f$ parameter within a $5\%$ accuracy up to scales $k=0.17\,h/{\rm Mpc}$ and up to scales $k=0.13\,h/{\rm Mpc}$ at $z=0.5$. Scoccimarro models with accurate implementations of the real-space power spectra such as 2L-SPT and 2L-RPT-${\cal N}_2$ largely over-predict $f$ at $k>0.05\,h/{\rm Mpc}$.
\begin{figure}

\centering

\includegraphics[clip=false, trim= 10mm 0mm 10mm 0mm,scale=0.28 ]{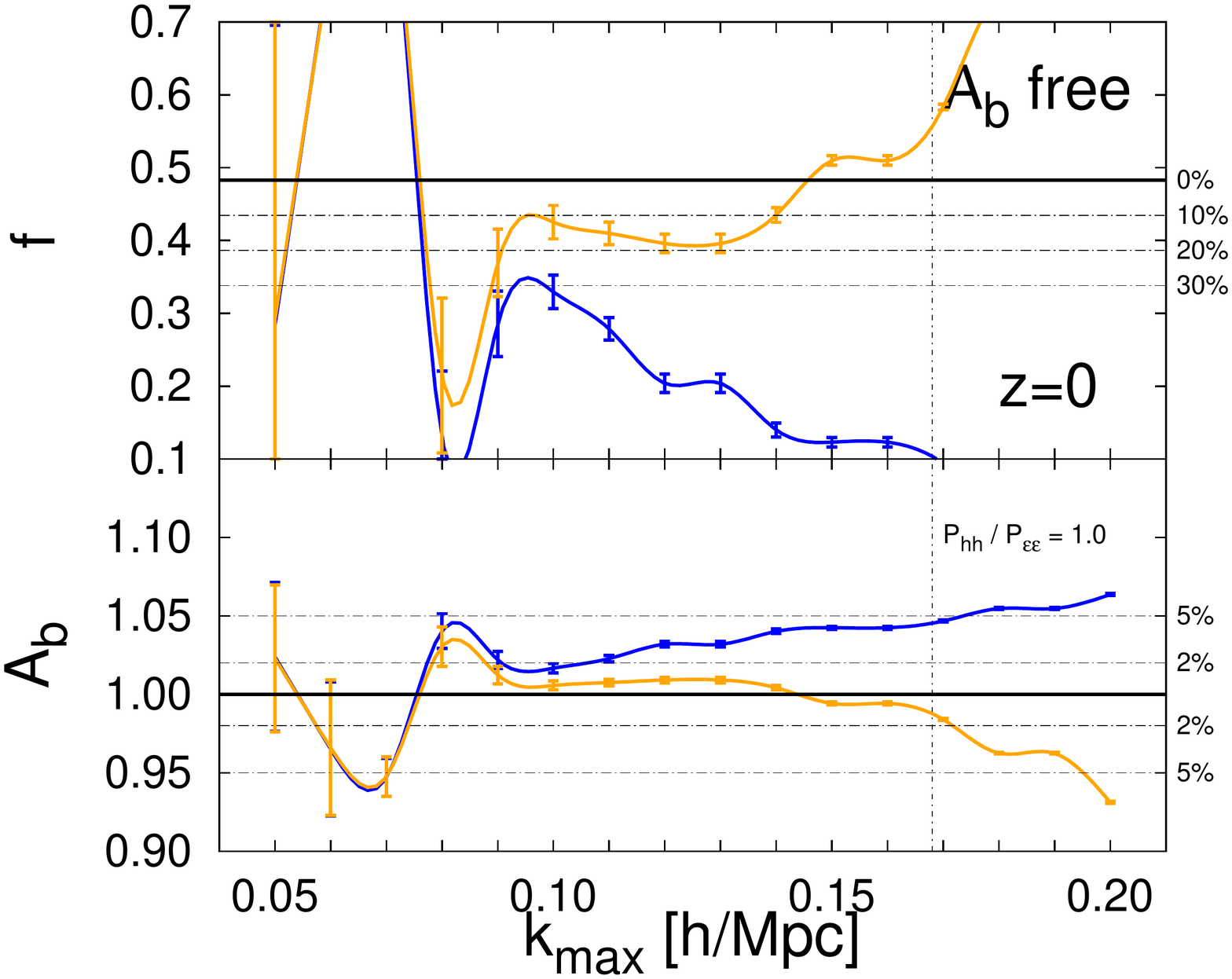}
\includegraphics[ clip=false, trim= 10mm 0mm 10mm 0mm,scale=0.28]{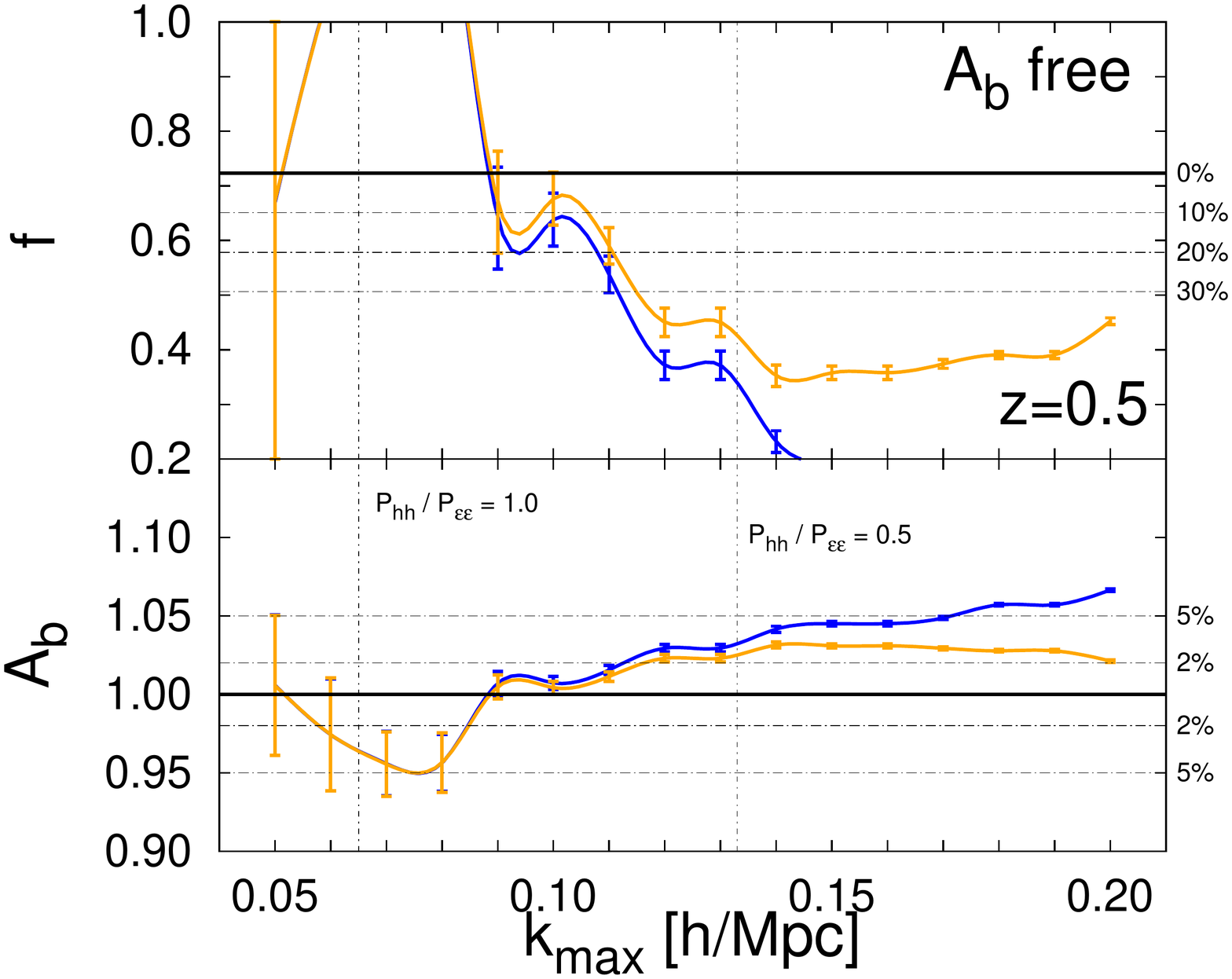}

\caption{Estimates for the parameter $f$ corresponding to halo monopole for $z=0$ (left panel) and $z=0.5$ (right panel), treating $f$ and $A_b$ as free parameters. Only the models TNS + 2L-SPT (solid blue line) and TNS + 2L-RPT-${\cal N}_2$ (solid orange line) are shown for simplicity. Errors correspond to $1-\sigma$, or $\Delta\chi^2=2.3$.}
\label{refind_f_haloes}
\end{figure}

\begin{figure}

\centering

\includegraphics[clip=false, trim= 10mm 0mm 10mm 0mm,scale=0.295 ]{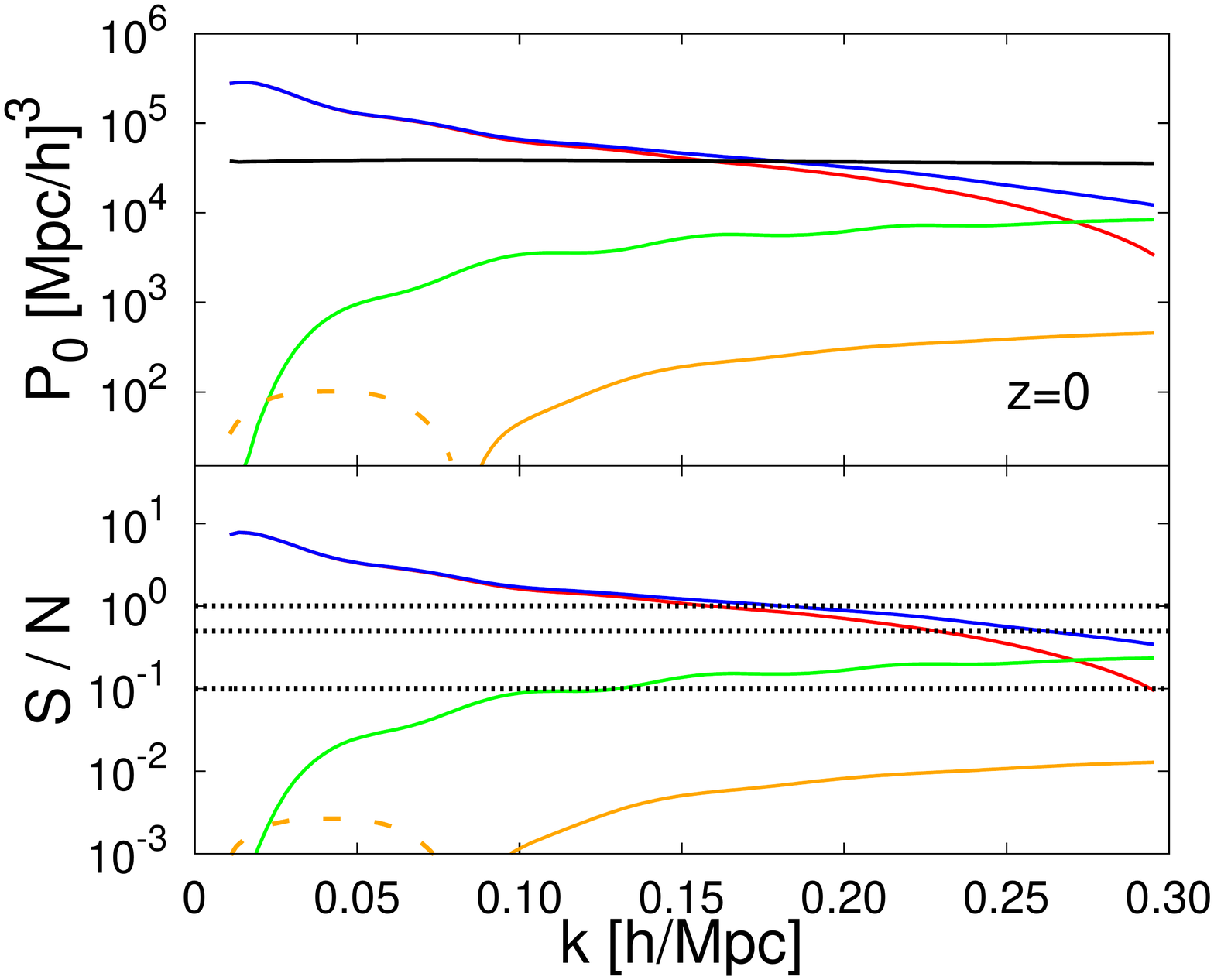}
\includegraphics[ clip=false, trim= 10mm 0mm 10mm 0mm,scale=0.295]{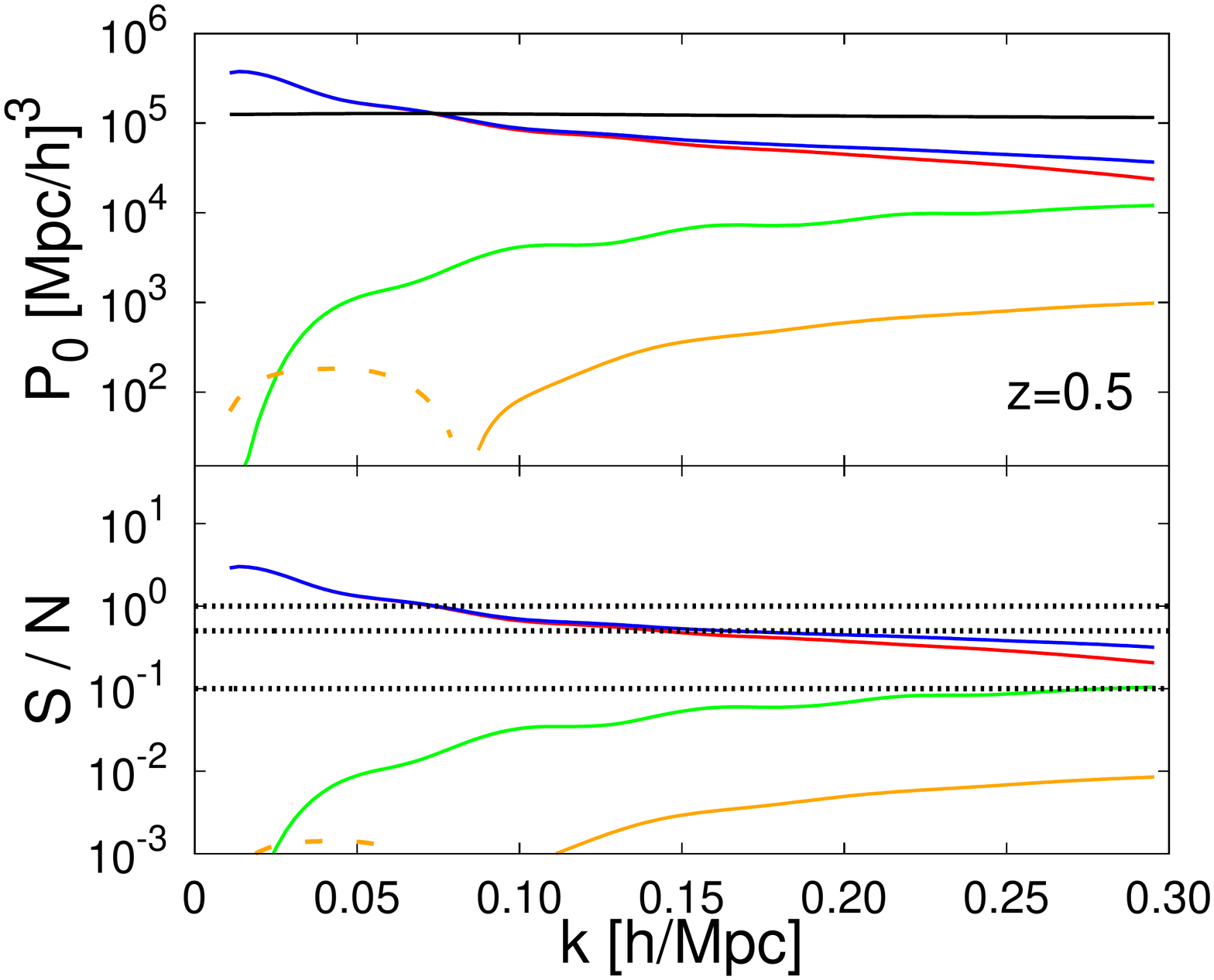}

\caption{Top subpanels: different contribution terms to the halo-monopole for 2L-RPT-${\cal N}_2$ model: Scoccimarro model (red line), TNS model (blue model), TNS-A function (green line), TNS-B function (orange line) and $P_{\epsilon\epsilon}$ (black line). Bottom subpanels: ratio between these models and $P_{\epsilon\epsilon}$, or the signal-to-noise (S/N). Dashed lines indicate a negative contribution. Horizontal dotted lines in bottom subpanels mark the reference quantities: S/N=1, 0.5, and 0.1. Left panels at $z=0$ and right panels at $z=0.5$.}
\label{Taruya_terms}
\end{figure}

Next we show in Fig. \ref{refind_f_haloes} the results of extracting both $A_b$ and $f$ from the halo monopole at the same time. We only show the two models that performed best at extracting $f$ when $A_b$ is known: TNS + 2L-SPT (solid blue line) and TNS + 2L-RPT-${\cal N}_2$ (solid orange line). Since now there are two free parameters, the $1-\sigma$ error bars correspond to contours of $\Delta\chi^2=2.3$ in the $f-b$ space. We see that in general, at $z=0$ and $z=0.5$, 2L-RPT-${\cal N}_2$ works better than 2L-SPT, as we have already observed in Fig. \ref{refind_beta_haloes}. However, 2L-RPT-${\cal N}_2$ tends to overestimate the bias amplitude and underestimate the $f$ parameter. In particular, at $z=0$ and $z=0.5$, at scales around $k\simeq 0.11\,h/{\rm Mpc}$, this model overestimate $A_b$ by $1\%$ and underestimate $f$ by about $15\%$. However, when $A_b$ was set 1, the  systematic error of  $f$ was less than $5\%$. Similar results were obtained by \cite{kwan}, when the TNS model was combined with the Closure perturbation theory of \cite{closure}. We have checked that this large underestimation in $f$ is due to the $1\%$ overestimation in the bias amplitude. In other words, if the bias amplitude is set by hand to a fixed value of $A_b=1.01$, we obtain similar results as in Fig \ref{refind_f_haloes}. Thus, this formalism tends to underestimate $f$ by $\simeq15\%$ while overestimating $A_b$ by only $\lesssim1\%$. Since the Scoccimarro model + 2-loop perturbation theory predictions were not able to predict $f$ when $A_b$ was assumed to be 1, the TNS terms $A$ and $B$ are the key ingredient of the TNS model to achieve a high accuracy recovering $f$ when the bias amplitude is assumed. However, the signal-to-noise ratio of these terms is low compared to the signal-to-noise of the whole monopole term. In Fig. \ref{Taruya_terms} we show this for $z=0$ (left panel) and $z=0.5$ (right panel). In top subpanels, the total contribution of Scoccimarro and TNS models is shown in red and blue lines respectively. In green and orange lines, the isolated contribution of $A$ and $B$ of the TNS model is shown respectively. Black line shows the stochastic noise, $P_{\epsilon\epsilon}$. In bottom subpanels, the ratio of all the signal terms with respect to $P_{\epsilon\epsilon}$ is shown. As a reference, the lines where ${\rm S/N}=1,\,0.5$ and 0.1 are also shown as black dotted lines. We see that the signal associated to the $A$ and $B$ terms is much less than for the $P_{\delta\delta}$, $P_{\delta\theta}$ and $P_{\theta\theta}$ term. In particular, we see that   for the scales of interest ($0.10\,h /{\rm Mpc} < k <0.15\,h/{\rm Mpc}$)   at $z=0$ the signal for $A$ and $B$ terms is about ${\rm S/N}=0.1$ and even less at $z=0.5$. This means that in Fig. \ref{refind_beta_haloes}, the crucial difference between red and blue lines (for a given PT model) comes from terms with low signal-to-noise ratio. In Fig. \ref{refind_f_haloes}, when $f$ and $A_b$ are allowed to vary, the low signal that the $A$ and $B$ terms have, has to be split to find one more parameter, and then, the accuracy recovering $f$ must decrease. In order to break the degeneracy between $f$ and $b$ from redshift space distortions, the signal-to-noise of the  non-linear A and B terms  must be as high as possible and it could be optimized by selecting haloes with suitable cuts in mass. The shot noise increases with the  mass cut  but also  does the bias. When optimizing for a measurement of the  angle-averaged power spectrum, for a fixed $P_{mm}$, the signal-to-noise scales  as a function of mass cut like $b^2/{\rm shot\, noise}$. If the shot noise can be approximated as Poissonian then the signal-to-noise  can be  roughly approximated as $b^2n$ with $n$ number density of tracers. When considering redshift space distortions, the signal  for the $A$ and $B$ terms scales like $b^3$ and $b^4$ suggesting  a different scaling of the signal-to-noise with mass  than for the real-space power spectrum,  which is favored by higher bias. These considerations might be useful when optimizing a  survey selection of targets, although the bias of observable tracers might not behave as the halo bias especially when multiple galaxies occupy the same halo.

\section{Summary and conclusions}\label{discussion_section}

Using a suite of 160 N-body simulations each with a volume of $V_{box}=13.8\, (Gpc/h)^3$, we have investigated the accuracy of analytic models in predicting the non-linear power spectrum of matter and dark-matter haloes in real and redshift space. The total simulated volume amounts to 2,200 (Gpc/h)$^3$, much larger than the volume surveyed by any forthcoming or planned survey ensuring that statistical errors in the determination of the simulation data points is negligible. We make the dark matter and halo power spectra publicly available and also provide the multipoles measured from these simulations for possible comparisons\footnote{\url{http://icc.ub.edu/~hector/Hector\_Gil\_Marin/Public.html}}.

We considered a number of theoretical schemes obtained by combining standard or resummed perturbation theory with analytical models for redshift-space distortions (based on the simplification of Eq. \ref{rsd_exact}). To predict power spectra in real space, we have employed 1- and 2-loop standard perturbation theory, and the resummed perturbation theory proposed by \cite{CS06A,MPTbreeze} that we have generalized to account for 2-loop correction terms in the resummed propagator. For the redshift-space power spectra, we have focused on the models proposed by Kaiser \citep{kaiser}, Scoccimarro \citep{SC04} and Taruya et al. \citep{taruya_model}.

At the level of dark matter in real space, increasing the order in loop corrections for the resummed propagator improves the theoretical predictions of the power spectrum. In particular, working at 2-loop correction in the resummed propagator, ${\cal N}_2$, an accuracy of $\lesssim1\%$ is achieved at different redshifts up to the following scales: $k=0.10h/\,{\rm Mpc}$ at $z=0$;  $k=0.15h/\,{\rm Mpc}$ at $z=0.5$; $k=0.20h/\,{\rm Mpc}$ at $z=1.0$ and $k=0.25h/\,{\rm Mpc}$ at $z=1.5$ for a 2-loop truncation of the infinite series. In general, working at 2-loop correction in the resummed propagator provides a more accurate description than working at 1-loop correction (as many of the public codes do \cite{MPTbreeze,Carlson,regpt}). 

Also, the price of working at 2-loop instead of 1-loop correction in the resummed propagator is not very high in terms of computational resources. It is true that evaluating ${\cal N}_2$  involves the 5-dimensional integration of $P_{15}$, but the angular part of this function can be either analytically  computed, or numerically precomputed for any shape of the linear power spectrum so, in the end, one ends up with a 2-dimensional integration,which can be easily performed.

For dark matter in redshift space, our results show that the model by Taruya et al. combined with a Lorentzian damping term for the FoG effect with $\sigma_0$ as a free parameter, is able to reproduce the multipoles from N-body simulations with high accuracy. In this paper, we have fit $\sigma_0$ as a function of $k_{\rm max}$ using monopole, quadrupole and hexadecapole data separately, which yields to 3 different values for $\sigma_0$. Although is possible to fit all these 3 multipoles with the same value of $\sigma_0$, the accuracy is expected to be reduced. When 3 different $\sigma_0$ are used, the high accuracy holds true for the monopole ($\lesssim 2\%$) and the quadrupole ($\lesssim 5\%$) irrespective of the flavor of perturbation theory adopted to compute the real-space power spectrum. For higher-order moments, such as the hexadecapole, the level of accuracy is more modest, $\sim10\%$ at $z=1$.
This suggests that  even on mildly non-linear scales at redshifts $z\lesssim 1.5$ the accuracy of the modeling of the redshift space distortions is more important than the modeling of the non-linear evolution of the real-space dark matter power spectrum.

 The difference between the Taruya et al. model  (which  attempts to include the density-velocity coupling at higher orders) and the other two models become more evident as the multipole order is increased, possibly suggesting that   non-linearities become more important for higher-order multipoles.  
 
While in linear theory only the monopole, quadruple and hexadecapole are non-zero, in principle non-linearities should ``excite"   all higher-order multipoles, and thus cosmological  signal could, in principle,  be  extracted from them. We find from the N-body simulations that the signal-to-noise decreases with increasing multipole, making the hexadecapole errors large even from  the large suite of simulations considered here. This suggests that for cosmological applications most of the signal-to-noise at these mildly non-linear scales --where analytic approaches can  provide a good modeling--  is still enclosed in the monopole and quadrupole.

The imprint of the baryon acoustic oscillations, is also visible in the multipoles.  We find that all models of redshift space distortions considered do not bias the wiggles location although the more linear models (Kaiser and Scoccimarro) over-predict their amplitude.
 These considerations  might be relevant  for recovering in an unbiased way  the angular and radial BAO information (separately) from forthcoming surveys.

Overall, the accuracy of the analytic description allows measurements of the logarithmic growth rate $f$ to percent level, when $\sigma_0$ is allowed to vary. In this case, when the Taruya et al. model is combined with the RPT prediction for the power spectrum in real space, $f$ is recovered within $\lesssim1\%$ up to $k_{\rm max}=0.15\,h/{\rm Mpc}$ for the monopole and quadrupole (separately) at $z=0$ and up to $k_{\rm max}=0.20\,h/{\rm Mpc}$ at $z=1$.  This indicates that the Taruya et al. model combined with the RPT real-space predictions is accurate enough to be used for precision cosmology.

Since most of the current and future redshift surveys target galaxies as tracers of the matter distribution, a more realistic way of estimating $f$ is to use dark-matter haloes instead of the dark matter density. The limited mass resolution of our N-body simulations, allows us to consider only cluster-sized haloes $M>10^{14}M_\odot/h$. Dealing with isolated haloes has the advantage of eliminating 
the imprint of FoGs from the power spectrum. However, the effect of the scale-dependent (and possibly non-linear) bias plays an important role. In this work, we have assumed that the bias is linear and that its $k$-dependence is known. Under these approximations, we have been able to recover $f$ with $\lesssim5\%$ when the amplitude of the bias is known {\it a priori}. 
For these massive haloes, the effect of bias is important: the degraded accuracy in recovering $f$ indicates that, at least for these massive haloes, the modeling of biasing is crucial.  
In particular, given the high shot noise that the statistics of these tracers have, a modeling of its behavior  both as a function of scale and halo mass is important at mildly non-linear scales.
 
When both the $f$ parameter and the bias amplitude are allowed to vary, we recover the bias amplitude to $1\%$ level in the best cases, but $f$ is underestimated by $10\%-20\%$ at $z=0$ and slightly more at $z=0.5$. Similar results are reported in the literature \citep{kwan}. Most likely this is not (or not only) due to a limitation of the model for the power spectrum, but also to the poor signal-to-noise ratio of the population of haloes used to extract $f$. It remains to be seen whether reducing the halo mass threshold increases the  signal-to-noise. The scaling with halo mass of the signal-to-noise  of the RSD terms that can break the degeneracy between $f$ and $b$, is different from the usual $nP\sim b^2n$ used for the angle-averaged, real-space power spectrum. Our simple considerations  indicate that a lower mass threshold increases the signal-to-noise, but if the number density is kept fixed, then the signal-to-noise for the A and B terms is favored by a higher bias (the signal-to-noise scales like $b^3n$ and $b^4 n$ rather than like $b^2n$).
These considerations might be useful when optimizing a  survey selection of targets, although one should keep in mind that bias of observable tracers might not behave as the bias of the host haloes.

The results found in this paper are in agreement with  those found in recent works. Kwan et al. \citep{kwan} found that, most of the RSD models fail at recovering $f$, underestimating its value even at large scales for $z=0$ and $z=0.5$. Considering the Taruya et al. model, relaxes the discrepancy (compared to Kaiser or Scoccimarro models), but does not completely fixes the problem. Other works, such as de la Torre \& Guzzo \citep{de_la_torre}, show that the Taruya et al. model (in configuration space) + numerical and phenomenological schemes to estimate the real-space spectra, are able to recover $f$ with $\lesssim5\%$ accuracy from different low-biased galaxy population. Okumura \& Jing \citep{okumura_jing} show that $\beta$ can be estimated accurately from very massive haloes ($M_h\geq10^{14}M_\odot/h$) using linear theory (Kaiser model) at large scales ($k\leq0.1h/\,{\rm Mpc}$); and $f$ can be also recovered with similar accuracy when the bias is assumed to be known, although the precision achieved is not very high.

Future surveys and missions will provide datasets about the distribution of galaxies on large scales. We envision that in order to extract useful information from these datasets, we will need more accurate theoretical models of structure formation. In this paper, we have shown that for extracting the growth of structure correctly,  accurate analytic models for both real and redshift space clustering are crucial. In particular 2L-RPT-${\cal N}_2$ in combination with the Taruya et al. formula, seems to be able to recover $f$ accurately from massive haloes, 
when the bias is assumed to be known or equivalently when one wants to recover the combination $\beta=f/b$.  However,  in our study we found that this approach fails when trying to recover the individual values of $f$ and $b$ simultaneously. This can be due to the low signal-to-noise ratio of this halo population, or a limitation in the model itself.  Should the  determination of $f$ and $b$ separately rather than in the $\beta$ combination  from RSD alone become a priority, studies will attempt to improve both the real and redshift-space models. In the case of real space, this can be done extending resummation theories both in Eulerian or Lagrangian spaces. In the case of redshift space, considering higher-order terms in the TNS formula should improve the model. Finally, it is also important to model correctly the stochasticity associated with the halo population and to determine correctly the scale dependence of the bias and its possible non-linearities.

\section{Acknowledgments}
H\'ector Gil-Mar\'in is supported by CSIC-JAE grant.
Licia Verde and Christian Wagner  are supported by FP7-IDEAS-Phys.LSS 240117.
Licia Verde and Raul Jimenez are supported by FPA2011-29678-C02-02.
Cristiano Porciani acknowledges support by the Deutsche Forschungsgemeinschaft through the SFB-Transregio 33 "The Dark Universe".
H\'ector Gil-Mar\'in thanks the Argelander Institut f\"ur Astronomie at the University of Bonn for hospitality.

\appendix

\section{Standard Perturbation Theory}\label{A}

In this section, we provide a short summary of the equations used to compute the 1- and 2-loop correction in Eulerian perturbation theory. For a detailed description of perturbation theory see \cite{SC98,bernardeau}.

According to standard perturbation theory (SPT) the power spectrum in real space can be expressed as a sum of loop corrections,
\begin{equation}
P(k)=P^{(0)}(k)+P^{(1)}(k)+P^{(2)}(k)\dots,
\end{equation}
where $P^{(0)}(k)=P^{\rm lin}(k)$ is the linear term. For Gaussian initial conditions, the different loop corrections read as, 
\begin{eqnarray}
 P^{(1)}(k)&=&2P_{13}(k)+P_{22}(k) \quad \mbox{1-loop correction},\\
P^{(2)}(k)&=&2P_{15}(k)+2P_{24}(k)+P_{33}(k)  \quad \mbox{2-loop correction},
\end{eqnarray}
where, as already mentioned in the main text, the subscripts $i$ and $j$ refer to the perturbative order of the terms $\delta({\bf k})$ used in eq. \ref{Pk} to compute the power spectrum $P_{ij}(k)$. 
For the case of matter-matter power spectrum, namely $P_{\delta\delta}$, these terms are \citep{fry94},
\begin{eqnarray}
\label {P13}P_{13}(k)&=&3P^{\rm lin}(k)\int \frac{d^3{\bf q}}{(2\pi)^3} F^{s}_3({\bf k},{\bf q}, {\bf -q}) P^{\rm lin}(q),\\
\label {P22}P_{22}(k)&=&2\int \frac{d^3{\bf q}}{(2\pi)^3}\, {F^{s}_2}^2({\bf q},{\bf k-q})P^{\rm lin}(q)P^{\rm lin}(|{\bf k-q}|),\\
\label{P15}P_{15}(k)&=& 15P^{\rm lin}(k)\int\frac{d^3q_1}{(2\pi)^3}\frac{d^3 q_2}{(2\pi)^3}F^s_5({\bf k},{\bf q}_1,-{\bf q}_1,{\bf q}_2,-{\bf q}_2)P^{\rm lin}(q_1)P^{\rm lin}(q_2)  \\
\label{P24}P_{24}(k)&=& 12\int \frac{d^3 q_1}{(2\pi)^3}\frac{d^3 q_2}{(2\pi)^3}F^s_2({\bf q}_1,{\bf k}-{\bf q}_1) F^s_4({\bf q}_1,{\bf k}-{\bf q}_1,{\bf q}_2,-{\bf q}_2)\times\\
\nonumber&\times& P^{\rm lin}(q_1)P^{\rm lin}(q_2)P^{\rm lin}(|{\bf k}-{\bf q}_1|), \\
\label{P33} P_{33}(k)&=&9P^{\rm lin}(k)\left[\int\frac{d^3q}{(2\pi)^3}F_3^s({\bf k},{\bf q},-{\bf q})P^{\rm lin}(q)\right]^2+\\
\nonumber&+&6\int\frac{d^3q_1}{(2\pi)^3}\frac{d^3q_2}{(2\pi)^3}{F^s_3}^2({\bf q}_1,{\bf q}_2,{\bf k}-{\bf q}_1-{\bf q}_2)P^{\rm lin}(q_1)P^{\rm lin}(q_2)P^{\rm lin}(|{\bf k}-{\bf q}_1-{\bf q}_2|).
\end{eqnarray}
In the 1-loop correction, $P_{22}$ accounts for the mode coupling between vectors with frequencies $\bf{k} -  \bf{q}$ and $\bf q$, whereas $P_{13}$  can be interpreted as the 1-loop  correction to the linear propagator. In a similar way, in the 2-loop correction term, only the second term of $P_{33}$ accounts for a full 2-loop mode coupling because is the only term that contains $P^{\rm lin}(|{\bf k}-{\bf q}_1-{\bf q}_2|)$. Also note that $P_{24}$ contains a term similar to a 1-loop coupling, $P^{\rm lin}(|{\bf k}-{\bf q}_1|)$, is similar to $P_{22}$. $P_{15}$ and the first term of $P_{33}$ contain no coupling between $\bf k$ and ${\bf q}_i$ and can be interpreted as a 2-loop propagators. 
In particular, the full $n$-propagator can be written as \citep{fry94}
\begin{equation}
\label{P_1n}P_{1n}(k)=n!! P^{\rm lin}(k) \int \frac{d^3 {\bf q}_1}{(2\pi)^3}\dots \frac{d^3 {\bf q}_x}{(2\pi)^3} F^s_n({\bf k}, {\bf q}_1, -{\bf q}_1,\ldots, {\bf q}_x, -{\bf q }_x) P^{\rm lin}(q_1)\ldots P^{\rm lin}(q_x),
\end{equation}
where $x=(n-1)/2$.

These similarities between these terms is the basis of the resummation process that is described in Appendix B. The kernels of Eqs. \ref{P13}-\ref{P33} are expressed as,
\begin{eqnarray}
\label{kernelF}F_n({\bf q}_1,\dots,{\bf q}_n)&=&\sum_{m=1}^{n-1} \frac{G_m({\bf q}_1,\dots,{\bf q}_m)}{(2n+3)(n-1)} \left[(2n+1)\alpha({\bf k},{\bf k}_1) F_{n-m}({\bf q}_{m+1},\dots,{\bf q}_n)+\right.\\
\nonumber &+&\left.2\beta({\bf k}, {\bf k}_1,{\bf k}_2)G_{n-m}({\bf q}_{m+1},\dots,{\bf q}_n)\right],\\
\label{kernelG}G_n({\bf q}_1,\dots,{\bf q}_n)&=&\sum_{m=1}^{n-1} \frac{G_m({\bf q}_1,\dots,{\bf q}_m)}{(2n+3)(n-1)} \left[3\alpha({\bf k},{\bf k}_1) F_{n-m}({\bf q}_{m+1},\dots,{\bf q}_n)+\right.\\
\nonumber &+&\left.2n\beta({\bf k}, {\bf k}_1,{\bf k}_2)G_{n-m}({\bf q}_{m+1},\dots,{\bf q}_n)\right],
\end{eqnarray}
with $F_1\equiv1$ and $G_1\equiv1$. Also, ${\bf k}_1\equiv {\bf q}_1+\dots+{\bf q}_m$, ${\bf k}_2\equiv {\bf q}_{m+1}+\dots+{\bf q}_n$, ${\bf k}\equiv{\bf k}_1+{\bf k}_2$ where the functions $\alpha$ and $\beta$ are defined as,
\begin{eqnarray}
\label{alpha}\alpha({\bf k}, {\bf k}_1)&\equiv&\frac{ {\bf k}\cdot{\bf k}_1}{k_1^2},\\
\label{beta}\beta({\bf k},{\bf k}_1,{\bf k}_2)&\equiv&\frac{k^2({\bf k}_1\cdot{\bf k}_2)}{2k_1^2k_2^2}.
\end{eqnarray}
The symmetrization process of the kernels is given by,
\begin{equation}
\label{sym_kernels} F^s_n({\bf q}_1,\dots,{\bf q}_n)=\frac{1}{n!}\sum_\pi F_n({\bf q}_{\pi (1)},\dots,{\bf q}_{\pi (n)}),
\end{equation}
where the sum is taken over all	 the permutations $\pi$ of	the set $\{1,\dots , n\}$. In particular, the expressions for $F_2^s({\bf k}_1,{\bf k}_2)$ and $G_2^s({\bf k}_1,{\bf k}_2)$ are,
\begin{eqnarray}
\label{F2s} F_2^s({\bf k}_1,{\bf k}_2)&=&  \frac{5}{7}+\frac{1}{2}\cos\theta\left(\frac{q}{k}+\frac{k}{q}\right)+\frac{2}{7}\cos^2\theta,\\
\label{G2s} G_2^s({\bf k}_1,{\bf k}_2)&=& \frac{3}{7}+\frac{1}{2}\cos\theta\left(\frac{q}{k}+\frac{k}{q}\right)+\frac{4}{7}\cos^2\theta,
\end{eqnarray}
where, $\cos\theta\equiv({\bf k}_1\cdot{\bf k}_2)/(k_1 k_2)$.

This SPT formalism presents some shortcomings. As noted by \cite{CS06A}, at large scales only the linear term contributes to the total power spectrum. However, at smaller scales, all loop corrections become of the same order with a significant cancellation among them. In particular, at low redshifts it can be seen that 1-loop correction overestimates the full power spectrum (from N-body simulations), whereas the 2-loop correction underestimates it. This is due to the fact that $P^{(1)}$ is negative on large scales while $P^{(2)}$ is positive, and both almost cancel out giving a remaining quantity which is close to the full power spectrum. In the same way, as we go to higher order, more cancellations come out among the different loop corrections. Thus, truncating at certain loop in SPT will naturally produce a systematic over- and under-prediction of the real-space power spectrum. A way to avoid this behavior is to resum some terms of the total SPT expansion. In \cite{CS06A,CS06B} a formalism for resumming part of these terms was proposed. The resulting expansion presents a more controlled behavior because each different loop contributes only positively to the total power spectrum and acts at different scales. In Appendix \ref{B} we present an alternative description (but mathematically identical) of the resummation presented by \cite{CS06A,CS06B}. As an extension of current works, we write not only the 1-loop resummed propagator, but we perform our computation up to 3-loops.

\section{Resummation in Standard Perturbation Theory}\label{B}

In this section we present the derivation of Eq. \ref{RPT}, \ref{theory_N1} and \ref{theory_N2}. We also show that the ${\cal N}_1$ expression is identical to the one used in \cite{MPTbreeze} when the propagator is perturbed at 1-loop. For completeness, we also show the result of perturbing the propagator for 3-loops. These equations come from resumming some terms in SPT under certain approximation in the kernels.  In particular, under the Zel'dovich approximation, the kernels read \citep{grinstein},
\begin{eqnarray}
\label{zeldovich}F_n^s({\bf k}_1,\,\dots,\, {\bf k}_n)&=&\frac{1}{n!}\frac{{\bf k}\cdot{\bf k}_1}{k_1^2}\dots\ \frac{{\bf k}\cdot{\bf k}_n}{k_n^2},\\
G_n^s({\bf k}_1,\,\dots,\, {\bf k}_n)&=&\frac{1}{n!}\frac{{\bf k}\cdot{\bf k}_1}{k_1^2}\dots\ \frac{{\bf k}\cdot{\bf k}_n}{k_n^2},
\end{eqnarray}
where ${\bf k}={\bf k}_1+\cdots+{\bf k}_n$.
As shown by  \cite{CS06A}, with this approximation the resummation process yields,
\begin{equation}
P(k,z)=\left[P^{\rm lin}(k,z)+P_{22}^{1L}(k,z)+P_{33}^{2L}(k,z)+\dots+P_{nn}^{(n-1) L}(k,z)+\dots\right]{\cal N}_0(k, z)^2,
\label{RPT_SC}
\end{equation}
where,
\begin{equation}
{\cal N}_0(k, z)\equiv\exp\left[-\frac{1}{2}k^2\sigma_v(z)^2\right],
\end{equation}
with
\begin{eqnarray}
\nonumber P_{nn}^{(n-1) {\rm L}}(k,z)&\equiv& n!\int \frac{d^3 q_1}{(2\pi)^3}\cdots\frac{d^3 q_{n-1}}{(2\pi)^3}\, {F_n^s}^2({\bf q}_1,\,\dots,\,{\bf q}_{n-1},\,{\bf k}-\sum_{i=1}^{n-1} {\bf q}_i)P^{\rm lin}(q_1,z)\dots P^{\rm lin}(q_{n-1},z) \times\\
&\times&P^{\rm lin}\left(\left|{\bf k}-\sum_{i=1}^{n-1} {\bf q}_i\right|,z\right),
\label{n_term}
\end{eqnarray}
and $\sigma_v$ is a characteristic scale defined as
\begin{equation}
\sigma_v^2(z)\equiv\frac{4\pi}{3}\int \frac{dq}{(2\pi)^3}\, P^{\rm lin}(q,z)\,.
\end{equation}
We will refer to Eq. \ref{RPT_SC} as RPT-${\cal N}_0$ model. With this technique the behavior of PT improves, since every new loop adds a positive term that only acts on a small range of scales. Therefore, using this technique  the oscillatory behavior observed in standard PT vanishes.

In this section we show that performing a slightly different approximation (not Zel'dovich) in the kernels we can end up with the same formula used in \cite{MPTbreeze}. Also, depending on how we `factorize' the kernels, we will end up with a 1-, 2- or higher-loop correction in the resummed propagator. These formulae present a notable improvement respect to Eq. \ref{RPT_SC} for a similar computational effort.
In this work we do not follow the approach of Feynman diagrams to resum the infinite terms as it is done in  \cite{CS06A,CS06B}. Alternatively, we present a different approach for doing this without requiring any knowledge of quantum field theory. We hope that this way of resumming is clearer for the reader who is no familiar with this kind of formalism. Furthermore, this approach allows us to easily compute the resummed propagator for higher-order loops.
Our method consists of rewriting the terms of  the $\ell$-loop correction (where $\ell\equiv(n+m)/2-1$), namely $P_{nm}$, as a sum of {\it subterms} which can be associated to lower loop corrections as we show below\footnote{The redshift dependence is understood for simplicity: it only appears through $P^{\rm lin}$.},
\begin{equation}
P_{nm}(k)=P_{nm}^{0{\rm L}}(k)+P_{nm}^{1{\rm L}}(k)+P_{nm}^{2{\rm L}}(k)+\dots +P_{nm}^{\ell {\rm L}}(k).
\label{subterms}
\end{equation}
The subterm with an index $0{\rm L}$ contains a $P^{\rm lin}(k)$ and corresponds to the linear power spectrum (no-loop correction), the subterm with  an index $1{\rm L}$ contains a $P^{\rm lin}(|{\bf k}-{\bf q}_1|)$ and therefore is similar to $P_{22}(k)$ (that corresponds to 1-loop correction). In the same way, the $2L$ subterm is similar to $P_{33}(k)$ (that corresponds to 2-loop correction) because contains a term $P^{\rm lin}(|{\bf k}-{\bf q}_1-{\bf q}_2|)$ and so on.
The generic way of writing these terms is the following\footnote{Note that the terms with $n$ odd and $m$ even (and vice versa) vanish for Gaussian initial conditions.}.

The 0-L subterm can be written,
\begin{enumerate}
\item for $n$ \&  $m$ even,
\begin{equation}
P_{nm}^{0L}(k)=0;
\end{equation}
\item for $n$ \&  $m$ odd,
\begin{eqnarray}
P_{nm}^{0{\rm L}}(k)&=&n!! m!!\int \frac{d^3q^n_1}{(2\pi)^3}\dots \frac{d^3q^n_{x_n}}{(2\pi)^3}\frac{d^3q_1^m}{(2\pi)^3}\dots \frac{d^3q_{x_m}^m}{(2\pi)^3} F_n^s({\bf k},{\bf q}_1^n,-{\bf q}^n_1,\dots,{\bf q}_{x_n}^n,-{\bf q}_{x_n}^n)  \times\\
\nonumber&\times&  F_m^s({\bf k},{\bf q}_1^m,-{\bf q}_1^m,\dots,{\bf q}_{x_m}^m,-{\bf q}_{x_m}^m) P^{\rm lin}(k) P^{\rm lin}(q_1^n)\dots P^{\rm lin}(q^n_{x_n}) P^{\rm lin}(q_1^m)\dots P^{\rm lin}(q^m_{x_m}),
\end{eqnarray}
where $x_i=(i-1)/2$.
\end{enumerate}

The 1-L subterm can be written,
\begin{enumerate}
\item for $n$ \& $m$ even,
\begin{eqnarray}
\nonumber P_{nm}^{1{\rm L}}(k)&=&\frac{1}{2}n(n-1)!!m(m-1)!!\int \frac{d^3q_1}{(2\pi)^3} \frac{d^3q_2^n}{(2\pi)^3}\dots \frac{d^3q_{x_n}^{n}}{(2\pi)^3}\frac{d^3q_2^m}{(2\pi)^3}\dots \frac{d^3q^{m}_{x_m}}{(2\pi)^3} P^{\rm lin}(|{\bf k}-{{\bf q}_1}|) P^{\rm lin}(q_1)  \times\\
\nonumber&\times& F_n^s({\bf q}_1,{\bf k}-{\bf q}_1,{\bf q}_2^n,-{\bf q}_2^n,\dots,{\bf q}^{n}_{x_n},-{\bf q}_{x_n}^n) P^{\rm lin}(q_2^n)\dots P^{\rm lin}(q^n_{x_n}) \times \\
&\times& F_m^s({\bf q}_1,{\bf k}-{\bf q}_1,{\bf q}_2^m,-{\bf q}_2^m,\dots,{\bf q}_{x_m}^m,-{\bf q}_{x_m}^m) P^{\rm lin}(q_2^m)\dots P^{\rm lin}(q^m_{x_m}),
\end{eqnarray}
where $x_i=i/2$;
\item for $n$ \& $m$ odd,
\begin{equation}
P_{nm}^{1{\rm L}}(k)=0.
\end{equation}
\end{enumerate}
The 2-L subterms can be written,
\begin{enumerate}
\item for $n$ \& $m$ even,
\begin{equation}
P_{nm}^{2{\rm L}}=0;
\end{equation}
\item for $n$ \& $m$ odd,
\begin{eqnarray}
\nonumber P_{nm}^{2{\rm L}}&=&\frac{n!! m!! }{6}(n-1)(m-1)\int \frac{d^3q_1}{(2\pi)^3} \frac{d^3 q_2}{(2\pi)^3} \frac{d^3 q_3^n}{(2\pi)^3} \dots \frac{d^3 q^n_{x_n}}{(2\pi)^3} \frac{d^3 q_3^m}{(2\pi)^3}\dots \frac{d^3 q^m_{x_m}}{(2\pi)^3}P^{\rm lin}(q_1)P^{\rm lin}(q_2)\times\\
\nonumber &\times& F^s_n({\bf q}_1,{\bf q}_2,{\bf k}-{\bf q}_1-{\bf q}_2, {\bf q}_3^n,-{\bf q}_3^n,\dots,{\bf q}_{x_n}^n,-{\bf q}_{x_n}^n)P^{\rm lin}(|{\bf k}-{\bf q}_1-{\bf q}_2|)P^{\rm lin}(q_3^n)\dots P^{\rm lin}(q_{x_n}^n)\times\\
&\times& F^s_m({\bf q}_1,{\bf q}_2,{\bf k}-{\bf q}_1-{\bf q}_2, {\bf q}_3^m,-{\bf q}_3^m,\dots,{\bf q}_{x_m}^m,-{\bf q}_{x_m}^m)P^{\rm lin}(q_3^m)\dots P^{\rm lin}(q_{x_m}^m),
\end{eqnarray}
with $x_i=(i-3)/2$.
\end{enumerate}
The 3-L subterms are,
\begin{enumerate}
\item for $n$ \& $m$ even,
\begin{eqnarray}
\nonumber P_{nm}^{3{\rm L}}(k)&=&\frac{1}{24}n(n-2)(n-1)!! m (m-2) (m-1)!! \int \frac{d^3 q_1}{(2\pi)^3} \frac{d^3 q_2}{(2\pi)^3} \frac{d^3 q_3}{(2\pi)^3} \frac{d^3 q^n_4}{(2\pi)^3}\dots \frac{d^3 q^n_{x_n}}{(2\pi)^3} \times\\
\nonumber&\times& \frac{d^3 q_4^m}{(2\pi)^3}\dots \frac{d^3 q^m_{x_m}}{(2\pi)^3}\,F_n^s({\bf q}_1, {\bf q}_2, {\bf q}_3, {\bf k}-{\bf q}_1-{\bf q}_2-{\bf q}_3, {\bf q}_4^n,-{\bf q}_4^n, \dots, {\bf q}_{x_n}^n,-{\bf q}_{x_n}^n)\times\\
\nonumber&\times&F_m^s({\bf q}_1, {\bf q}_2, {\bf q}_3, {\bf k}-{\bf q}_1-{\bf q}_2-{\bf q}_3, {\bf q}_4^m,-{\bf q}_4^m, \dots, {\bf q}_{x_m}^m,-{\bf q}_{x_m}^m) P^{\rm lin}(q_1)P^{\rm lin}(q_2) \times\\
\nonumber&\times&P^{\rm lin}(q_3) P^{\rm lin}(|{\bf k}-{\bf q}_1-{\bf q}_2-{\bf q}_3|) P^{\rm lin}(q^n_4)\dots P^{\rm lin}(q^n_{x_n})P^{\rm lin}(q^m_4)\dots P^{\rm lin}(q^m_{x_m}),\\
\end{eqnarray}
with $x_i=(i-4)/2$;
\item for $n$ \& $m$ odd,
\begin{equation}
P_{nm}^{3{\rm L}}(k)=0,
\end{equation}
\end{enumerate}
and similarly for higher-order subterms. In particular, it is important to note that when $n=m$ the subterm with the highest loop correction (${\ell }=n-1$) is expressed as,
\begin{eqnarray}
\nonumber P_{nn}^{(n-1) {\rm L}}(k)&=&n!\int \frac{d^3 q_1}{(2\pi)^3} \frac{d^3 q_2}{(2\pi)^3} \dots \frac{d^3 q_{n-1}}{(2\pi)^3}\, {F_n^s}^2({\bf q}_1, {\bf q}_2,\cdots,{\bf q}_{n-1},{\bf k}-\sum_{i=1}^{n-1} {\bf q}_i)P^{\rm lin}(q_1)\dots P^{\rm lin}(q_{n-1}) \times\\
&\times& P^{\rm lin}(|{\bf k}-\sum_{i=1}^{n-1} {\bf q}_i|),
\end{eqnarray}
which is the same term used in Eq. \ref{RPT_SC}. This indicates that the resummation of terms described in \cite{CS06A} corresponds to resumming the terms $P_{nm}^{\ell L}$ for $\ell<(n+m)/2-2$.
In order to make possible the resummation we perform an approximation in the kernels.
\subsection{1-loop factorization}
If we want to end up with a resummed propagator of  1-loop correction, the prescription in factorizing the kernels is the following,
\begin{enumerate}
\item for 0-L subterms,
\begin{equation}
\label{N1_kernela}F_n^s({\bf k},{\bf q}_1,-{\bf q}_1,\ldots,{\bf q}_{x_n},-{\bf q}_{x_n})\simeq\frac{1}{n!} \left[3!F_3^s({\bf k}, {\bf q}_1,-{\bf q}_1)\right]\cdot \ldots \cdot\left[3!F_3^s({\bf k}, {\bf q}_{x_n},-{\bf q}_{x_n})\right];
\end{equation}
\item for 1-L subterms,
\begin{eqnarray}
\label{N1_kernelb} F_n^s({\bf q}_1,{\bf k}-{\bf q}_1,{\bf q}_2,-{\bf q}_2,\dots,{\bf q}_{x_n},-{\bf q}_{x_n}) &\simeq&\frac{1}{n!}\left[2!F^s_2({\bf q}_1,{\bf k}-{\bf q}_1)\right]\times\\
 \nonumber &\times& \left[3!F^s_3({\bf k}, {\bf q}_2, -{\bf q}_2)\right]\cdot\ldots\cdot \left[3!F^s_3({\bf k}, {\bf q}_{x_n}, -{\bf q}_{x_n})\right];
\end{eqnarray}
\item for 2-L subterms,
\begin{eqnarray}
\label{N1_kernelc}F^s_n({\bf q}_1, {\bf q}_2, {\bf k}-{\bf q}_1-{\bf q}_2,{\bf q}_3, -{\bf q}_3,\ldots,{\bf q}_{x_n},-{\bf q}_{x_n})&\simeq&\frac{1}{n!}\left[3!F^s_3({\bf q}_1, {\bf q}_2, {\bf k}-{\bf q}_1-{\bf q}_2)\right]\times\\
\nonumber&\times&\left[3!F_3^s({\bf k},{\bf q}_3,-{\bf q}_3)\right]\cdot\ldots\cdot\left[3!F_3^s({\bf k},{\bf q}_{x_n}, -{\bf q}_{x_n})\right];
\end{eqnarray}
\end{enumerate}
and similarly for higher-order loops. We factorize the $n$-order kernel in a product of 2- and 3-order kernels (which correspond to 1-loop correction terms) keeping the sum ${\bf q}_1+{\bf q}_2+\dots={\bf k}$ in all the 2- and 3-order kernels.
Under these approximations we rewrite the subterms of Eq. \ref{subterms}. 
\begin{enumerate}
\item For 0-L subterms with $n$ \& $m$ odd,
\begin{equation}
P_{nm}^{0{\rm L}}(k)\simeq\frac{P^{\rm lin}(k)}{x_n! x_m!}\left(\frac{P_{13}(k)}{P^{\rm lin}(k)}\right)^{x_n+x_m},
\end{equation}
with $x_i=(i-1)/2$.
\item For 1-L subterms with $n$ \& $m$ even,
\begin{equation}
P_{nm}^{1{\rm L}}(k)\simeq\frac{P_{22}(k)}{x_n! x_m!}\left(\frac{P_{13}(k)}{P^{\rm lin}(k)}\right)^{x_n+x_m},
\end{equation}
with $x_i=(i-2)/2$.
\item For 2-L subterms with $n$ \& $m$ odd,
\begin{equation}
P_{nm}^{2{\rm L}}(k)\simeq\frac{P_{33}^{2L}(k)}{x_n! x_m!}\left(\frac{P_{13}(k)}{P^{\rm lin}(k)}\right)^{x_n+x_m},
\end{equation}
with $x_i=(i-3)/2$;
\end{enumerate}
and similarly for higher-order subterms. Now we can proceed with the resummation of Eq. \ref{subterms}. Reordering the terms we write,
\begin{equation}
P(k)=\sum_{n=1}^{\rm odd}\sum_{m=1}^{\rm odd}P_{nm}^{0{\rm L}}(k)+\sum_{n=2}^{\rm even}\sum_{m=2}^{\rm even}P_{nm}^{1{\rm L}}(k)+\sum_{n=3}^{\rm odd}\sum_{m=3}^{\rm odd}P_{nm}^{2{\rm L}}(k)+\ldots .
\end{equation}
The first term is,
\begin{equation}
\sum_{n=1}^{\rm odd}\sum_{m=1}^{\rm odd}P_{nm}^{0{\rm L}}(k)=P^{\rm lin}(k)\sum_{x_n=0}^{\infty}\sum_{x_m=0}^\infty\frac{1}{x_n!x_m!}\left[\frac{P_{13}(k)}{P^{\rm lin}(k)}\right]^{x_n+x_m}=P^{\rm lin}(k)\exp\left[2P_{13}(k)/P^{\rm lin}(k)\right].
\end{equation}
The second term is,
\begin{equation}
\sum_{n=2}^{\rm even}\sum_{m=2}^{\rm even}P_{nm}^{1{\rm L}}(k)=P_{22}(k)\sum_{x_n=0}^{\infty}\sum_{x_m=0}^\infty\frac{1}{x_n!x_m!}\left[\frac{P_{13}(k)}{P^{\rm lin}(k)}\right]^{x_n+x_m}=P_{22}(k)\exp\left[2P_{13}(k)/P^{\rm lin}(k)\right].
\end{equation}
The third term is,
\begin{equation}
\sum_{n=3}^{\rm odd}\sum_{m=3}^{\rm odd}P_{nm}^{2{\rm L}}(k)=P_{33}^{2{\rm L}}(k)\sum_{x_n=0}^{\infty}\sum_{x_m=0}^\infty\frac{1}{x_n!x_m!}\left[\frac{P_{13}(k)}{P^{\rm lin}(k)}\right]^{x_n+x_m}=P_{33}^{2L}(k)\exp\left[2P_{13}(k)/P^{\rm lin}(k)\right],
\end{equation}
and the same for higher-order terms.
Therefore, after this resummation we can express the power spectrum as,
\begin{equation}
P(k)=\left[P^{\rm lin}(k)+P_{22}^{1{\rm L}}(k)+P_{33}^{2{\rm L}}(k)+\ldots\right]{\cal N}_1(k)^2,
\label{RPT_hector}
\end{equation}
with,
\begin{equation}
{\cal N}_1(k)\equiv\exp\left[P_{13}(k)/P^{\rm lin}(k)\right].
\end{equation}
Here, $P^{1{\rm L}}_{22}(k)\equiv P_{22}(k)$ is given by Eq \ref{P22}, whereas $P^{2{\rm L}}_{33}(k)$ is given by the second term of Eq. \ref{P33},
\begin{equation}
\label{P332L}P_{33}^{2{\rm L}}(k)=6\int\frac{d^3q_1}{(2\pi)^3}\frac{d^3q_2}{(2\pi)^3}{F^s_3}^2({\bf q}_1,{\bf q}_2,{\bf k}-{\bf q}_1-{\bf q}_2)P^{\rm lin}(q_1)P^{\rm lin}(q_2)P^{\rm lin}(|{\bf k}-{\bf q}_1-{\bf q}_2|).
\end{equation}
The factor in the exponential, $P_{13}(k)/P(k)$, can be partially computed, because the integral over the angular part of Eq. \ref{P13} can be performed for any shape of $P^{\rm lin}$. For doing this, is convenient to express the symmetrized $F_3^s$ kernel as,
\begin{eqnarray}
6F_3^s({\bf k}, {\bf q},-{\bf q})&=&\frac{1}{9}G_2^s({\bf k},{\bf q})\left[ 7\alpha({\bf k},{\bf k}+{\bf q})+4\beta({\bf k},{\bf k}+{\bf q},-{\bf q})\right]\\
\nonumber &+&\frac{1}{9}G_2^s({\bf k},-{\bf q})\left[ 7\alpha({\bf k},{\bf k}-{\bf q})+4\beta({\bf k},{\bf k}-{\bf q},{\bf q})\right]\\
\nonumber &+&\frac{7}{9}\alpha({\bf k},{\bf q})\left[ F_2^s({\bf k},-{\bf q})-F_2^s({\bf k},{\bf q})\right].
\end{eqnarray}
Taking into account Eqs. \ref{alpha}-\ref{F2s}, the angular dependence is now explicit through $\cos\theta$, and the angular integration of Eq. \ref{P13} can be performed. This computation yields,
\begin{eqnarray}
\label{f_sc}2\frac{P_{13}(k)}{P^{\rm lin}(k)}&=&\int_0^\infty\frac{4\pi}{504k^3q^3}\left[6k^7q-79k^5q^3+50q^5k^3-21kq^7+\right.\\
\nonumber&+&\left.\frac{3}{2}(k^2-q^2)^3(2k^2+7q^2)\ln\left|\frac{k-q}{k+q}\right|\right]P^{\rm lin}(q)\, dq,
\end{eqnarray}
which is the expression proposed by \cite{MPTbreeze} as the resummed propagator. We will refer to the model of Eq. \ref{RPT_hector} as RPT-${\cal N}_1$ model.
\subsection{2-loop factorization}
It is also possible to obtain a 2-loop resummed propagator if we factorize the kernels in a different way. In that case, we want to split the kernels in pieces of $F_3^s$ and $F_5^s$ kernels,
\begin{enumerate}

\item for 0-L subterms with $n=1,5,9,13,\dots$,
\begin{eqnarray}
\label{N2_kernela}F_n^s({\bf k},{\bf q}_1,-{\bf q}_1,&\ldots&,{\bf q}_{x_n},-{\bf q}_{x_n})\simeq\frac{1}{n!} \left[5!F_5^s({\bf k}, {\bf q}_1,-{\bf q}_1,{\bf q}_2,-{\bf q}_2)\right]\cdot \ldots \times\\
\nonumber&\times&\ldots\cdot\left[5!F_5^s({\bf k}, {\bf q}_{x_n-1},-{\bf q}_{x_n-1},{\bf q}_{x_n},-{\bf q}_{x_n})\right];
\end{eqnarray}

\item for 0-L subterms with $n=3,7,11,15,\dots$,
\begin{eqnarray}
\label{N2_kernelb}F_n^s({\bf k},{\bf q}_1,-{\bf q}_1,&\ldots&,{\bf q}_{x_n},-{\bf q}_{x_n})\simeq\frac{1}{n!} \left[3!F_3^s({\bf k},{\bf q}_1,-{\bf q}_1)\cdot5!F_5^s({\bf k}, {\bf q}_2,-{\bf q}_2,{\bf q}_3,-{\bf q}_3)\right]  \\
\nonumber&\times&\ldots\cdot\left[5!F_5^s({\bf k}, {\bf q}_{x_n-1},-{\bf q}_{x_n-1},{\bf q}_{x_n},-{\bf q}_{x_n})\right];
\end{eqnarray}

\item for 1-L subterms with $n=4,8,12,16,\dots$,
\begin{eqnarray}
\label{N2_kernelc} F_n^s({\bf q}_1,{\bf k}-{\bf q}_1,{\bf q}_2,-{\bf q}_2,&\dots&,{\bf q}_{x_n},-{\bf q}_{x_n}) \simeq\frac{1}{n!}\left[2!F^s_2({\bf q}_1,{\bf k}-{\bf q}_1)\right]\cdot\left[3!F^s_3({\bf k}, {\bf q}_2, -{\bf q}_2)\right]\times\\
 \nonumber &\times&\left[5!F^s_5({\bf k}, {\bf q}_{3}, -{\bf q}_{3},{\bf q}_{4}, -{\bf q}_{4})\right]\cdot\ldots\cdot \left[5!F^s_5({\bf k}, {\bf q}_{x_n-1}, -{\bf q}_{x_n-1},{\bf q}_{x_n}, -{\bf q}_{x_n})\right];
\end{eqnarray}

\item for 1-L subterms with $n=2,6,10,14,\dots$,
\begin{eqnarray}
\label{N2_kerneld} F_n^s({\bf q}_1,{\bf k}-{\bf q}_1,{\bf q}_2,-{\bf q}_2,&\dots&,{\bf q}_{x_n},-{\bf q}_{x_n}) \simeq\frac{1}{n!}\left[2!F^s_2({\bf q}_1,{\bf k}-{\bf q}_1)\right]\times\\
 \nonumber &\times& \left[5!F^s_5({\bf k}, {\bf q}_2, -{\bf q}_2,{\bf q}_3, -{\bf q}_3)\right]\cdot\ldots\cdot \left[5!F^s_5({\bf k}, {\bf q}_{x_n-1}, -{\bf q}_{x_n-1},{\bf q}_{x_n}, -{\bf q}_{x_n})\right];
\end{eqnarray}

\item for 2-L subterms with $n=1,5,9,13,\dots$,
\begin{eqnarray}
\nonumber F^s_n({\bf q}_1, {\bf q}_2, {\bf k}-{\bf q}_1-{\bf q}_2,{\bf q}_3, -{\bf q}_3,&\ldots&,{\bf q}_{x_n},-{\bf q}_{x_n})\simeq\frac{1}{n!}\left[3!F^s_3({\bf q}_1, {\bf q}_2, {\bf k}-{\bf q}_1-{\bf q}_2)\right]\cdot\left[3!F_3^s({\bf k},{\bf q}_3,-{\bf q}_3)\right]\times\\
\nonumber&\times&\left[5!F_5^s({\bf k},{\bf q}_{4}, -{\bf q}_{4}, {\bf q}_{5}, -{\bf q}_{5})\right]\cdot\ldots\cdot\left[5!F_5^s({\bf k},{\bf q}_{x_n-1}, -{\bf q}_{x_n-1}, {\bf q}_{x_n}, -{\bf q}_{x_n})\right];\\
\label{N2_kernele}
\end{eqnarray}

\item for 2-L subterms with $n=3,7,11,15,\dots$,
\begin{eqnarray}
\label{N2_kernelf}F^s_n({\bf q}_1, {\bf q}_2, {\bf k}-{\bf q}_1-{\bf q}_2,{\bf q}_3, -{\bf q}_3,&\ldots&,{\bf q}_{x_n},-{\bf q}_{x_n})\simeq\frac{1}{n!}\left[3!F^s_3({\bf q}_1, {\bf q}_2, {\bf k}-{\bf q}_1-{\bf q}_2)\right]\times\\
\nonumber&\times&\left[5!F_5^s({\bf k},{\bf q}_3,-{\bf q}_3,{\bf q}_4,-{\bf q}_4)\right]\cdot\ldots\cdot\left[5!F_5^s({\bf k},{\bf q}_{x_n-1}, -{\bf q}_{x_n-1}, {\bf q}_{x_n}, -{\bf q}_{x_n})\right];
\end{eqnarray}

\end{enumerate}
and similar for higher-order loops. As before, with this approximation we can perform an exact resummation of all terms yielding to,
\begin{equation}
P(k)=\left[P^{\rm lin}(k)+P_{22}^{1L}(k)+P_{33}^{2L}(k)+\ldots\right]{\cal N}_2(k)^2,
\label{RPT_hector2}
\end{equation}
with,
\begin{equation}
\label{N2}{\cal N}_2(k)\equiv \cosh\left[  \sqrt{ \frac{2P_{15}(k)}{P^{\rm lin}(k)} }     \right] + \frac{P_{13}(k)}{P^{\rm lin}(k)}\sqrt{\frac{P^{\rm lin}(k)}{2P_{15}(k)}}\sinh\left[  \sqrt{ \frac{2P_{15}(k)}{P^{\rm lin}(k)} }     \right].
\end{equation}
This is the general form for ${\cal N}_2$. However, at large scales, $P_{15}<0$, and this expression becomes,
\begin{equation}
{\cal N}_2(k)= \cos\left[  \sqrt{ \frac{2|P_{15}(k)|}{P^{\rm lin}(k)} }     \right] + \frac{P_{13}(k)}{P^{\rm lin}(k)}\sqrt{\frac{P^{\rm lin}(k)}{2|P_{15}(k)|}}\sin\left[  \sqrt{ \frac{2|P_{15}(k)|}{P^{\rm lin}(k)} }     \right].
\end{equation}
We will refer to Eq. \ref{RPT_hector2} as RPT-${\cal N}_2$ model. 

Note that if we perform on $P_{15}(k)$ in Eq. \ref{N2} the approximation, 
\begin{equation}
F_5^s({\bf k}, {\bf q}_1, -{\bf q}_1,{\bf q}_2,-{\bf q}_2)\simeq \frac{1}{5!} 3!F_3^s({\bf k}, {\bf q}_1,-{\bf q}_1) 3!F_3^s({\bf k}, {\bf q}_2,-{\bf q}_2),
\end{equation}
we obtain that $2P_{15}\rightarrow\left(P_{13}/P^{\rm lin}\right)^2P^{\rm lin}$ and therefore ${\cal N}_2\rightarrow{\cal N}_1$. In the same way, when we apply the Zel'dovich approximation on the kernel of $P_{13}$,
\begin{equation}
F_3^s({\bf k}, {\bf q}, -{\bf q})\simeq\frac{1}{3!}\frac{{\bf k}\cdot{\bf q}}{q^2}\cdot\frac{-{\bf k}\cdot{\bf q}}{q^2},
\end{equation}
we obtain that $2P_{13}/P^{\rm lin}\rightarrow -k^2\sigma_v^2$, and therefore, ${\cal N}_1\rightarrow{\cal N}_0$.
Note that for the computation of ${\cal N}_2(k)$ one needs to compute $P_{15}(k)$ which requires the knowledge of $F_5^s$. This computation is a 6-dimensional integral that reduces trivially to 5-dimensional exploiting rotational invariance. In principle, one could integrate analytically the remaining 3 angles and reduce the computation of $P_{15}$ to a 2-dimensional integral in the same way $P_{13}$ is reduced to a 1-dimensional integral in Eq. \ref{f_sc}. However this is  hard, because the symmetrized kernel $F_5^s$ is the sum of $5!=120$ different cyclic permutations. A possible alternative, is to precompute the angular part of $P_{15}$ as a 3-dimensional integral for a wide range values of $k$, $q_1$ and $q_2$, and then use this to compute $P_{15}$ as a 2-dimensional integral for any shape of $P^{\rm lin}$. 
Nevertheless, for practical reasons, in this paper $P_{15}$ is always computed numerically as a 5-dimensional integral.
For completeness we also report the expression for the 3-loop resummed propagator, ${\cal N}_3(k)$ function as function of the full-propagator terms $P_{13}$, $P_{15}$ and $P_{17}$,
\begin{eqnarray}
\label{N3}{\cal N}_3(k)\equiv\frac{1}{3}\left\{  A\left[   \sqrt[3]{\frac{6P_{17}(k)}{P^{\rm lin}(k)}}   \right]  \right. &+& \left. \frac{P_{13}(k)}{P^{\rm lin}(k)}  \sqrt[3]{\frac{P^{\rm lin}(k)}{6P_{17}(k)}}B\left[   \sqrt[3]{\frac{6P_{17}(k)}{P^{\rm lin}(k)}}   \right] + \right. \\
\nonumber &+& \left. \frac{2P_{15}(k)}{P^{\rm lin}(k)}\sqrt[3]{\left( \frac{P^{\rm lin}(k)}{6P_{17}(k)} \right)^2}C\left[   \sqrt[3]{\frac{6P_{17}(k)}{P^{\rm lin}(k)}}   \right] \right\},
\end{eqnarray}
where the functions $A$,$B$ and $C$ are given by,
\begin{eqnarray}
A(x)&\equiv& \exp(x)+2\exp\left(-\frac{x}{2}\right)\cos\left(\frac{\sqrt{3}}{2}x\right),\\
B(x)&\equiv& \exp(x) - \exp\left(-\frac{x}{2}\right)\left[  \cos\left(\frac{\sqrt{3}}{2}x\right)+\sqrt{3}\sin\left( \frac{\sqrt{3}}{2}x \right) \right],\\
C(x)&\equiv& \exp(x) - \exp\left(-\frac{x}{2}\right)\left[  \cos\left(\frac{\sqrt{3}}{2}x\right)-\sqrt{3}\sin\left( \frac{\sqrt{3}}{2}x \right) \right].
\end{eqnarray}
We do not use this function in this paper, because  it requires the computation of $P_{17}$ which is a 8-dimensional integral (after exploiting rotational invariance), which goes beyond the scope of this paper. We leave the analysis of this function for a future work.

\end{document}